\documentclass{elsarticle}
\usepackage[utf8]{inputenc}
\usepackage[T1]{fontenc}
\usepackage{lineno,hyperref}
\modulolinenumbers[5]

\journal{Journal of Information Sciences}

\usepackage{latexsym}
\usepackage{amsmath}
\usepackage{tabularx}
\usepackage{color,colortbl}
\definecolor{Gray}{gray}{0.9}
\usepackage{multirow}
\usepackage{setspace}
\usepackage{rotating}
\usepackage{algorithm}
\usepackage{algpseudocode}
\usepackage{todonotes}
\usepackage{doi}

\graphicspath{{./}{./fig/}}
\newcommand{\con}{\wedge} 
\newcommand{\imp}{\Rightarrow} 
\usepackage[normalem]{ulem}
\newcommand{\new}[1]{#1} 
\newcommand{\era}[1]{} 
\newcommand{\newrr}[1]{#1} 
\newcommand{\newrrr}[1]{#1} 
\newcommand{\erarrr}[1]{} 
\begin{document}
\hspace{-2.5cm}
\framebox{
\begin{minipage}{1.3\linewidth}
\begin{center}
{\large
\vspace{.5cm}
This document is a pre-print copy of the accepted article for\\ 
\textbf{\emph{Information Sciences}}\\
2024,
vol.\ 653,
no.\ 119772,
pp.\ 1--31\\
\mbox{}\\
\textcolor{blue}{\texttt{\url{https://www.sciencedirect.com/science/article/abs/pii/S0020025523013579}}}
\vspace{.5cm}
\mbox{}\newline
\textcolor{blue}{\texttt{\doi{10.1016/j.ins.2023.119772}}}
\vspace{.5cm}
}
\end{center}
\end{minipage}
}\\

\pagenumbering{gobble}
\setcounter{page}{0}

\begin{frontmatter}

\title{Ambient-aware continuous aid for mountain rescue activities}

\author{Rados{\l}aw Klimek}
\address{AGH University of Krakow,\\ al.\ A.~Mickiewicza 30, 30-059 Krakow, Poland}


\cortext[mycorrespondingauthor]{Corresponding author}
\ead{rklimek@agh.edu.pl}


\begin{abstract}
Ambient-awareness in conjunction with pervasive computing is a significant challenge for system designers. 
It follows the necessity of gathering raw, massive and heterogeneous environmental data \newrrr{which we} obtained, 
while middleware processes must merge context modelling and reasoning seamlessly.
We proposed a system supporting mountain rescuers 
which is demanding due to the large number of environmental objects interacting,
as well as high data variability.
We presented complex context processing embedded in 
the proposed context life cycle and implemented it \erarrr{following a proposed workflow for a demanding}\newrrr{in a difficult} mountain environment.
We introduced five weather scenarios which are a basis for contextual and perceptual processing
during the validation of our model.
The system \erarrr{binds together} \newrrr{merges} a message streaming broker for massive data transport,
low and high-level processing algorithms, repositories and a logical SAT solver.
It constitutes a Context-Aware-as-a-Service (CAaaS) system,
offering advanced support for mountain rescue operations.
The provided software model defines middleware components
which act on a predicted context and transform in situ sensor data into smart decisions,
and which could operate as a platform-based cloud computing model.
It is an enabler yielding a synergy effect with different software components orchestration 
when providing pro-activeness and non-intrusiveness concerning smart decisions.
\end{abstract}


\end{frontmatter}

\pagenumbering{arabic}

\section{Introduction}
\label{sec:introduction}

\subsection{Stage planning and motivations}
\label{sec:motivation}

Complex and networked intelligent information systems have become a norm for computing.
Context-aware applications adapt behaviours to changing environments, 
which is substantial \erarrr{to}\newrrr{for} monitored objects.
They operate on a wide range of contextual data~\cite{Dey-Abowd-1999,Zimmermann-etal-2007},
for example, 
time point or timestamp, 
location, 
device used, 
identity used, 
user name, 
role played, 
due and applicable level of privilege, 
type of activity, 
type of task, 
type of process
and many others,
which can constitute Intelligent Environments (IEs)~\cite{Augusto-etal-2013}
implementing Ambient Intelligence (AmI)\erarrr{~[56]} systems.

\new{Intelligent systems with context-awareness play an important role in information science and the reasons are as follows: 
improved relevance and personalization (filtering and prioritizing information based on users' context, saving time and effort while delivering more intelligent results), 
enhanced decision-making (leveraging context-awareness to support decision-making processes while delivering more accurate and tailored contextual data), 
adaptation to changing environments (sensing changes in the context and adjusting their behaviour accordingly), 
improved user experience (anticipating user needs and \erarrr{proactively} providing relevant support \newrrr{proactively}, 
\erarrr{leading}\newrrr{ensuring a} more seamless user experience and reducing its cognitive load)
and efficient resource utilization (considering \newrrr{the} contextual factors of a system, we can \erarrr{intelligently} manage resources and reduce costs \newrrr{intelligently}). 
These points provide a clear justification that context-awareness and intelligent systems are critical for information science.}

This also applies to mountain environments and rescue operations which are challenging.
The notion of a context, 
\era{first} defined \new{for the first time} in a seminal paper by Dey and Abowd~\cite{Dey-Abowd-1999},
means, in short, any information useful in describing the surroundings of the object,
which does not meet the expectations of the present times and needs.
Its practical use requires at least categorisation,
see Zimmerman et al.~\cite{Zimmermann-etal-2007} as an example,
which allows to interpret and manage the context complexity.
We \era{have} introduced categorisation and a context life cycle for the considered and demanding mountain environments.
Decisions supporting mountain rescuers are taken autonomously and pro-actively
and follow various in situ situations and recommended rescue scenarios.
The raw and heterogeneous sensor data is transformed through system operations into smart decisions.
All algorithms and data transformations are performed in \newrrr{the} middleware,
and all \newrrr{the} processing is invisible to the sensed hiking tourists.
In this way, 
\newrrr{the} middleware provides a \emph{Context-Aware-as-a-Service (CAaaS)} system.
This service can be \era{rented} \new{leased} or delivered as an integrated solution, 
for example, \erarrr{through}\newrrr{by means of} network connections.
There are \erarrr{many}\newrrr{numerous} advantages of such an approach,
for \erarrr{example}\newrrr{instance}: 
increased efficiency,
lower costs,
limited risk,
easy management,
and others,
see Subsection~\ref{sec:service} for additional information in this regard.
Hence, 
it increases the attractiveness of the proposed approach.

To sum up:
our motivation is a system that is characterised by 
pro-activeness, 
non-intrusiveness and 
scalability for demanding mountain environments, 
which both supports mountain rescuers and recognises hiking tourist threats,
while the system receives large, raw and heterogeneous data streams from various sources,
for example:
weather stations, avalanche threats,
BTS (Base Transceiver Station) tourist geolocations, 
GPS (Global Positioning System) tourist geolocations, 
GPS animal geolocations, 
and some other control signals.

\subsection{Contributions}

The first \newrrr{objective and} contribution\erarrr{,
as well as the objective,}
is to propose and organise the processing of 
massive and heterogeneous contextual data streams in 
a demanding mountain environment to obtain \erarrr{the}\newrrr{a} system that aids rescue operations.
The appropriate context life cycle \era{has been} \new{was} established and
adapted to \newrrr{the} rescue system requirements and the specificity of the \erarrr{considered} domain \newrrr{under consideration}.
It is expressed in the proposed workflows including contextual data categorisation.
Categorisation is an enabler in understanding the specificity of the mountain rescuers' activity
and in reducing data complexity.
Moreover, the introduced categorisation is characterized by contextual data encapsulation, 
which, combined with the heterogeneity of data streams, 
enables the parallelization of a context life cycle and better \erarrr{tasks}\newrrr{task} organisation.
The proposed context life cycle defines an integrated framework which is deployed by its workflow.
The approach also allows to prioritise activities,
including system requirements,
assigning threat levels to particular categories and introduced hierarchy.
The achievement for the proposed contextual data processing is 
the \newrrr{on-line detection of} hiking tourist threats \erarrr{on-line detection} with 
the opportunity of \newrrr{acquiring tourist} behavioural traces\erarrr{ acquisition}.
The identified threats create a formal language,
which, owing to \era{managing the complexity of} \new{the management of} contextual data \new{complexity},
becomes a regular language.
Behavioural traces are another regular language.
We \era{have} also \erarrr{showed}\newrrr{demonstrated} the disciplined semantic transformations (metaphors) of contextual data between domains.
All these findings were verified theoretically or experimentally.
We think that the contextual data processing presented in this article 
can be a benchmark for other smart space projects, 
not to mention the fact that, 
to the best of our knowledge, 
this is the first such an approach in relation to mountain environments and rescuers.

The second objective and contribution is to verify holistically the designed and implemented system 
supporting mountain rescue operations.
It is accomplished through a series of intensive experiments.
Only individual and hypothetical system operations 
(message brokers and SAT solvers~\cite{Klimek-2018-Access} operations) 
were tested previously,
moreover, it was not based on real data,
but on datasets generated randomly and separately for particular components.
Recently,
apart from the supporting system itself, 
a prototype of a mountain environment simulator has been built.
It mirrors the key essential aspects of typical mountain environments and hiking tourists. 
In our opinion, 
we obtained a lot of ``real'', even \erarrr{if}\newrrr{though} generated artificially, 
input data streams,
which reflect real mountain conditions.
They are based on five predetermined but arbitrary weather scenarios.
The generated source data streams \era{have been} \new{were} subjected to intense experimentations,
including, first of all, 
a comprehensive evaluation of the implemented system supporting mountain rescuers' operations. 
Consequently, 
what is important for this target validation,
we can examine many interesting properties typical for context-aware systems.
These properties may \era{be} \new{include}: 
redundancy,
spatial proximity,
context transition,
context sharing, 
and others.
Such properties and features are essential and typical for intelligent systems, 
and the image obtained proves 
the naturalness of the system operations and the correctness of the research and results,
which is one of the \newrrr{objectives of this} article\erarrr{ objectives}.

The third \newrrr{objective and} contribution \erarrr{and objective} is to propose and to demonstrate 
the operational capacities of an intelligent, context-aware and pro-active system supporting rescue operations, 
which operates following the cloud computing concept as a CAaaS component
and within a software middleware layer. 
By providing a specific CAaaS service, we can offer \erarrr{rescuers} potentiality \newrrr{to rescuers} on which 
they can build their configurations for \newrrr{the} actual use, 
for example, their specific mountain environments.
The adopted context life cycle is encapsulated in CAaaS.
The CAaaS approach is characterised by 
the high versatility and the numerous possibilities of further applications. 

Summarising this subsection,
we are going to show that a sensor-based context-aware aiding system for demanding mountain environments,
when taking sensor weather data streams into account,
along with the sensor geolocations of \newrrr{the} monitored tourists calculated via different data sources,
can model and utilize effectively the massive and heterogeneous contextual data,
including its hierarchy, 
recognising human threats and behaviours,
\erarrr{to support}\newrrr{supporting} rescue activities effectively,
\newrrr{when} operating as a CAaaS system.
To put it \erarrr{another way}\newrrr{in other words}, 
this research covers human activity recognition and behaviour understanding,
with the objectives to detect threats within the \erarrr{resulting}\newrrr{resultant} much safer environment.
The system can work in large-scale mountain ranges with (multi-stakeholders or) many tourists.
We \era{conclude this} \new{draw this conclusion} because we \era{have} \new{did} not \era{encountered} \new{encounter} any efficiency problems during the simulation; 
therefore, 
increasing simulation parameters (the number of tourists, more routes, more sensory data, etc.)
should not limit the system. 
Not to mention that it is acceptable and natural for rescuers to receive the output current situation for 
the monitored area every half a minute or a minute, not more frequently,
and the typical threat detection lasts seconds, or shorter.
Thus, 
we \era{conclude} \new{presume} that the system is feasible -- all the technical aspects are achievable,
the system is also reliable -- consistently good in performance,
the system is vital -- necessary for \newrrr{the assistance of} mountain rescue operations\erarrr{ assistance},
and to sum up,
all these aspects are authenticated.
We think that
both the supporting system and the article has also an educational value,
showing a great potential for analysing and managing complex contextual data. 

\erarrr{In other words,}\newrrr{By way of explanation,} 
\newrr{our context-aware system operating pervasively in a demanding mountain environment processes 
(collects, filters, analyses, prioritises, infers, broadcasts) 
rich contextual information supporting decision-making operations and rescuers' activities 
depending on the tourists' geolocations and the prevailing circumstances. 
It underscores the pivotal role of contextually analysed data in influencing reasoning procedures that 
bolster the capabilities of rescuers. 
It harnesses contextual insights to steer rescuers through capricious weather variations. 
We think that it spotlights noteworthy research findings,} \erarrr{showcasing}\newrrr{promoting} \newrr{both 
the pragmatic and intriguing facets of the study.}

\new{Hence,
our research fits well into the classic issues of information science,
as defined by Stock and Stock~\cite{Stock-Stock-2013},
that is
the analysis (e.g.\ Figures~\ref{fig:automaton}, \ref{fig:simulation-context-transitions-time}--\ref{fig:simulation-context-sharing}), 
collection (e.g.\ Figures~\ref{fig:context-life-cycle-basic}, \ref{fig:context-life-cycle-workflow}),
classification (e.g.\ Figure~\ref{fig:context-model-categories}), 
manipulation (e.g.\ Figures~\ref{fig:context-life-cycle-basic}, \ref{fig:context-life-cycle-workflow}, \ref{fig:SAT-solver-responses}, \ref{fig:summary-processing}), 
storage (e.g.\ Figures~\ref{fig:data-transformations}, \ref{fig:context-life-cycle-workflow}), 
retrieval (e.g.\ Figures~\ref{fig:context-life-cycle-basic}, \ref{fig:context-life-cycle-workflow}), 
movement (e.g.\ Figure~\ref{fig:context-data-transformation})
and 
dissemination (e.g.\ Figure~\ref{fig:context-data-transformation})
of information.
The article deals with the processing of contextual data streams which is crucial for intelligent environments.
The research objectives are concise and clearly stated and 
the methodology employed is carefully planned and explained, thus it can be replicated successfully.
The experiments carried out were based on reliable data discussed and explained with domain experts or
based on acquired in situ real data.
The findings, as well as the implications and applications are evaluated in detail.
A meta-analysis provides a comprehensive overview of the state of knowledge
concerning contextual processing, 
identifying patterns and discrepancies.}

\erarrr{Finally yet importantly, 
surrounding this system's capabilities should cover a wide range of areas benefiting from 
its approach and implementation.}\newrrr{And finally yet importantly, the potential for employing the presented approach and system appears to be promising.}
\newrr{Notably, it holds the potential in intelligent disaster management. 
For example, in safeguarding forests from wildfires, 
the system's adaptability and predictive capabilities that follow 
context analysis could play a pivotal role when identifying fire-prone zones.} 
\newrr{Similarly, in maritime disaster management, such as oil spills, 
the real-time contextual data analysis aids rapid decision-making using information from 
such sources as satellite imagery and others. 
This highlights the system's potential in different contexts, 
recognizing its capacity to transform to various sectors}. 
\newrr{Both examples underscore its adaptability and smart decision-making for 
diverse issues.} \erarrr{and acknowledge its potential to remodel various sectors addressing critical challenges beyond the immediate context.}
\newrrr{It also acknowledges the potential to remodel various sectors, or cases, addressing critical challenges that arise from these cases.}

\subsection{Overview}

The main issues addressed in this paper are as presented below.
Section~\ref{sec:preliminaries} 
reiterates all the basic assumptions and objectives connected with 
the idea of monitoring mountain areas to support
the everyday work of mountain rescuers,
defining threats within their hierarchy and threat reaction levels.
Then, 
two main sections of the paper follow,
that is Section~\ref{sec:context-modeling-utilization} and~\ref{sec:mountain-and-simulation}.
Firstly,
Section~\ref{sec:context-modeling-utilization}
\era{shows} \new{presents} all the issues related to context processing.
It includes the basic system architecture,
context life cycle,
its workflows,
regular language relating to threats,
behaviour understanding,
some preliminary simulation experiments concerning the proposed workflow,
as well as the idea of closing the whole processing as CAaaS.
Subsequently,
Section~\ref{sec:mountain-and-simulation}
shows all the primary simulation experiments done for the designed system supporting rescue operations. 
After presenting the basic assumptions, 
the \era{obtained} results \new{obtained} were shown and discussed carefully. 
They illustrate both the simulation process performance, 
as well as \era{showing} the data which is characteristic for context-aware systems, 
\newrrr{and} which may be a valuable source of information for further analyses. 
Section~\ref{sec:discussion} 
highlights and summarizes the most important results obtained.
Section~\ref{sec:related-works}
discusses a considerable volume of literature which has been published on the topics covered.
Section~\ref{sec:conclusions} 
concludes the article and outlines research trends for the future.

\begin{figure*}[!htb]
	\centering
\begin{tabular}{lcr}
\begin{minipage}[c]{2.4cm}Mountain\\ rescuers\end{minipage}
& 
\begin{minipage}[c]{6cm}\includegraphics[width=1.0\columnwidth]{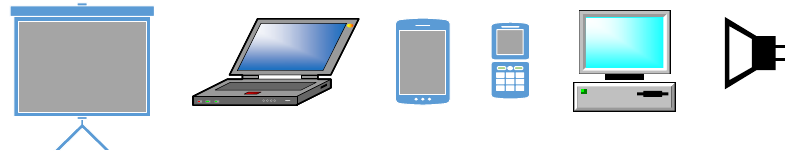}\end{minipage}
&
\begin{minipage}[c]{2.8cm}External\\ clients\end{minipage}\\
\\
\hline
\\
\begin{minipage}[c]{2.4cm}Apps\end{minipage}
& 
\begin{minipage}[c]{6cm}\includegraphics[width=1.0\columnwidth]{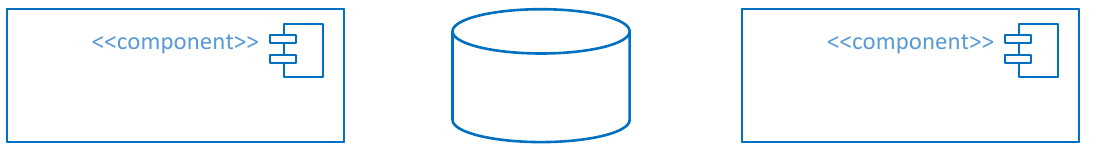}\end{minipage}
&
\begin{minipage}[c]{2.8cm}Context reasoning\\ engine\end{minipage}\\
\\
\hline
\begin{minipage}[c]{2.4cm}Network\end{minipage}
& 
\begin{minipage}[c]{6cm}\includegraphics[width=1.0\columnwidth]{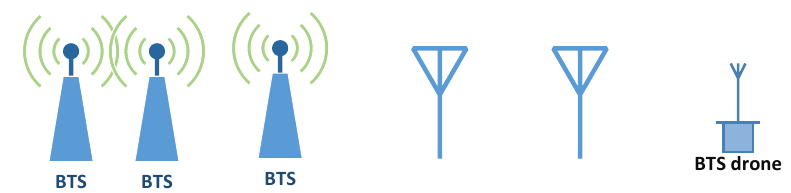}\end{minipage}
&
\begin{minipage}[c]{2.8cm}Publish/subscribe\\ mechanism\end{minipage}\\
\\
\hline
\\
\begin{minipage}[c]{2.4cm}Mountain\\ environments\end{minipage}
& 
\begin{minipage}[c]{6cm}\includegraphics[width=1.0\columnwidth]{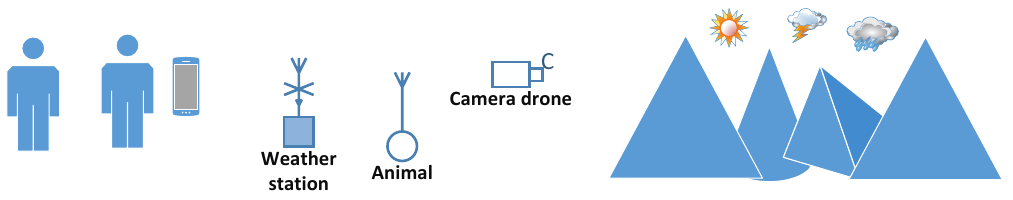}\end{minipage}
&
\begin{minipage}[c]{2.8cm}Sensors \&\\ devices \end{minipage}
\end{tabular}    
    \caption{Initial research project perspective or a layered view of the problem as its motivation.
    (Mountain environments which are saturated with sensors \& devices, 
    a network which provides a publish/subscribe mechanism,
    apps which deploy a context reasoning engine and
    mountain rescuers which are external clients)}
    \label{fig:introduction}
\end{figure*}

\new{And} last but not least,
the article is \erarrr{held together by}\newrrr{summarized in} two figures:
Figure~\ref{fig:introduction} shows the starting point and motivation 
for this research,
while Figure~\ref{fig:summary-processing} (in Section~\ref{sec:discussion})
constitutes a kind of a \erarrr{summary}\newrrr{holistic view} and the end point \newrrr{of the project}.

We use the following notations in the~article:
\begin{itemize}
\item
components diagrams of UML, 
see~\erarrr{[11]}\newrrr{\cite{Pender-2003}} -- 
Figures~\ref{fig:caaas} and~\ref{fig:components-two},
\item
workflows and data flow diagrams,
see~\erarrr{[46,69]}\newrrr{\cite{White-2004}} -- 
Figures~\ref{fig:context-life-cycle-basic}--\ref{fig:context-life-cycle-workflow},
\item
regular expressions, automaton and formal language,
see~\cite{Hopcroft-etal-2006} -- 
Equations~\ref{for:algorithms-loop} and~\ref{for:threat-language}, 
Subsection~\ref{sec:threats},
Figure~\ref{fig:automaton}.
\end{itemize}

\section{Preliminaries}
\label{sec:preliminaries}

The main assumptions of the system,
especially the ones connected with the adopted context model,
as well as the proposed threat levels will be presented.
Some pieces of information provided below in this section are based 
on~\cite[Section III]{Klimek-2018-Access},	
where it is possible to find more details.
On the other hand, 
the information presented here \era{has been} \new{was} clarified. 
The context model is more concise. 

\begin{figure*}[!htb]
	\centering
    \includegraphics[width = 0.7\columnwidth]{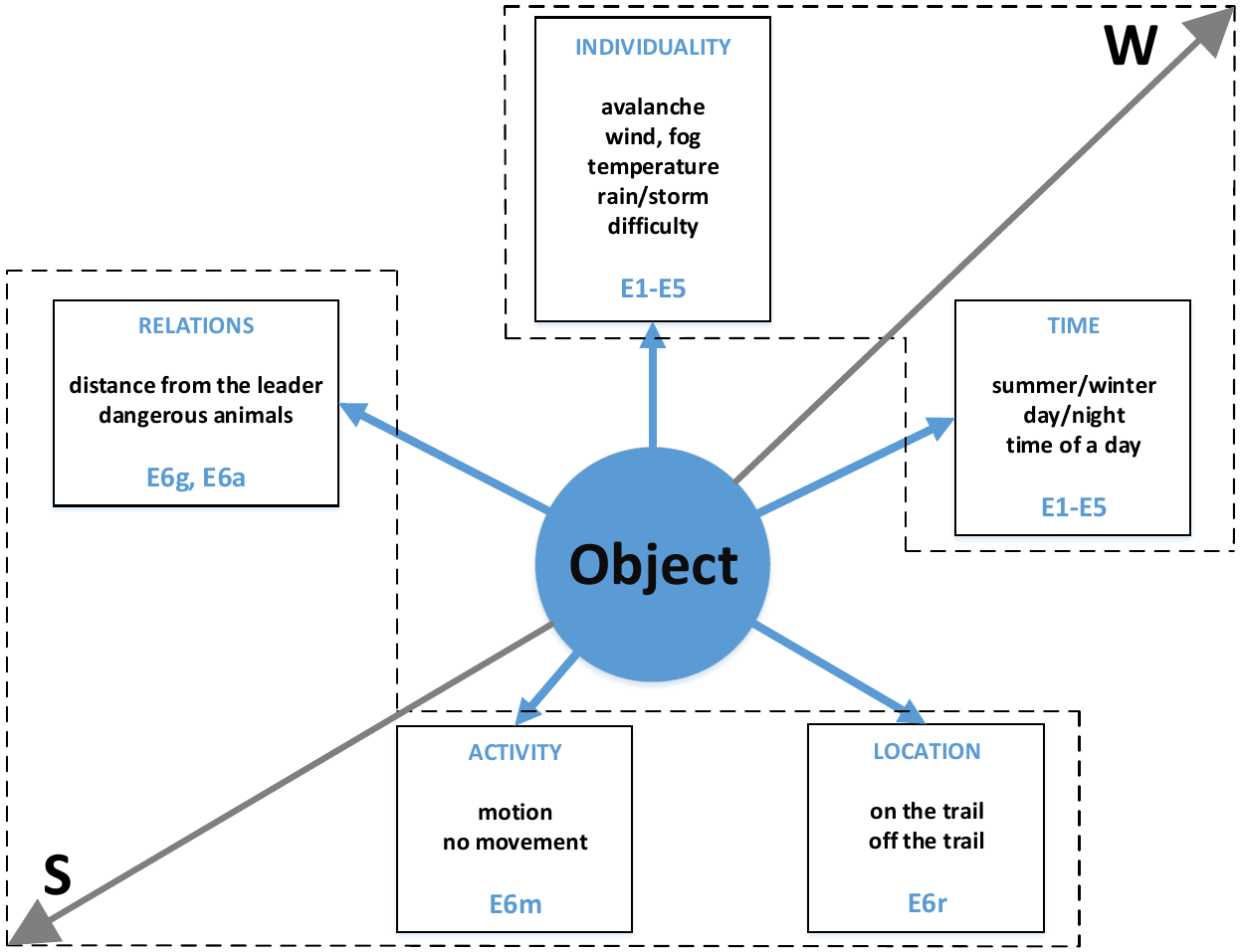}
    \caption{The assumed context model or the categories of \new{contextual} information
    (with the basic threat levels presented, 
    see Tab.~\ref{tab:threat-levels}),
    see also~\cite{Klimek-2018-Access},
    as well as Fig.~\ref{fig:preliminary-simulation-influence-categories},
    within two constituents $W$ and $S$}
    \label{fig:context-model-categories}
\end{figure*}

Figure~\ref{fig:context-model-categories} illustrates the basic context model of the system,
see also a simplified and essentially different case in~\cite{Klimek-2020-Sensors}.
Thus,
the \emph{context} 
is any information which describes the situation of a monitored and analysed object. 
The \emph{category} is a group of data that have similar features.
Firstly,
we establish all \newrrr{the} \era{of the} information categories which influence the object and 
the course of the system operations,
and which also affect the interaction between them.
When analysing each category, 
the potential threats were \era{clearly} indicated \new{clearly}.
On the other hand, 
the relations between the categories are rather weak, 
which is an intended designing process to get a strong encapsulation,
see Table~\ref{tab:context-categories-relationship}.

\begin{table}[!htb]
\caption{The assumed levels of difficulties for weather contextual data $W$, see also~\cite{Klimek-2018-Access}}
\centering
\setstretch{1.05}
\begin{tabularx}{1\columnwidth}{l|l|c|X}
\hline
\multicolumn{2}{c|}{Considered context} &
Labels/Tags &
Description\\
\hline\hline
\multicolumn{2}{c|}{Avalanche}&
\textbf{A1 -- A5} &
$\underrightarrow{\mbox{Increasingly difficult conditions}\strut}$\\
\hline
\multirow{4}{*}{
\begin{sideways}
\begin{tabular}{c}
Weather\\
conditions
\end{tabular}
\end{sideways}
} &
Wind force &
\textbf{W1 -- W3} &
\multirow{4}{*}{
\hspace{-.5cm}
\begin{tabularx}{6.1cm}{X}
$\underrightarrow{\mbox{Increasingly difficult conditions}\strut}$
\end{tabularx}
}\\
\cline{2-3}
&
Fog, visibility &
\textbf{F1 -- F3} &
\\
\cline{2-3}
&
Temperature &
\textbf{T1 -- T3} &
\\
\cline{2-3}
&
Rain, storm &
\textbf{R1 -- R3} &
\\
\hline
\multicolumn{2}{c|}{Difficulty levels} &
\textbf{D1 -- D4} &
$\underrightarrow{\mbox{Increasingly difficult trails}\strut}$\\
\hline
\end{tabularx}
\label{tab:weather-context}
\end{table}

We \era{have} separated and, consequently, 
divided all the contextual data into two groups or classes.
The first class is \emph{weather context} $W$ and 
the second one is \emph{non-weather context} $S$.
The weather context consists of many variables related to weather conditions, 
see Table~\ref{tab:weather-context}. 
A set of tourist trails was predefined: 
H1, H2, $\ldots$ 
In general, 
there is no limit 
\erarrr{when it comes to}\newrrr{in} the number of trails which are inside the monitored mountain area. 
One must note that 
the crucial aspect of each trail is its difficulty. 
Difficulties are pre-defined at difficulty levels. 
Rescuers, as domain experts, 
classify difficulty levels related to each trail. 
Thus, $d(h)$ is a \emph{difficulty level} for each fixed trail $h$ in the monitored area, 
that is
$\{H1, H2, ...\} \rightarrow \{D1, D2, D3, D4 \}$. 
Avalanche risks are set manually by rescuers. 
All the contextual data which follows immediately from raw sensor data, 
after the filtration process, is available for further reasoning purposes.

\begin{table}[!htb]
\caption{The assumed levels of threats, \erarrr{shown increasingly}\newrrr{ascending} in the case of weather
(all predefined weather threats are illustrated with a symbol, colour and danger name, 
all predefined non-weather threats are illustrated with a symbol, colour and surrounding shape).
The meanings of threat symbols: E1 -- normal monitoring,
E2 -- disadvantageous conditions, monitoring, 
E3 -- notification, camera drone, intensified monitoring,
E4 -- possibility of a rescue operation,
E5 -- rescue action necessary,
E6g -- too far from the leader, notification,
E6a -- too close to a dangerous animal, notification,
E6m -- prolonged time without any \era{movement} \new{motion}, notification, camera drone, rescuers intervention,
E6r -- off-trail hiking, notification, camera drone, rescuers intervention.}
\centering
\begin{tabularx}{1\columnwidth}{l|lllll}
\hline
\multirow{3}{*}{
\begin{tabularx}{2.8cm}{X}
Weather
\end{tabularx}
}
        & \textbf{E1} & \textbf{E2} & \textbf{E3} & \textbf{E4} & \textbf{E5}\\
        & \textcolor{green}{green} & \textcolor{yellow}{yellow} & \textcolor{orange}{orange} & \textcolor{red}{red} & \textcolor{black}{black}\\
        & low & medium & increased & high & very high\\
\hline\hline
\multirow{3}{*}{
\begin{tabularx}{2.8cm}{X}
Non-weather
\end{tabularx}
}
&            & \textbf{E6g} & \textbf{E6r} & \textbf{E6m} & \textbf{E6a}\\
&            & \textcolor{violet}{violet} & \textcolor{violet}{violet} & \textcolor{violet}{violet} & \textcolor{violet}{violet}\\
&            & pentagon & circle & square & triangle\\
\hline
\end{tabularx}
\label{tab:threat-levels}
\end{table}

There are the levels of particular threats or difficulties established, 
which are intended for both predefined context categories, $W$ and $S$, 
in Table~\ref{tab:threat-levels}.
The pictograms of people who are on the route are marked and visualised
(usually as a colour dot, colours are shown in the table)
on all the devices available to the rescuers,
see also Figure~\ref{fig:primary-simulation-screen-and-scenarios}.
Only E1 signifies the lack of any threat and it is referred to as a normal situation. 
Threats require taking a proper action stipulated by procedures. 
The final decisions are made by the rescuers,
as a result of monitoring the system prompts.
The rescue teams may also influence 
the means of interpreting the pieces of data by using a reasoning component, 
especially in relation to trail difficulties.  
Another example of interpretation is
when it can take place within ``Individuality'' and ``Time'' categories. 
The time of the day (day/night) or the season of the year (summer/winter),
as the elements of the ``Time'' category, 
may influence the ``Individuality'' category by more or less 
restrictive interpretation of weather data.

\section{Context modelling and utilization}
\label{sec:context-modeling-utilization}

The essential aspects concerning context data processing are presented in this section.
Hence,
\begin{itemize}
\item
upon the short presentation of changes and improvements in respect to the former version of the system supporting rescuers,
we demonstrated the internal architecture and organisation of the supporting system as CAaaS (Subsection~\ref{sec:organisation-CAaaS});
\item
we proposed the entire context life cycle and its specification through its detailed workflow (Subsection~\ref{sec:context-processing});
\item
then, we conducted the semantic analysis of contextual data
and we indicated the preservation of some selected properties (Subsection~\ref{sec:context-processing});
\item
we observed that threats detected by the supporting system constitute a kind of a formal language
and we proved this language to be regular;
this language is also a good opportunity to understand the concept of tourist behaviour as presented in this article (Subsection~\ref{sec:context-processing});
\item
we also \erarrr{carried out the}\newrrr{performed an} interesting simulation of the previously proposed workflow, 
and called it a preliminary simulation, 
though within a narrower scope (Subsection~\ref{sec:preliminary-simulation}).
\end{itemize}

The supporting system design~\cite{Klimek-2018-Access} \erarrr{has been}\newrrr{was} implemented \era{now}\new{in this article}.
It \erarrr{has been}\newrrr{was} also improved, 
especially \erarrr{when we determine}\newrrr{for determining} tourist geolocations \newrrr{with the interpretations of geolocations}\erarrr{system} and 
\erarrr{in relation to}\newrrr{for} the coverage of trails with weather stations\erarrr{ and data interpretation}.
When it comes to tourist tracking within the monitored area, 
\erarrr{now, we have}\newrrr{there are} two methods \newrrr{available}.
Apart from setting their geolocations, 
as done previously, 
that is on the basis of data from BTS (Base Transceiver Station) stations, 
another possibility was also introduced by determining 
their location \erarrr{on the basis of}\newrrr{based on} data from a GPS (Global Positioning System) system. 
\erarrr{Simply speaking,}\newrrr{To put it plainly,} now, \erarrr{we have}\newrrr{there are} three possible situations: 
\begin{enumerate}
\item
tourists have modern phones (smartphones) and 
agree to install (for the \erarrr{duration of} monitoring \newrrr{duration}) 
an application which sends their exact GPS position to the system;  
\item
tourists have phones with a GPS module but 
do not agree to this form of tracking -- 
their location is determined by the use of BTS data; 
\item
tourists have normal phones, 
that is without a GPS module -- 
their location is determined by the use of BTS data. 
\end{enumerate}
This kind of improvement seems to be very important
due to its positive influence on the quality of the many system operations.
\erarrr{Another}\newrrr{The remaining} mentioned system changes relate to the means of setting the weather context
through the weather stations' readings, for a particular tourist, 
and it will also be presented below,
see the paragraph right after Formula~(\ref{for:algorithms-loop}).

\subsection{Internal organisation of CAaaS}
\label{sec:organisation-CAaaS}

Figure~\ref{fig:middleware-architecture}
shows the overall architecture of the entire supporting system, 
especially in relation to \newrrr{the} middleware,
when it is organised as a CAaaS component. 
This type of the system organisation is related to work~\cite{Henricksen-etal-2005}.
The second layer from the bottom,
data \erarrr{repositoring}\newrrr{repository},
can be decomposed into 
a separate layer of data collection and a separate pre-processing layer. 
Those types of detailing changes can also appear in the case of other layers. 

\begin{figure}[!htb]
\centering
\setstretch{1.5}
{\normalsize
\begin{tabularx}{1.1\columnwidth}{|l|c|l|X}
\cline{1-1}\cline{3-3}
\cellcolor{lightgray}End-user apps & & \cellcolor{lightgray}Layer4: web-system components & \\
\cline{1-1}\cline{3-3}
\multirow{3}{*}{
\begin{tabular}{l}
\hspace{-.3cm}Middleware
\end{tabular}}
           & & Layer3: context reasoning & A3, A4, SAT solver, A5\\
\cline{3-3}           
           & & Layer2: context repository & $Repository$, $Alerts$\\
\cline{3-3}           
           & & Layer1: context preprocessing & Msg.\ broker, A1, A2\\
\cline{1-1}\cline{3-3}
\cellcolor{lightgray}Physical env. & & \cellcolor{lightgray}Layer0: sensor/BTS/GPS dev. &\\
\cline{1-1}\cline{3-3}
\end{tabularx}
}
\caption{Middleware architecture of the supporting system.
(Right side and bottom-up description:
Msg. broker -- messages/raw data transporting,
A1 -- $PhoneManagement$, 
A2 -- $WeatherReading$, 
$Repository$ -- context data,
$Alerts$ -- predefined reaction/threat levels,
A3 -- $ThreatDetection$, 
A4 -- $CheckThreat$, 
SAT solver -- logical reasoning engine,
A5 -- $ThreatManager$)}
\label{fig:middleware-architecture}
\end{figure}

$Layer0$ sends data from weather stations, 
BTS stations and GPS data from phones. 
The data is transported by a message broker, 
for example RabbitMQ \erarrr{[53]}\newrrr{(see: \url{https://www.rabbitmq.com})},
to $Layer1$ and gathered as raw data. 
Here, 
it undertakes initial processing. 
Algorithm $A1$ establishes a precise geolocation of every phone, 
records the time of the arrival at and the departure from the monitored area, etc. 
(All \new{the} algorithms mentioned here, 
that is A1--A5, 
are \era{carefully} designed \new{carefully} and discussed in~\cite{Klimek-2018-Access},
here they \era{have only been} \new{are only} summarized in Table~\ref{tab:algorithms-summary}.)
\begin{table}[!htb]
\caption{Summary of the algorithms used}
\centering
\begin{tabularx}{1\columnwidth}{l|X}
\hline
Msg.\ broker 
&
raw data transportation\\
\hline
A1
& objects' geolocation with timestamp and route identification, 
arrival and departure times for tourists\\
\hline
A2
& downloading and pre-processing of weather data\\
\hline
A3
& detecting non-weather threats and calls for weather threats\\
\hline
A4
& encoding data for logical reasoning for weather threats\\
\hline
SAT solver
& SAT-based logical reasoning\\
\hline
A5
& initiation of the system, managing other processes\\
\hline
\end{tabularx}
\label{tab:algorithms-summary}
\end{table}
The changes, acknowledged at the beginning of this section, 
were introduced in relation to geolocation. 
The new algorithm also solves the problem of data redundancy in situations 
when the system, in relation to a single phone, 
receives data from both BTS and GPS stations. 
In this case, 
the pieces of data which enable a much more precise geolocation of the phone are chosen. 
At the same time,
BTS data for each phone is always downloaded,
and may be useful in case the GPS data flow interrupts, or stops, for unknown reasons. 
Algorithm $A2$ is responsible for downloading and processing 
weather data, 
which is fundamental in setting a weather context. 

\begin{figure*}[!htb]
	\centering
    \includegraphics[width = .9\columnwidth]{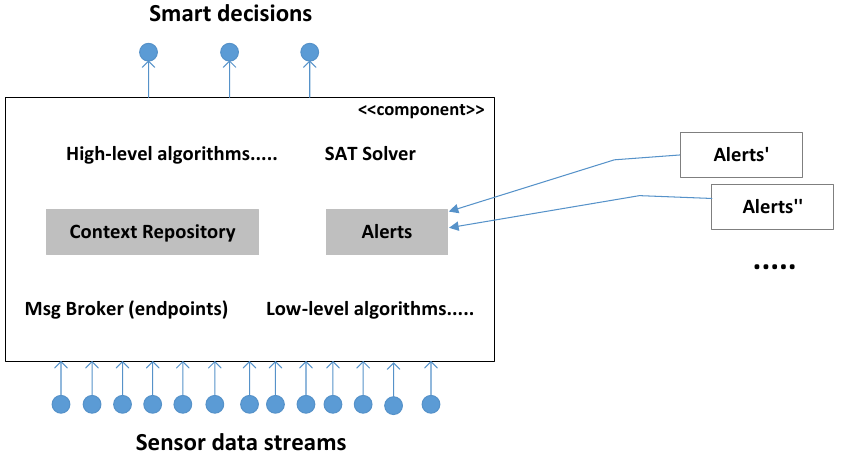}
    \caption{Context-Aware-as-a-Service CAaaS
             or from sensor data streams to smart decisions.
             (For denotations see also Fig.~\ref{fig:middleware-architecture},
              predefined $Alerts'$, $Alerts''$, etc.\ constitute the different sets of data
              or constraints,
              which can be injected, 
              and afterwards they can affect the work of the component providing different alerts)}
    \label{fig:caaas}
\end{figure*}

$Layer2$ manages the repository of the system. 
The $Repository$ possesses all \newrrr{the} context data collected within the monitored area. 
Those pieces of data are updated regularly.
The \newrrr{system} administrator \erarrr{of the system} decides about their frequency but 
it is presumed that it should take place every minute. 
$Alerts$ are determined by rescuers.
They select and define a context level, as a form of trigger, 
in relation to each route difficulty,
which has to be signified by a threat alert. 
It is also possible to set the diversified levels and contexts of reaction, 
different for daylight and night conditions, 
or the changing seasons of the year,
for the same route difficulties,
see also different alerts shown in Figure~\ref{fig:caaas}. 

$Layer3$ is responsible for reasoning processes. 
Algorithm $A3$ detects threats caused by a weather or non-weather context, 
see Table~\ref{tab:threat-levels}. 
In the case of a weather context, it is necessary to 
evoke algorithm $A4$ which prepares data for logical reasoning. 
The data is later sent to a SAT solver,
for example, Sat4j \erarrr{[40]}\newrrr{(see: \url{http://www.sat4j.org})},
and the reasoning process enables us to detect particular threats, 
if present.
The reasoning process is based on the well-known 
\emph{deduction theorem}~\erarrr{[28,29]}\newrrr{\cite{Kleene-2002}}, 
to put it simply, 
for $S \models A$ we have $S \imp A$,
where the implication predecessor is data from weather readings, 
and the successor is an alert, 
see Figure~\ref{fig:caaas}, 
predefined and selected by rescuers.
Then we use a \emph{modus ponens} rule~\cite{Kleene-2002}
to obtain $(S\imp A) \con S$,
which is the basis of logical reasoning processes,
see~\cite[Subsection V.C]{Klimek-2018-Access} for details. 
We have four SAT solver calls for weather threats, 
sequentially for eventually occurring threats from E5 to E2, 
informally we have the following colours: black, red, orange and yellow, respectively,
see Table~\ref{tab:threat-levels} and Figure~\ref{fig:primary-simulation-screen-and-scenarios};
other tourists are in a normal weather situation (E1, green).
Algorithm $A5$ is responsible for the initiation of 
the whole system and it is used later for threat detections, 
which are performed in an infinite loop of calls (infinite looping standby). 
The exemplary sequence of calls, without showing an infinite loop directly, 
can be presented as follows:  
\begin{eqnarray}
A5(\ldots,MsgBroker,\ldots
A1,A2,\ldots,A3(\ldots,
A4(\ldots,SATsolver,\ldots),\nonumber\\
A4(\ldots,SATsolver,\ldots),
A4(\ldots,SATsolver,\ldots),\nonumber\\
A4(\ldots,SATsolver,\ldots),\ldots),\ldots)
\label{for:algorithms-loop}
\end{eqnarray}

It needs to be emphasised that 
the significant improvement of the supporting system design is also connected 
with algorithm $A4$ in relation to weather data. 
Now, the weather context, 
for a reasoning process concerning a particular object moving along the route, 
is determined by finding data from the nearest weather station, 
but not from a single station assigned to the whole route.  
It is currently \erarrr{being} assumed that there is 
a number of weather stations distributed within the monitored area. 
If there are two stations on different routes for 
a moving object within its circle of influence, 
the data from the tourist's route is taken into consideration, 
as long as they do not change their route. 
If there are two stations on the same route, 
the data of the nearest station is analysed. 
If the stations are located in a comparable distance, 
then the data of origin from the station that the tourist is moving towards is used for the analysis. 

\newrr{To sum up, 
it should be noted that intricate decision-making processes within $Layer3$ rely on thorough context analysis and reasoning. 
As a consequence, 
these processes facilitate the accentuation and choice of a specific decision-making scenario aimed at aiding rescuers. 
This, in turn, involves utilizing the contextual information by rescuers in supporting through unexpected weather fluctuations, 
effectively proving the system's practical efficacy.}

$Layer4$ is responsible for presenting the data as 
smart decisions in a clear and easily interpreted form. 
\era{It is usually} \new{Usually, it is} provided  as visualisations and maps,
where the positions of tourists are marked, 
along with any existing threats.

To sum up,  
Figure~\ref{fig:caaas} 
gives \erarrr{both}
a synthetic account on Context-Aware-as-a-Service CAaaS concept, 
which could be located in \newrrr{the} middleware. 
CAaaS is understood as a method of providing services
which are appropriate for cloud computing models. 
This type of processing, 
via generally available networks, 
increases the possibilities of the supporting system by providing 
a service from\erarrr{, and for,} any mountain location \newrrr{and for such a location}. 
Sensory data streams are transported using a message broker which is 
the bloodstream of the system. 
After initial processing and filtering out, 
data is stored in a repository. 
For this service, 
it is possible to use \newrrr{the} different sets of \emph{alert} data
($Alert'$, $Alert''$, $Alert'''$, etc.), 
defining the numerous levels of the weather danger reactions, 
when considering a difficulty level for each trail and local weather conditions,
as well as adding new sets.\erarrr{, which} \newrrr{They} could be predefined on-line by rescuers,
if necessary.
All \newrrr{the} context data, together with alerts, 
becomes the input for a reasoning engine which is the heart of the whole system. 
The process of logical reasoning enables us to achieve smart decisions. 
The process of downloading new sensor data, 
its pre-processing and logical reasoning, 
takes place cyclically in an infinite loop, 
as long as enforcing actions appear.

\subsection{Context processing}
\label{sec:context-processing}

\subsubsection{Context life cycle}

\begin{figure*}[!htb]
	\centering
    \includegraphics[width = 1\columnwidth]{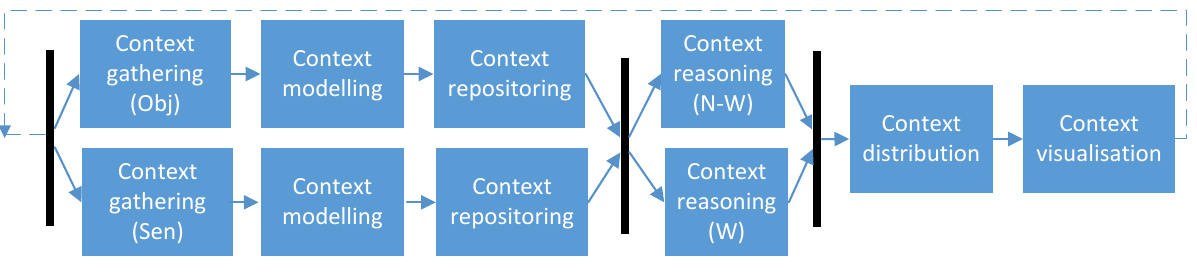}
    \caption{Established context life cycle for the supporting system.
    Gathering geolocation data for monitored objects (Obj) and from sensor stations (Sen),
    and reasoning for non-weather (N-W) and weather (W) threats}
    \label{fig:context-life-cycle-basic}
\end{figure*}   

\erarrr{It is well known}\newrrr{It is common knowledge} that software and its development processes 
form a software life cycle\erarrr{~[60]}, 
but also in the case of contextual data, we can observe a \emph{context life cycle}.
By such a cycle, 
we understand a structure consisting of a sequence of steps or phases, 
related to the processing and transformation of contextual data, 
during which \era{they undergo} \new{it undergoes} a planned metamorphosis,
see Figure~\ref{fig:context-life-cycle-basic}.
Particular stages transform context data
according to the designed algorithms contained in our CAaaS component.
All \newrrr{the} pieces of data are gathered, pre-processed or modelled.
During these operations they are located in repositories.
When logical inference is finished,
they are distributed in various locations and visualised, if necessary.
After a period of time, 
the context data should be updated and the whole process starts again.
Thus,
the context data is updated periodically.

\begin{figure*}[!htb]
	\centering
    \includegraphics[width = 1\columnwidth]{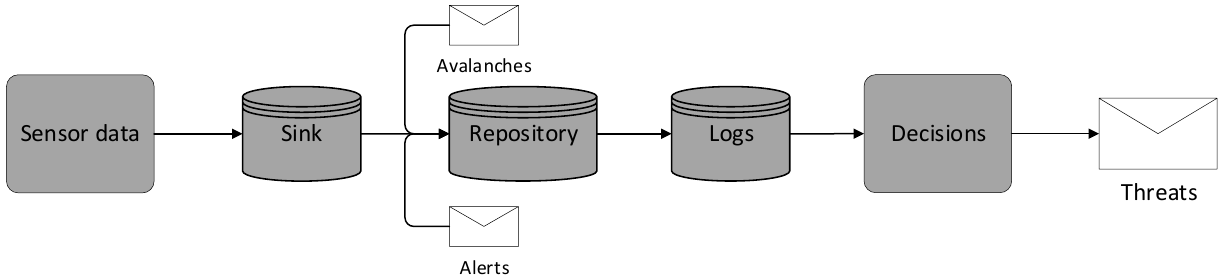}
    \caption{Data transformation diagram of a designed supporting system}
    \label{fig:data-transformations}
\end{figure*}

Figure~\ref{fig:data-transformations}
illustrates graphically the successive transformations of all the key data,
that is conversion from raw sensor data to the output threat information.
Information is stored in multiple intermediate databases which are enriched and, 
after recombination and transformation into new formats, 
information is passed to the final reasoning processes.

\begin{figure*}[!htb]
	\centering
    \includegraphics[width = 1\columnwidth]{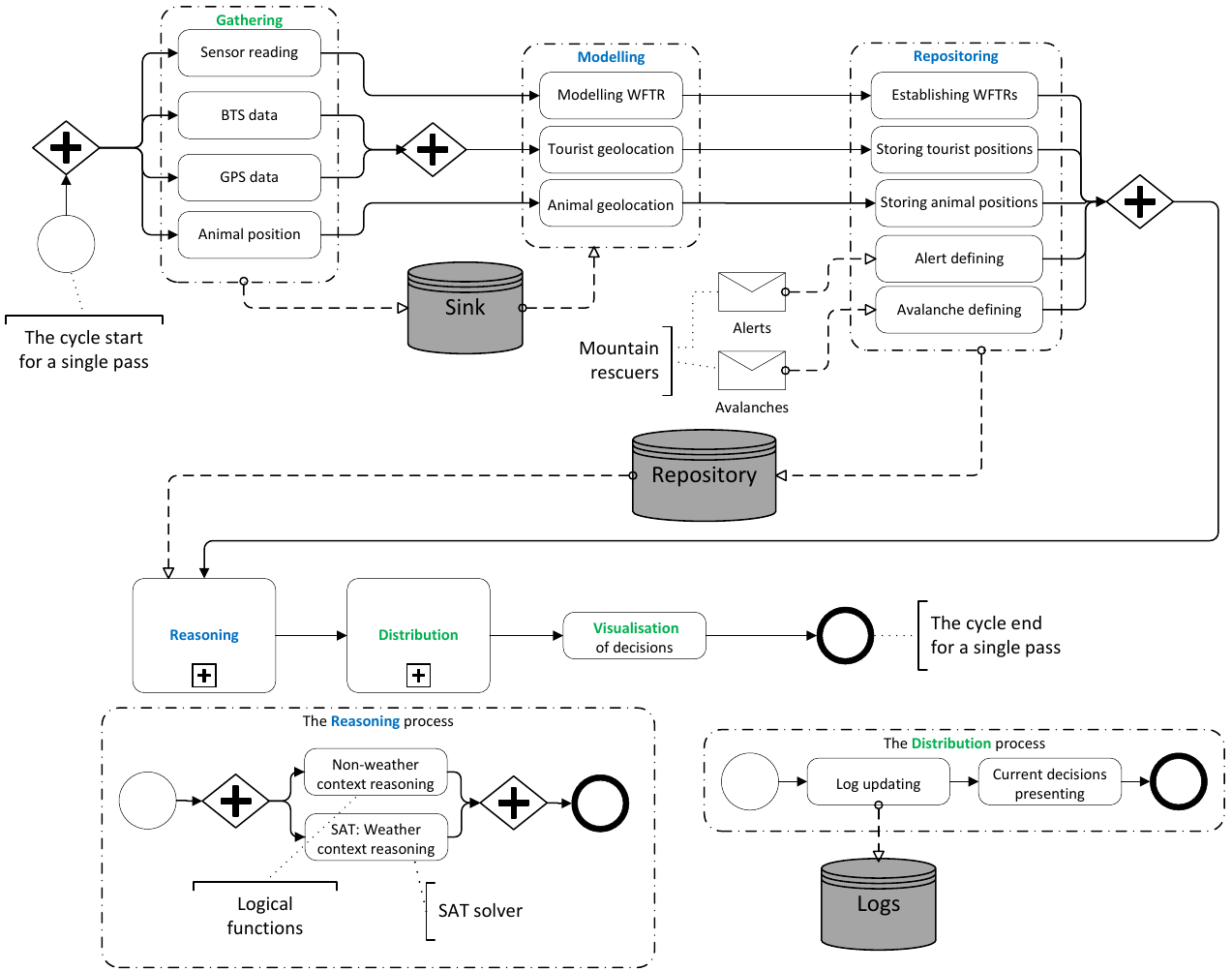}
    \caption{Context processing control loop 
    or a detailed workflow for a single pass for the contextual data processing 
    and its iteration gives a full contextual data lifecycle,
    where
    green names indicate contextual data collection-dissemination activities and
    blue names indicate contextual data domain logic activities.
    (Establishing WFTRs means determining tags for wind, fog, temperature and rain,
             see Tab.~\ref{tab:weather-context})}
    \label{fig:context-life-cycle-workflow}
\end{figure*}

\emph{Workflow} is the series of activities involved to complete a task.
Figure~\ref{fig:context-life-cycle-workflow} 
shows a basic workflow~\erarrr{[46,69]}\newrrr{\cite{White-2004}} 
focused on the activities of contextual data processing.
The graph depicts only those activities which are significant for
understanding the manner in which context data is converted 
and transformed in the whole system.
It presents single transitions in the entire context life cycle.
Figure~\ref{fig:context-life-cycle-workflow}
should be regarded as the detailing of Figure~\ref{fig:context-life-cycle-basic}.
The proposed workflow is focused on all the important aspects of context data processing,
while some others resulting from, for example, Formula~(\ref{for:algorithms-loop}) \era{have been}\new{were} skipped.

As already mentioned, 
there are some databases with various forms of contextual data.
$Sink$ is intended for storing different raw data,
for example, data from weather stations, 
and further geolocations of tourists from neighbouring BTS stations, 
but also the geolocations of tourists obtained from GPS devices, 
and geolocation data from GPS devices mounted on animals.
(Some animals, especially dangerous animals, like bears or wolves, 
may have GPS devices fitted by the employees of the national park. 
It increases the safety of both animals and hiking  tourists.
The source of such an idea \era{were} \new{was} the author's own experience related to 
an unexpected and disturbing encounter with a bear in the mountains,
\era{on the other hand,} \new{moreover,} many park managers started to adopt such solutions.)
Before being moved and tabulated in $Repository$,
this data is both filtered and modelled,
through assigning weather reading ranges
(wind, fog, temperature, rain),
geolocating tourists and animals.
The effective performance of these operations precedes defining\erarrr{, in literal terms, 
see Table~\ref{tab:weather-context},} 
weather conditions regarding each of them \newrrr{in literal terms, see Table~\ref{tab:weather-context},}
and to each established mountain route. 
On the other hand, 
there is also a calculated strict route location of each tourist,
that is every tourist is assigned to a specific mountain route.
(For instance,
weather condition for H2 trail may be described by means of 
weather factor symbols,
e.g.\ W2, F1, T3, R1,
for: wind, fog, temperature and rain, respectively.
What is more, 
each tourist under observation is assigned to a given trail, 
e.g.\ H2.)
Additionally, 
mountain rescuers establish manually the levels of alerts and avalanches.
$Repository$ stores pre-processed contextual data in a tabular form. 
It is the most convenient form of data storage. In the succeeding step 
specific values could constitute the input data for the automatic logical inference, 
that is they produce smart decisions.
Some decisive processes are implemented through the designed dedicated logical functions
(checking the animal proximity, being off the trail, etc.);
nevertheless,
their substantial portion is performed with the use of the embedded SAT solver,
as described in Subsection~\ref{sec:organisation-CAaaS}.
The decisive process results are saved in system logs
but they are also being broadcasted for each tourist in the monitored area,
obtaining separately its individual threat level, if any,
see labels in Table~\ref{tab:threat-levels}.
The different and fitted forms of visualizations cover all the devices available to rescuers,
for example, monitors and terminals in the headquarter, 
as well as the individual smartphones of rescuers.
It is also possible to send simple text messages to \era{at-risk} tourists \new{at risk}.

\subsubsection{Contextual data transformations}
\label{sec:metaphor}

\begin{figure*}[!htb]
	\centering
    \includegraphics[width = .8\columnwidth]{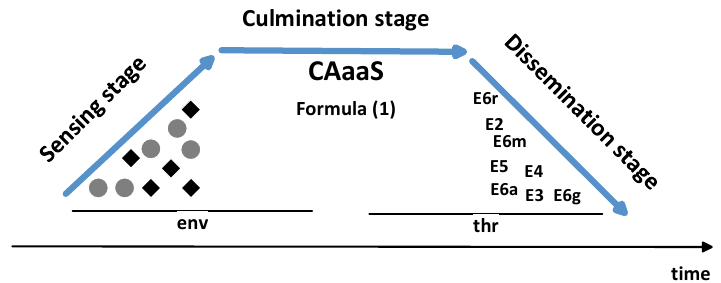}
    \caption{Contextual data transformations.
    (From left to right:
    two shapes symbolise the two types of raw data, 
    weather sensory and geolocation sensory, collected during the sensing stage,
    the processing in CAaaS during the culmination stage,
    and possible various threat signals produced during the dissemination stage.
    Two domains: environmental $env$ and threat $thr$)}
    \label{fig:context-data-transformation}
\end{figure*}

Contextual data is constantly transformed.
\emph{Conceptual metaphor}, or simply \emph{metaphor}, 
invokes and refers to the comprehension of one idea, in a considered domain, in terms of another.
Figure~\ref{fig:context-data-transformation}
shows the transformation of contextual data,
from a raw data stream to more complex data forms.
Metaphors that are transformed from one field to another
are discussed in article~\cite{Aggeri-etal-2007},
see also~\cite{Lakoff-Johnson-1980},
and we will adapt the approach to our context-oriented processing case.
The source domain is environmental data and 
the target domain is threat data.
The source domain $env$ includes \erarrr{the} raw data up to the initial processing.
The target domain $thr$ includes \erarrr{the} data after pre-processing,
\erarrr{by}\newrrr{intended for} further processing, 
until the threat signals are obtained.
Let us consider the following transformation properties to examine semantic properties.
\emph{Causation}
means the ability of mapping some event relation $cause$
between every event $e_i$ and $e_j$,
that is between the source, or environment, domain $env$ and 
the target, or threat, domain $thr$,
which is expressed as:
\begin{eqnarray}
\forall e_i, e_j\,\,
(cause(e_i, e_j)_{env} \mapsto cause(e_i, e_j)_{thr})
\end{eqnarray}
\emph{Velocity}
means that
the qualitative progress rate for a certain event in 
the source domain expressed by relation $velocity$ maps identically
the qualitative progress rate in the target one,
that is
\begin{eqnarray}
\forall e, r\,\,
(velocity\,(e,r)_{env} \mapsto velocity\,(e,r)_{thr})
\end{eqnarray}
where $r$ is a certain rate.
\emph{Time-order}
means that
the sequence of every event $e_i$ and $e_j$,
when $e_i$ $precedes$ $e_j$,
in the source domain is 
the same as in the target domain,
which is expressed \new{by}:
\begin{eqnarray}
\forall e_i, e_j\,\,
(precede(e_i, e_j)_{env} \mapsto precede(e_i, e_j)_{thr})
\end{eqnarray}
\emph{Value-evaluation}
means that the level of goodness evaluated using relation $valuation$
is identical in the source and target domains,
that is
\begin{eqnarray}
\forall e,v\,\, 
(valuation\,(e,v)_{env} \mapsto valuation\,(e,v)_{thr})
\end{eqnarray}
where $v$ is the assumed level of a value.
The semantic interpretation presented above concerning contextual data, 
and the relations therein, 
could be continued and extended with subsequent properties.

It must be stated here that all four properties mentioned above 
(causation, rate, time-order, value-evaluation)
are satisfied in the proposed contextual data processing,
see Figure~\ref{fig:middleware-architecture} or Formula~(\ref{for:algorithms-loop}),
because:
\begin{enumerate}
\item
If causation occurs in its primary environment,
it must also appear after data processing, 
whose goal is to determine such causation in other categories and 
forms of contextual data, starting with raw data processing. 
The entire processing is only \erarrr{to obtain}\newrrr{aimed at obtaining} some information on threats.
\item
Velocity is maintained, which is identical in both domains --
changing a pace and faster or slower occurrence of 
respective threats does not depend on the form of contextual data; 
therefore, 
velocity is the same and the increase or decrease in the rate of 
specific events occurrence is reflected instantly 
in the identical increase or decrease of detection in both domains.
\item
Time-order is obviously preserved in both domains, 
which results from the fact that certain phenomena may take place one after another 
and the form of contextual data is of no importance here --
such phenomena expressed in data are only obtained from them and, 
consequently, the sequence of occurrence transforms from one domain to another.
\item 
Finally, 
value-evaluation is preserved as well, 
because when a given phenomenon takes place in the primary domain, 
then, after contextual data processing, 
it will also occur in the target domain;
if it does not occur, 
it means that it will not occur in both domains at the same time. 
\end{enumerate}

\subsubsection{Threats as a language}
\label{sec:threats}

\begin{figure*}[!htb]
	\centering
    \includegraphics[width = .35\columnwidth]{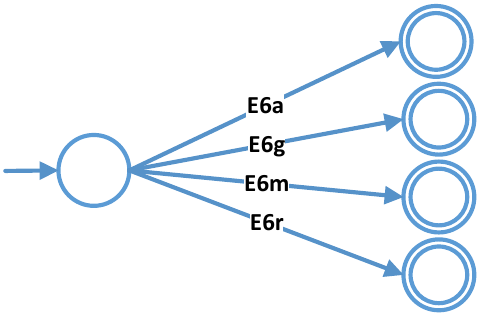}
    \qquad
    \includegraphics[width = .35\columnwidth]{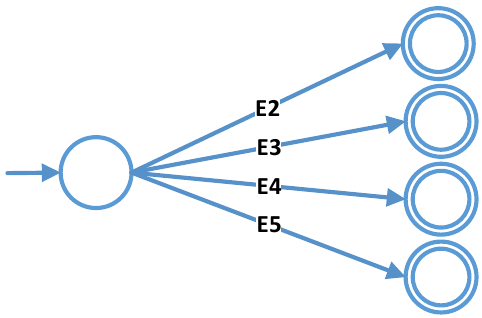}\vspace{4mm}\\
    \includegraphics[width = .7\columnwidth]{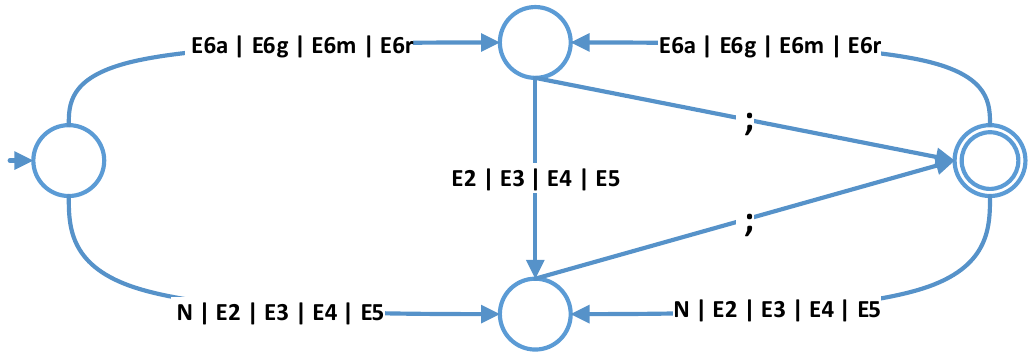}
    \caption{Automaton as an acceptor of the threat language for contextual data processing:
    top left -- for regular expression $S$,
    top right -- for regular expression $W$,
    bottom -- for the result regular expression $L_i$.
    (Initial states with arrows without input and final states in double circles)}
    \label{fig:automaton}
\end{figure*}

We can consider the entire process of monitoring tourists, 
and thus generating warnings concerning potential threats from a quite different point of view.
Each tourist is tracked individually, 
and thus his/her trace is described by various threats, if any. 
The \emph{individual threat trace} of \era{a} tourist $t_i$ is expressed by \era{a} formal language
$L_i (t_i) \equiv \{(E ;)^{+}\}$,
defined by a \emph{regular expression}~\cite{Hopcroft-etal-2006},
where the semicolon separates every workflow pass,
see Figure~\ref{fig:automaton},
but $E$ is defined by other regular expression
$E\equiv N | S | W | S \!\cdot\! W$,
where
every $N$ denotes the absence of a threat, 
yet $S\equiv E6a | E6g | E6m | E6r$
and also
$W\equiv E2 | E3 | E4 | E5$,
see Figure~\ref{fig:automaton} top left and top right, respectively.
A sample sentence for $L_i$ language is a finite sequence: 
$N$; $E6g$; $N$; $E6a$; $N$; $N$; $E6mE3$; $E2$; $\ldots$
It terminates if and only if the tracked tourist leaves the monitored area.
On the other hand, 
the workflow in Figure~\ref{fig:context-life-cycle-workflow}
produces threat warnings, if any, for every individual tourist.
As a result of this, 
we obtain an \emph{(entire) threat language} $L \equiv L_1 \cup L_2 \cup \ldots L_n$,
as a sum of individual languages, 
where $n$ is the total number of tourists, 
that is, 
it includes both tourists currently staying in 
the monitored area and tourists who have already left it.
To sum up,
every $L_i$ is a regular language, 
and $L$ is also a regular language.
In other words,
all these languages are generated by a type-3 grammar~\cite{Hopcroft-etal-2006}.
Thus, 
every regular expression $E$ was converted to the form of a finite automaton,
which was subjected to determinisation (via a powerset construction) and minimisation by
means of familiar algorithms~\cite{Hopcroft-etal-2006}.
Consequently,
we obtained a resultant automaton,
see Figure~\ref{fig:automaton}, bottom,
well-depicting the operating principle of the system under consideration,
\erarrr{and strictly speaking}\newrrr{namely},
the automaton accepting the threat language resulting from contextual data processing
in the designed pro-active and context-aware system.

\subsubsection{From threats to behaviours}
\label{sec:behaviours}

The designed supporting system not only detects threats to hiking tourists in the monitored area,
but also makes it possible to understand their activity better.
It is intuitively recognizable when
the relevant notions are introduced more formally through the extension of the threat language.
Let us assume that a certain tourist stays in the monitored area for three hours,
then assessing the threats every half a minute
(see~\cite[Theorem~4]{Klimek-2018-Access})
gives a sequence of three hundred and sixty \emph{threat assessment points},
and this is what the supporting system produces step by step, or point by point, 
for every hiking tourist,
as a result of the \era{performing} \new{performance} workflow from Figure~\ref{fig:context-life-cycle-workflow}.
A sample, any length but finite prefix of a sequence is 
\begin{eqnarray}
N;N;E6mE2;E6m;N;N;E2;E2;E2;E3;N;N;N;E6a;E6a;N;N;\ldots\label{for:threat-language}
\end{eqnarray}
which is an individual threat trace.
Each point,
let us denote it as $th$, 
can be extended easily with other data, 
if necessary, such as 
the hiking tourist i$d$,
the current geolocation data of a tourist $geo$, 
timestamp $ts$, 
ongoing hiking trail $h$ and its difficulty $d$, etc.
(Let us note,
that these pieces of data are obtained naturally in the processing shown 
in Figure~\ref{fig:context-life-cycle-workflow}, 
see also Formula~\ref{for:algorithms-loop} and Figure~\ref{fig:caaas}.)
\emph{Individual behaviour}, or \emph{behavioural trace}, $B_{i}$ of hiking tourist $t_i$ is 
a sequence of an extended threat assessment point,
that is every $B_i$ consists of individual points $B_{i,1}$, $B_{i,2}$, $\dots$, $B_{i,j}$, $\ldots$,
where every
\begin{eqnarray}
B_{i,j} \equiv \langle id(t_i), th, geo, ts, h, d\rangle
\end{eqnarray}
where $th, geo, ts, h, d$ are calculated or established for tourist $t_i$ at $j$-th point.
For example,
the ``E3'' threat point from Formula~(\ref{for:threat-language}) 
for tourist $t_i$,
could be extended to 
\begin{eqnarray}
B_{i,j} = \langle id(t_i), E3, (49^{\circ}34'24''N, 19^{\circ}31'46''E), (19.09.2021,14.30), H2, D3 \rangle
\end{eqnarray}
which means that a certain tourist $t_i$ was in the E3 threat being in specific geospatial position, 
at specific time, being on the H2 trail with the D3 difficulty level.
We could consider other extensions, that is with more or other information, if necessary.
An \emph{(entire) behaviour language} $B \equiv B_1 \cup B_2 \cup \ldots B_n$
is the sum of all the individual behaviours of tourists who have stayed in the monitored area,
see also Figure~\ref{fig:summary-processing}.
Every $B_i$, as well as $B$, is also a regular language,
which could be generated by a type-3 grammar~\cite{Hopcroft-etal-2006}.
Such data sets, involving the vast amounts of data,
and	 showing, for example, behaviours throughout the whole tourist season,
or subsets covering the winter or summer seasons separately,
or consecutive seasons,
can be analysed using clustering 
algorithms.\erarrr{~[71,62]}
This would allow for understanding the characteristics of hiking tourists, taking into account many local aspects of a national park.
As a result, it may \erarrr{allow for}\newrrr{enable} the calibration of tourism models by national parks' employees.
\erarrr{However, this}\newrrr{This} topic exceeds the size of this article,
but it seems to be a good idea for further research.

\subsection{Preliminary simulation}
\label{sec:preliminary-simulation}

Before considering the objective, and primarily, simulation
(see the next section)
we perform a preliminary simulation to observe the generation of threats. 
It is related to and it follows the proposed \new{context life cycle} workflow, 
see Figure~\ref{fig:context-life-cycle-basic}.
However,
it is also based on our idea of how the threat generations\erarrr{, 
without any detailed analysis of contextual data,} should look like \newrrr{without any detailed analysis of contextual data}.
At least it is not related to the simulator prototype shown in Figure~\ref{fig:components-two}. 
When the preliminary simulation is performed,
then there will be an opportunity to check whether the results are consistent with 
the primary simulation, which \era{is} \new{will be} related to complete contextual data processing. 
Moreover,
\era{learning about} \new{just knowing} the structure of activities and events occurring in threat detections will be another benefit.

\begin{figure*}[!htb]
	\centering
    \includegraphics[width = .45\columnwidth]{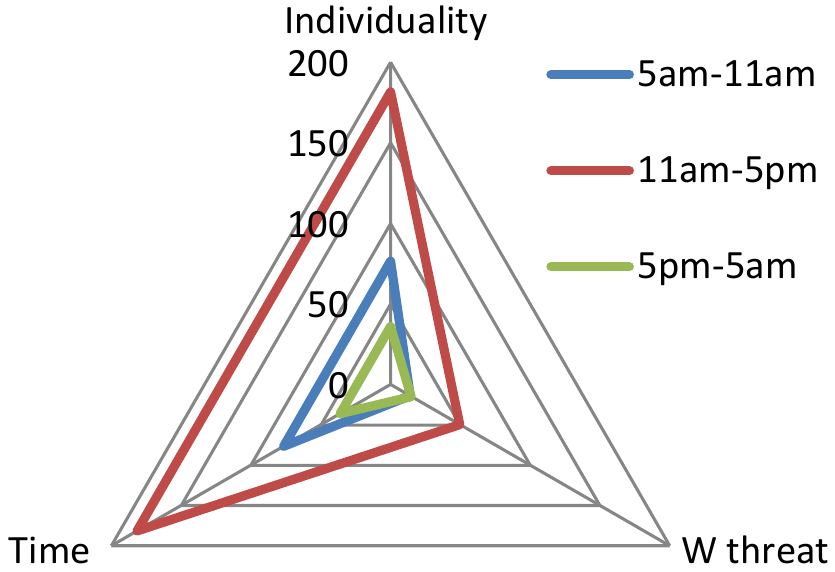}
    \includegraphics[width = .48\columnwidth]{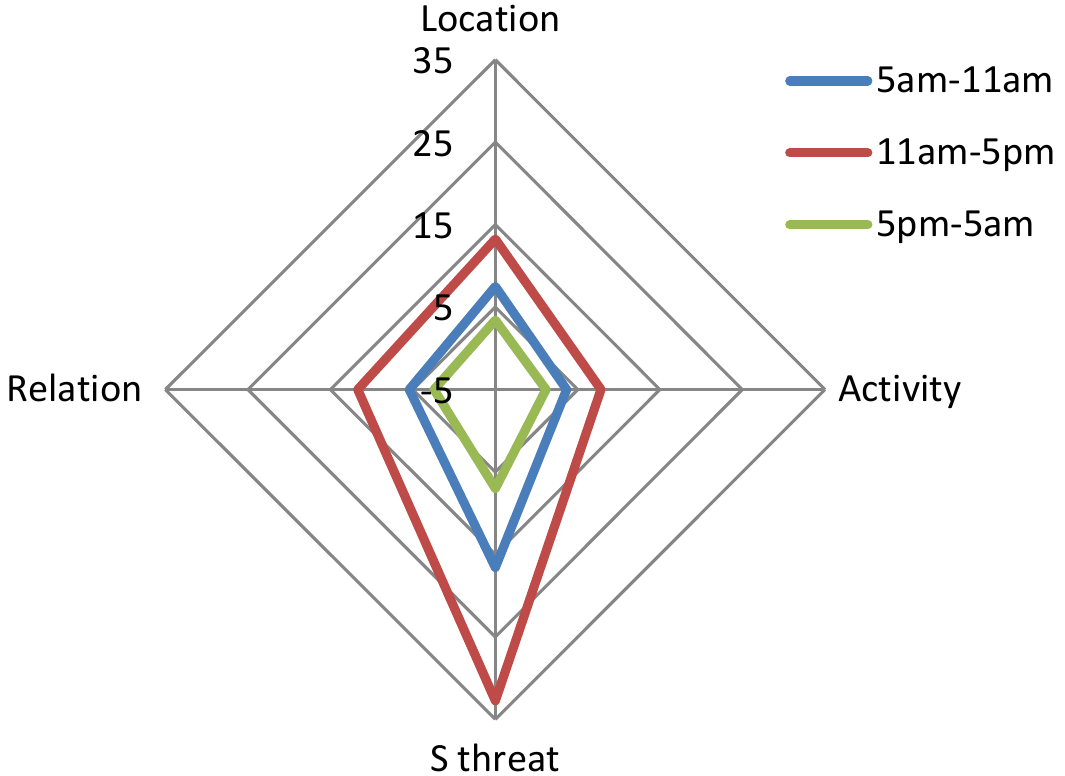}\vspace{2mm}\\
    \includegraphics[width = .40\columnwidth]{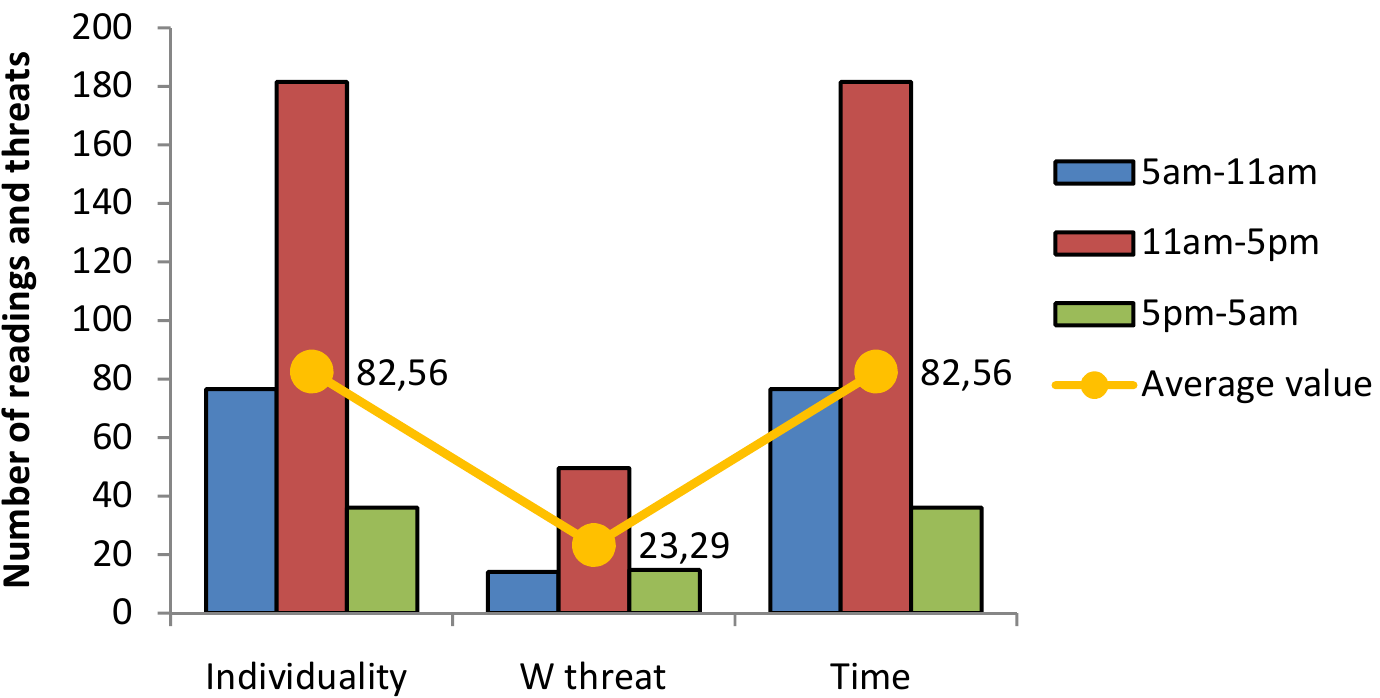}
    \includegraphics[width = .56\columnwidth]{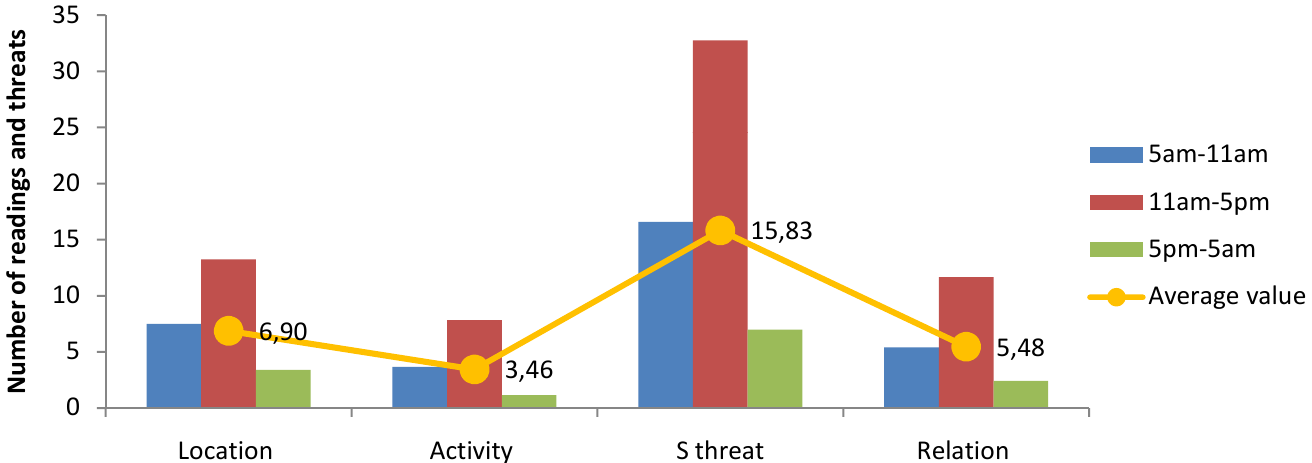}
    \caption{A preliminary simulation experiment and its results,
    the numbers of data readings and the number of threat occurrences are shown:
    top left -- weather contextual data with its ingredients,
    top right -- non-weather contextual data with its ingredients,
    bottom -- average values for all the time intervals considered together}
    \label{fig:preliminary-simulations}
\end{figure*}

Figure~\ref{fig:preliminary-simulations} shows the preliminary simulation results.
The simulation mirrored the 24-hour monitoring of the mountain area, that is day and night.
All pieces of data were established every half an hour,
thus we had 48 repetitions of the operation for downloading threats that occurred.
Concerning tourists staying in the area, 
we assumed that the peak of attendance took place at 2~p.m. with 200 people, 
and with a normal distribution and a standard deviation of 4 hours.
We consider the following periods:
\begin{enumerate}
\item the morning interval -- that is a period from 5:00~a.m.\ to 11:00~a.m., 
\item the noon-afternoon interval  -- that is a period from 11:00~a.m.\ to 5:00~p.m.,
\item the evening-night interval -- that is a period from 5.00~p.m.\ to 5.00~a.m. 
\end{enumerate}
The probability of weather threats for the 1st time interval is 20\%, 
no matter which \new{threat} \newrrr{it is}, from E2 to E5. \era{threat} \erarrr{it is.}
However, it increases by 20\% in the 2nd time interval and by 100\% in the 3rd time interval.
The probability of non-weather threats \erarrr{goes}\newrrr{is} as follows:
we have 30\% of tourists following mountain trails in groups
and the probability of E6g is 5\%, 
\erarrr{and the probability of}\newrrr{for} E6m and E6a \newrrr{it} is 5\% each, 
and for E6r it is 10\%. 
The \era{probabilities} \new{events resulting from the assumed probabilities} for both context groups, $W$ and $S$, are calculated separately;
on the other hand, 
if the \era{probability} \new{event} of threats for $S$ is identified, 
the calculation for $W$ is omitted.

\begin{figure*}[!htb]
	\centering
    \includegraphics[width = .68\columnwidth]{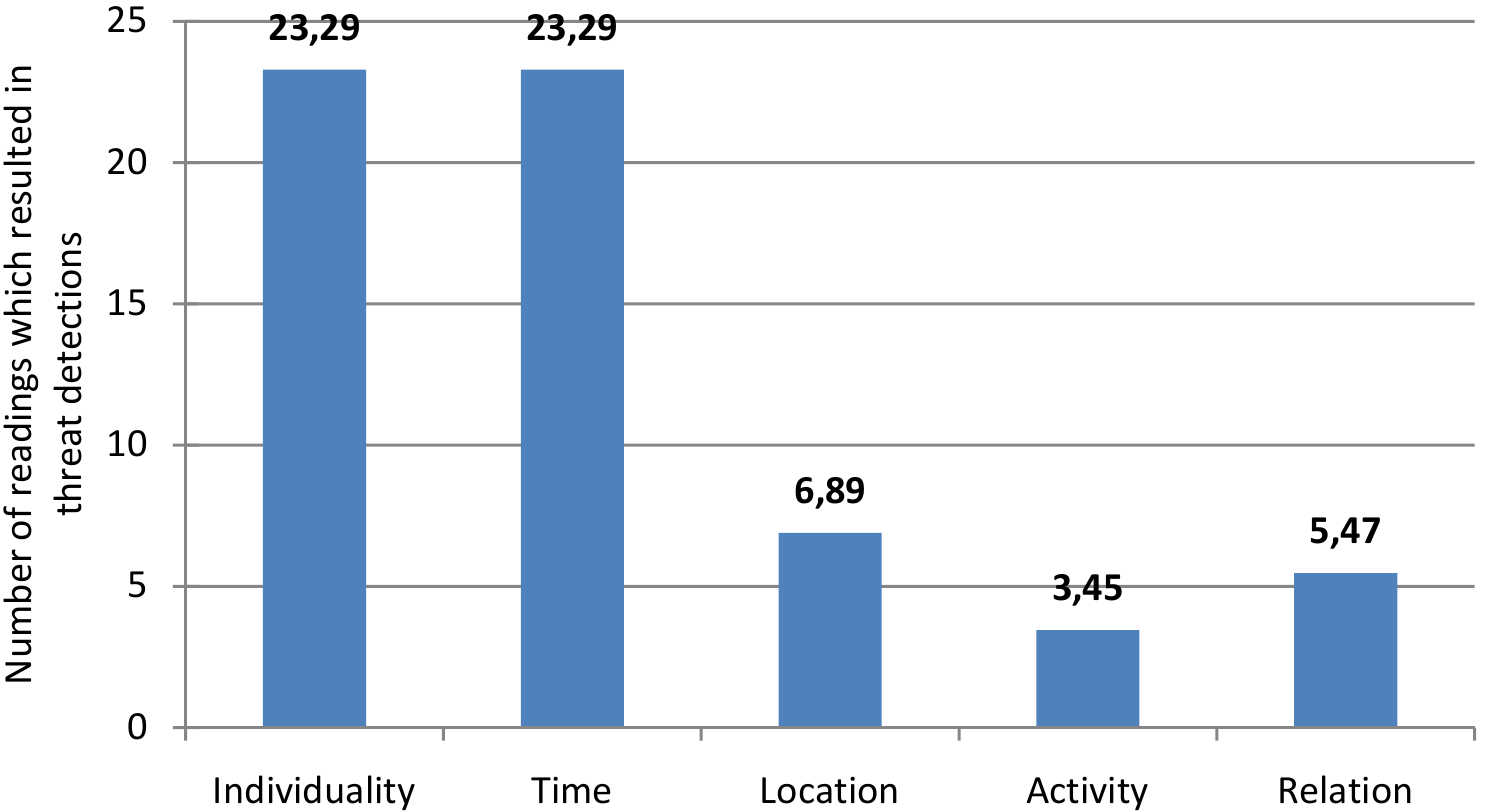}
    \caption{Influence of the particular contextual pieces of data \new{readings} (average values \new{for all the time intervals considered together}) 
    on both detected threat categories in the preliminary simulation,
    see also Fig.~\ref{fig:context-model-categories}
    }
    \label{fig:preliminary-simulation-influence-categories}
\end{figure*}

Figure~\ref{fig:preliminary-simulation-influence-categories}
shows the \era{results} \new{influence} for individual contextual data \new{categories} belonging to both groups,
\new{that is $W$ and $S$, for threat detections. 
(The categories in 
Figure~\ref{fig:preliminary-simulation-influence-categories} and 
Figure~\ref{fig:context-model-categories} are the same.)
The contextual pieces of data are related to the categories, 
the readings of which\erarrr{, together with their necessary further interpretation,}
are the basis for possible detection of threats\newrrr{, together with their necessary further interpretation}.
The assumed threats are declared in Figure~\ref{fig:context-model-categories},
see E2-E5, E6g, E6a, E6m, E6r.}

Individuality and Time have equal values because they are equally involved in 
determining the situation of the type-$W$ threat.
Categories ``Relation'', ``Activity'' and ``Location'' determine the type-$S$ threat, 
but category ``Relation'' is more commonly used because of E6a and E6g.
\era{On the other hand,}
\new{However,}
the type-$W$ threats frequency depends both on 
the presence of tourists in the area, but also on the time of the day.
On the one hand, 
the most tourists are \erarrr{on}\newrrr{present in} the routes in the 11:00~a.m.--5:00~p.m.\ interval, 
and on the other hand, 
in the 5~p.m.--5 a.m. interval, that is at night, which is the longest interval,
there are the fewest tourists 
but the probability of threats doubles. 
It results in a large number of type-$W$ threats. 
Whereas, 
the sizes of \erarrr{categories} ``Individuality'' and ``Time'' \newrrr{categories} are large because 
they apply to every tourist, 
regardless of the time of the day when type-$W$ threats are determined. 
The number of tourists is low at night, 
so the values of type-$S$ threats are also low.

\section{Mountain surroundings and simulation processes}
\label{sec:mountain-and-simulation}

This section contains the objective simulation of the entire developed system supporting rescue operations,
that is the primary simulation for this article, 
including the contextual data processing.
Hence,
\begin{itemize}
\item
we outlined the prototype of the mountain environment simulator briefly 
(Subsection~\ref{sec:simulator}); 
and then
\item
we discussed the fundamentals of the planned simulation process
covering five weather scenarios
(Subsection~\ref{sec:simulation});
\item
we \erarrr{showed}\newrrr{presented} the simulation results and observations characteristic for 
phenomena occurring in a mountain area with a large number of hiking tourists
(Subsection~\ref{sec:simulation-results});
as well as
\item
contextual data processed in the established context life cycle, 
including its properties and interrelations (Subsection~\ref{sec:simulation-results}).
\end{itemize}

\subsection{Basic assumptions}

\subsubsection{Simulator}
\label{sec:simulator}

\begin{figure*}[!htb]
	\centering
    \includegraphics[width = .9\columnwidth]{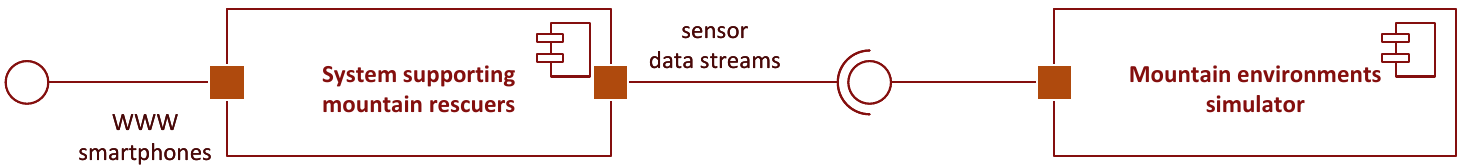}
    \caption{Component diagram for a supporting system,
    that is the \emph{Mountain rescuers supporting system} (or \emph{MoSup} for short) and 
    a simulator prototype,
    that is the \emph{Mountain environments simulator} (or \emph{MoSim} for short)}
    \label{fig:components-two}
\end{figure*}

We \era{have} built a prototype of a simulator~\cite{Indyka-2022}
to reflect the real mountain surroundings and conditions accurately, 
see Figure~\ref{fig:components-two}.
This is a significant difference concerning article~\cite{Klimek-2018-Access}, 
where only single hypothetical system components (message brokers and SAT solvers) were tested separately, 
and the supporting system itself had not yet been implemented. 
Currently, we can test the developed and implemented supporting system,
which receives data from a mountain environments simulator, 
which, \newrrr{expressing it} informally\erarrr{ speaking}, feeds it with various data imitating a mountain environment.
\new{This data, 
that is sensor data streams,
transferred via \emph{required} and \emph{provided} interfaces}~\erarrr{[11,49]}\newrrr{\cite{Pender-2003}}\new{,
respectively,
include: 
readings from weather stations (wind, fog, temperature, rain), 
geolocation of tourists (both GPS- and BTS-based), 
geolocation of animals, 
and other data (avalanches, alerts, etc.).}
\erarrr{However,}\newrrr{Nevertheless} it should be emphasized that testing the supporting system is 
one of the most important objectives in this article \newrrr{in order}\erarrr{,
so as} to verify our thinking about the proposed contextual data processing model.
And this is what \erarrr{we devote the most}\newrrr{receives our highest} attention\erarrr{ to}.

\begin{figure*}[!tb]
	\centering
    \includegraphics[width = 1\columnwidth]{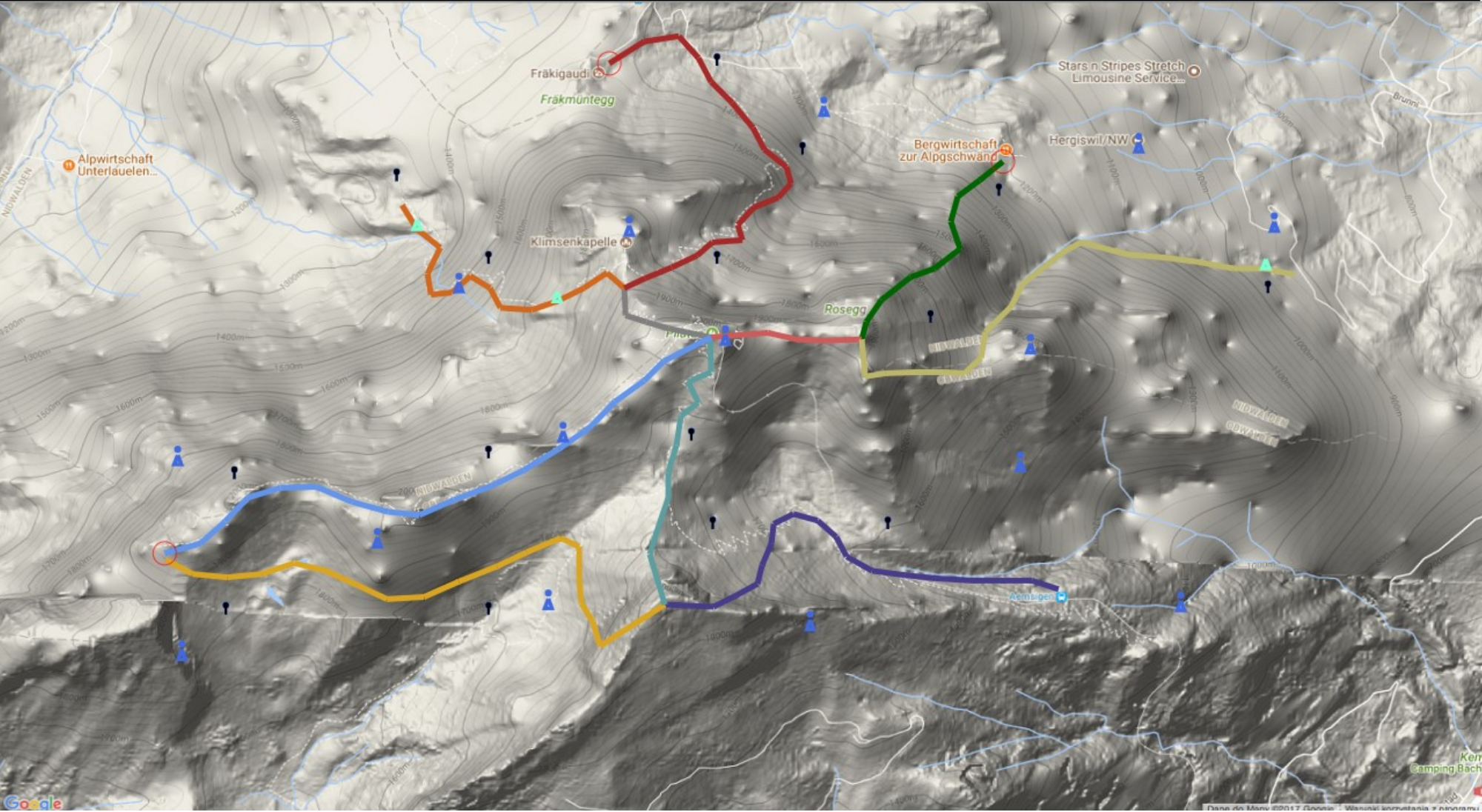}\vspace{1mm}\\
    \includegraphics[width = 0.7\columnwidth]{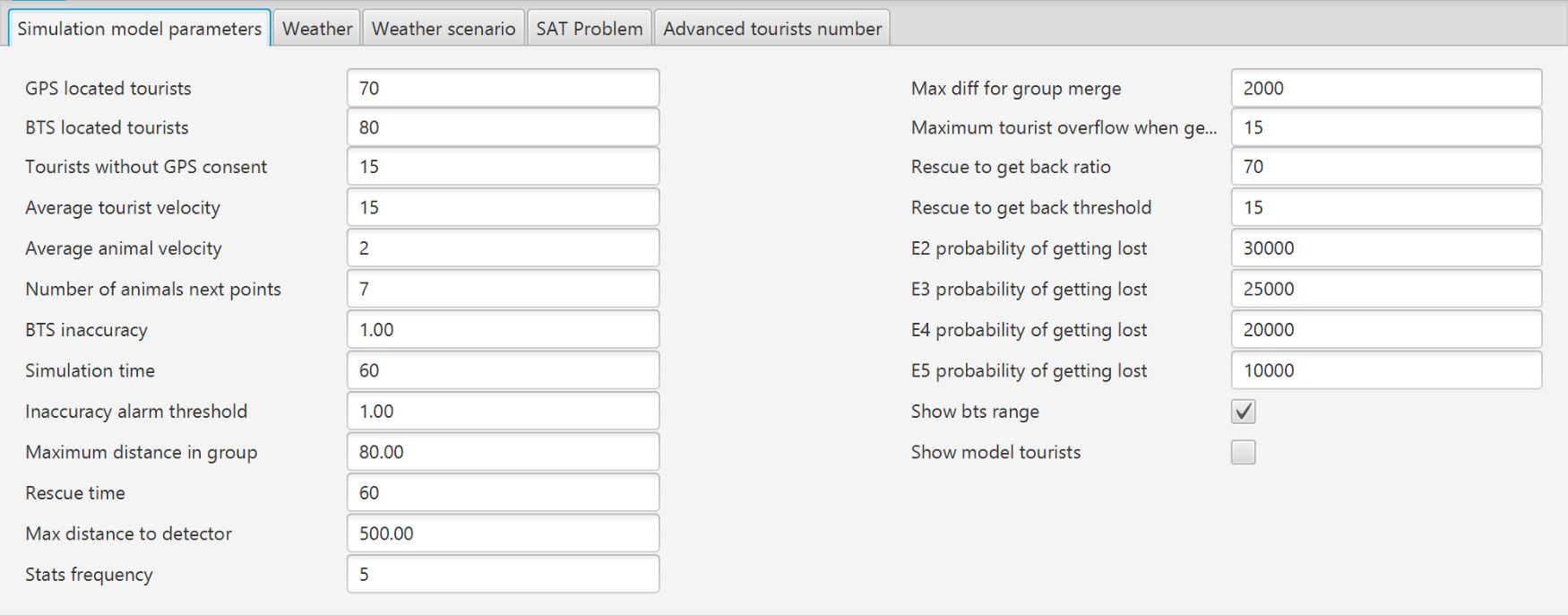}\\
    \includegraphics[width = 0.7\columnwidth]{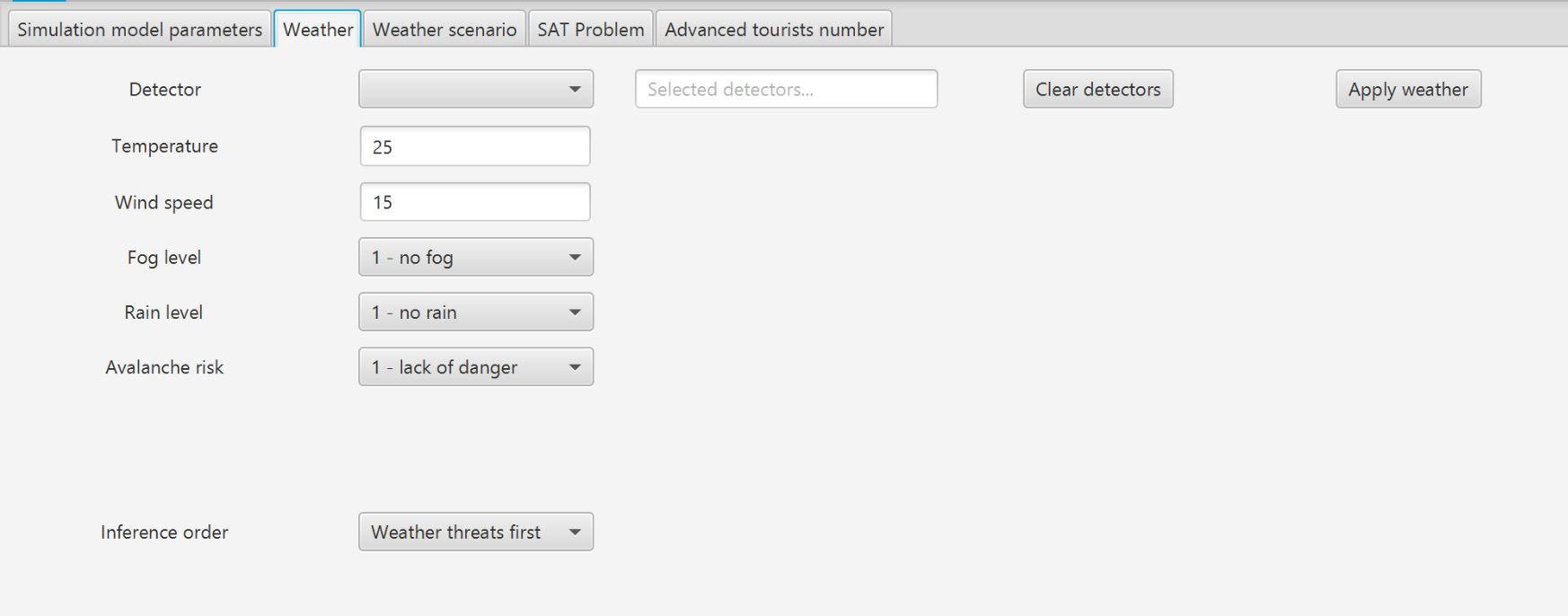}
    \caption{Mountain environments simulator:
     top -- the map of a sample monitored area with established tourist routes,
     fixed weather stations~\includegraphics[width=.85mm]{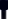} (black pictograms),
     BTS points~\includegraphics[width=1.5mm]{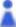} (blue pictograms),
     special places~\includegraphics[width=1.8mm]{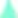} (light green triangles), 
     where tourists may stay longer and it is not treated as E6m threat,
     and route entry points (circles);
     the situation before the start of the tourist visits is shown
     (the basemap of the area originates from \textit{Google Maps}
     but all the routes are pictured with the use of QGIS\erarrr{[52]})~\cite{Indyka-2022}, 
     that is the routes are empty,
     \era{bottom left}\new{middle} -- the simulation tab of a sample panel for admin,
     \era{bottom right}\new{bottom} -- a weather tab of a sample panel~\cite{Indyka-2022}}
    \label{fig:simulator-screenshot-start}
\end{figure*}

Let us present some details about the prototype simulator.
It mirrors fixed tourist routes,
hiking tourist traffic,
weather station deployment,
various methods of obtaining tourist geolocation data, etc.
Figure~\ref{fig:simulator-screenshot-start}
\erarrr{presents}\newrrr{depicts} an exemplary mountain tourist region with established routes
as screenshots from the simulator system.
Tourist routes, for the purposes of this simulation, 
were generated in a QGIS system~\erarrr{[52]}\newrrr{(see:~\url{https://www.qgis.org})} by 
defining the sequence of points which create particular routes. 
The routes generated in this manner were superimposed on a map of the mountain area. 
The environmental picture obtained is revised constantly by the supporting system,
the tourist pictograms are updated,
and the new tourist geolocations with detected threats, if any, are shown.

The mountain environments simulator
enables the generation of data within the monitored area,
while mirroring real mountain weather and non-weather conditions,
and it applies to its numerous weather and non-weather aspects. 
First of all, 
weather conditions, 
that is, 
temperature and its time fluctuations, 
fog, rain and wind force are taken into consideration. 
Those conditions can apply to the whole area but also to 
sub-areas covering the circle described by its radius. 
Weather data is sent by meteorological stations. 
It is necessary to point out that 
the simulator enables the preparation of different weather scenarios. 
The system configuration can involve many details,
for example:
the location of the trail entrance to the monitored area, 
the speed of walking, 
the probability of a trail change for an event related to the crossing of trails,
the probability of a tourist getting lost, 
in such cases, also the probability of an autonomous return to 
the correct route or the necessity of being assisted by rescuers 
and taken outside the monitored area. 
All these parameters are determined by \era{specifying} specific numerical values.
The simulator’s work includes the presence of animal migrations and their velocity.
The \erarrr{animal data} pieces \newrrr{of data concerning animals} are obtained from GPS receivers. 
In this case a significant difficulty is that animals 
do not walk along designated tourist routes but 
intersect them randomly. 
Moreover, in reality, 
animals avoid direct contacts with people so the probability of changing an animal route, 
in the case of being in the direct vicinity of people, 
was implemented as well. 
The next important aspect of the simulator’s work is the work of BTS stations, 
\erarrr{precisely speaking}\newrrr{namely}, 
the possibility of localising a mobile phone owner according to 
the methods \newrrr{mentioned in article~\cite{Klimek-2018-Access}}\erarrr{and precision described in 
the article~[14]}. 
In the case of tourists, if they give their consent,
GPS receivers can also be used.  
Figure~\ref{fig:simulator-screenshot-start} 
presents the exemplary screenshots of different admin panels \era{for} 
concerning the different aspects of the simulation process. 
They allow to influence mountain weather and non-weather conditions.
\erarrr{As a result of this,}\newrrr{Consequently,} it provides rich system datasets.  
A detailed description of the simulator’s work extends 
the aims and size of this paper,
and our goal is to present the processing of contextual data,
that is the supporting system;
however, 
some further interesting details will be presented in 
the next parts of the article.

\subsubsection{Simulation}
\label{sec:simulation}

Let us analyse five different weather circumstances,
for each case a separate simulation process will be performed.
The source data \era{has been} \new{was} prepared based on actual weather conditions in the Babia G{\'{o}}ra National Park, 
which has a very popular mountain peak in Poland
(see: \url{http://en.wikipedia.org/wiki/Babia\_Gora}),
and is known not only for its landscape values, 
but for the great weather variability, and not so rare fatal accidents.
This knowledge and the data obtained are mapped into the following scenarios:
\begin{description}
\item[Scenario \#1] --
summer season, 
we assume rather bad weather conditions, 
possible periodical fluctuations when conditions improve; 
\item[Scenario \#2] --
summer season, 
we assume excellent weather conditions;
nevertheless,
they deteriorate significantly at \erarrr{some}\newrrr{a certain} time point
and also remain unchanged until the end of the entire simulation;
\item[Scenario \#3] --
summer season, 
we assume excellent weather conditions,
they deteriorate significantly for short time during the simulation process; 
\item[Scenario \#4] --
winter season, 
we assume difficult weather conditions, 
possible periodical fluctuations when conditions improve; 
\item[Scenario \#5] --
winter season, 
we assume very difficult weather conditions, 
but they improve significantly at \erarrr{some}\newrrr{a certain} time point
and also remain unchanged until the end of the entire simulation.
\end{description}

\begin{figure*}[!tb]
	\centering
    \includegraphics[width = .7\columnwidth]{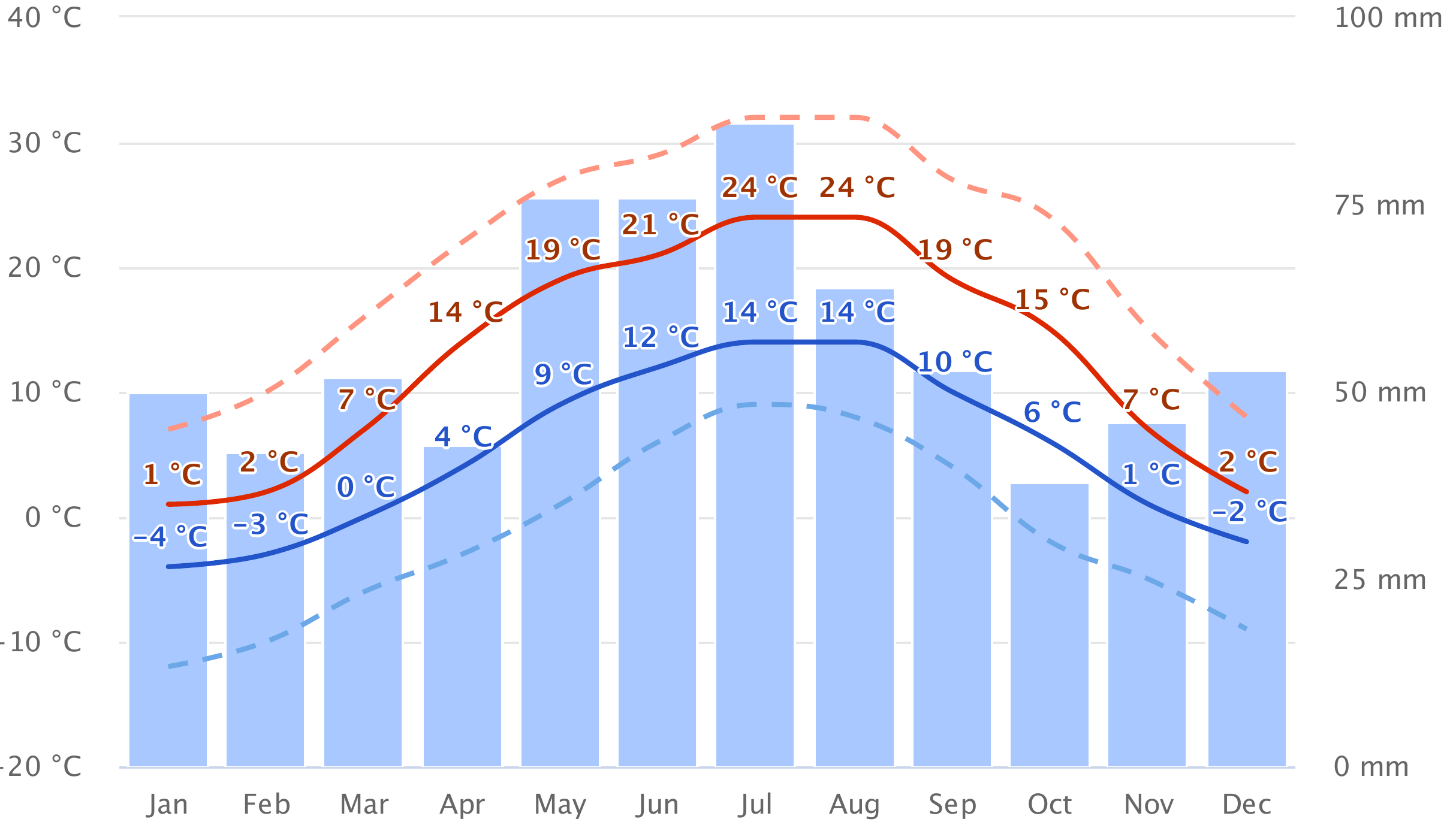}
    \caption{Average month temperatures (the left side) and precipitation (the right side) for the Babia G{\'{o}}ra National Park, Poland, in 2021.
             Solid red line or solid blue line mean the maximum or minimum temperature of an average day for every month, respectively;
             dashed red and blue lines refer to multi-year averages and light blue bars indicate precipitations.
             (The data is taken from \textit{meteoblue AG}, 
             \url{https://www.meteoblue.com/en/weather/historyclimate/climatemodelled/babia-g\'{o}ra_polska_3104023})}
    \label{fig:Babia-Gora-weather}
\end{figure*}

Figure~\ref{fig:Babia-Gora-weather}
shows the sources of our weather data using temperature as an example.
The temperature is taken as a value that fluctuates around the mean value,
that is the average between red and blue solid lines.
If the weather improves, 
we take the maximum values (solid red) 
and if the weather worsens, 
we take the minimum values (solid blue).
Both rules apply separately to summer (July and August) and winter (January and February) scenarios.
When walking along mountain trails, 
some tourists may get lost, 
go off the trail 
or stop moving for a longer period of time 
(see threats E6r and E6m in 
Figure~\ref{fig:context-model-categories}
and Table~\ref{tab:threat-levels}).
Among those tourists, 
some of them may, 
after a certain period of time, 
come back to the route themselves, 
but others may require the \newrrr{assistance of} rescuers. 
For this latter group, 
some tourists are guided back to 
the trail by rescuers,
but the rest of them need more serious interventions, 
and they are evacuated by rescuers from the monitored area.

\begin{figure*}[!tb]
	\centering
    \includegraphics[width = 1\columnwidth]{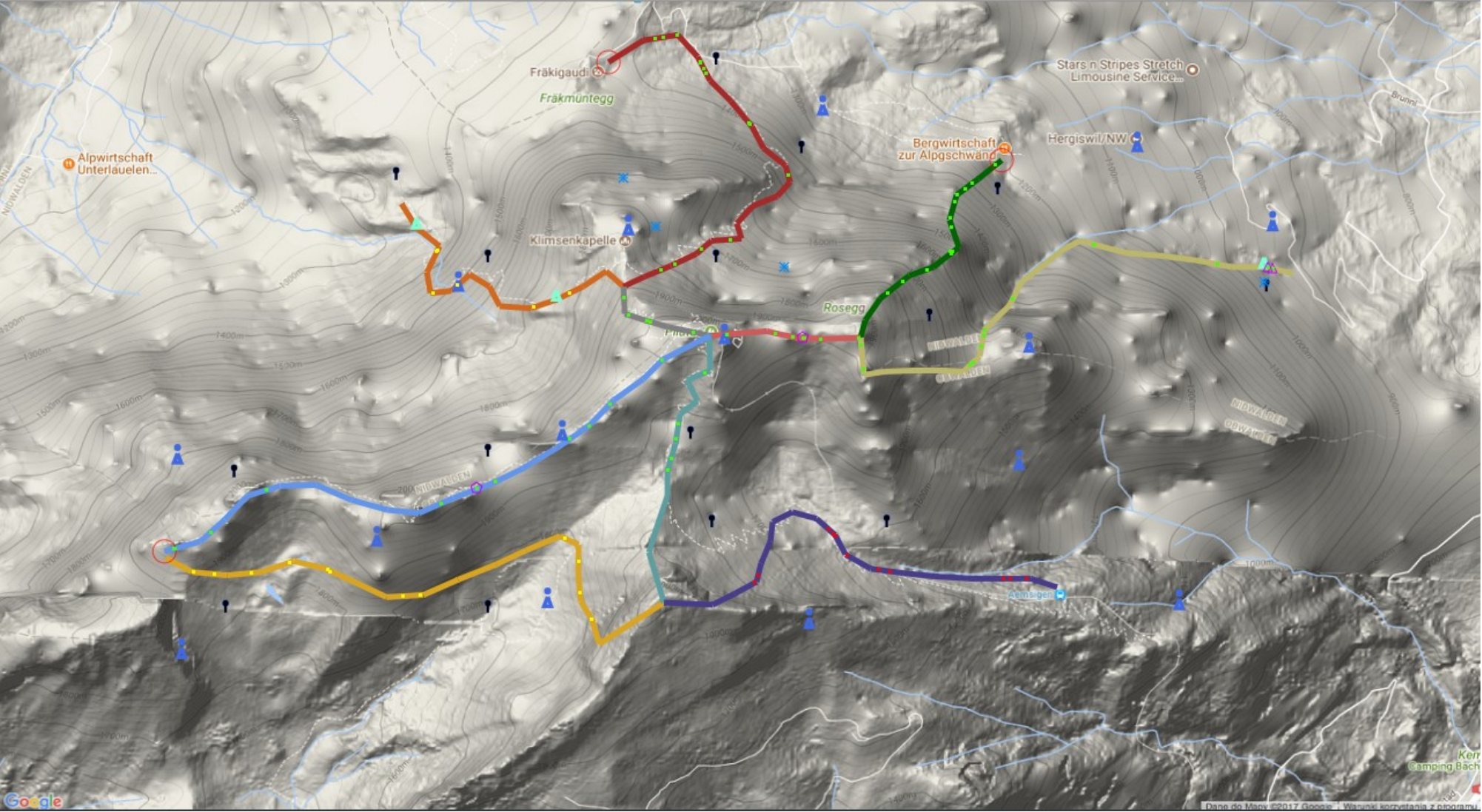}
    \vspace{.1cm}\\
    \includegraphics[width=.8\columnwidth]{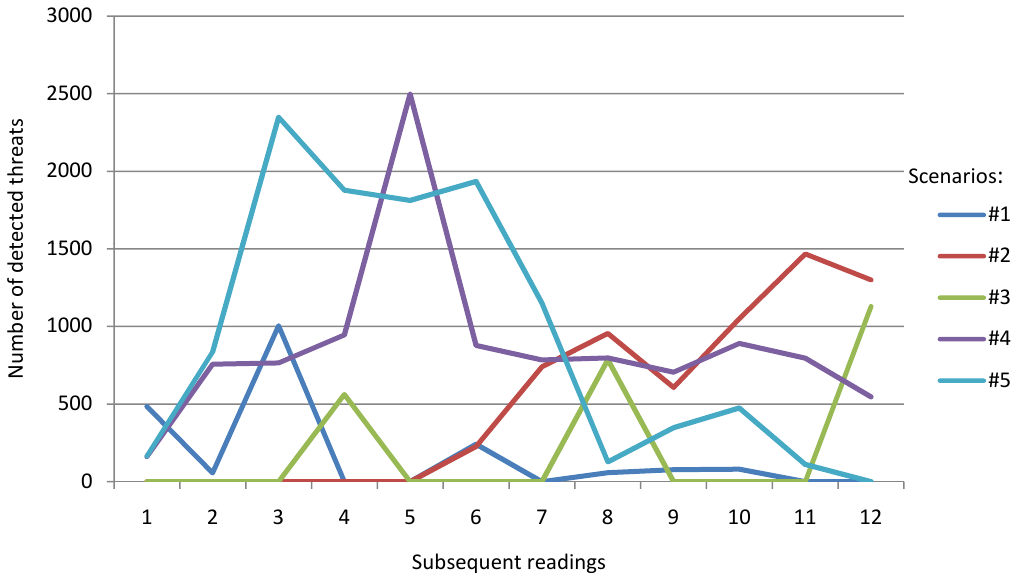}
    \caption{The runs of particular predefined weather scenarios:
    top --
     a sample screenshot of the map for the monitored area during the tourist presence
     (the basemap of the area originates from \textit{Google Maps}
     but all the routes are pictured with the use of QGIS\erarrr{[52]})~\cite{Indyka-2022}, 
     where colour dots on the trails mark hiking tourists (green, yellow and red, that is E1/normal, E2 and E4 threats, respectively),
     blue stars symbolize wandering animals
     and two pentagons and two triangles symbolize non-weather threats (that is E6g and E6a, respectively),
     see~Tab.~\ref{tab:threat-levels},
    bottom -- 
     the total number of threats for five predefined weather scenarios,
     which are a consequence of changing aura factors.
             (The simulation data is downloaded twelve times at the same time interval.)}
    \label{fig:primary-simulation-screen-and-scenarios}
\end{figure*}   

We planned the entire simulation process for one hour.
However, 
it is necessary to differentiate between 
the process of simulation itself 
and the time the process would take in real life, 
which is represented by simulation. 
Simulation goes around 25 times faster,
or let us state equivalently:
each of the five planned scenarios corresponds roughly to thirty hours in a natural mountain environment
and it is
equal to a one-hour simulation. 
Figure~\ref{fig:primary-simulation-screen-and-scenarios} 
shows adopted weather scenarios. 
The drawing has to be supplemented with important remarks. 
The following method was applied to obtain the charts: 
\begin{itemize}
\item
the cumulative numbers of threats are shown for each scenario
in a further discussion
(E2, E3, E4 and E5 represent particular threats and E1 is a no-threat situation);
\item
all the simulation data, including contextual data,
\newrr{process information, environment variables, etc.\ all considered as a \emph{memory dump},}
is collected every five minutes of the simulation, 
which gives twelve data readings
(it is also possible to set a different frequency of data collection, both higher and lower),
\newrr{however, 
a distinction should be made between the system's real-time responsiveness,
which is about half a minute,
see the first paragraph in Subsection~\ref{sec:behaviours},
and the time of every memory dump,
which is a time-consuming process due to the memory size and the detailed data it contains.
However, executing a memory dump does not affect the system operation correctness in terms of its reactivity to threats,
and we can skip it, or manage without, it.
\erarrr{On the other hand,}\newrrr{Nevertheless} every memory dump along with rich system logs provide a valuable analytical data,
see for example Figures~\ref{fig:groups-redundancy}--\ref{fig:simulation-context-sharing} and other tables,
therefore it would be disadvantageous to resign from it};
\item
it is obvious that
the deterioration of weather conditions entails the increase of 
the threat detection recurrence as regards objects present on the area monitored;
\item
we presume that tourists in routes will be hiking at a similar speed \era{similar} \new{close} to that of an adult human;
\item
weather changes resulting from the adopted scenarios occur immediately throughout the area.
\end{itemize}

Scenarios~\#1, \#2 and~\#3 relate to summer. 
Scenario~\#1 shows weather fluctuations, 
and from a certain point \era{it stabilizes} \new{they stabilize}.
Scenarios~\#2 and~\#3 have very good conditions initially.
In the case of Scenario~\#2, they suddenly deteriorate.
In the case of Scenario~\#3, there are rather local disturbances. 
Scenarios~\#4 and~\#5 relate to winter. 
Scenario~\#4 shows ordinary fluctuations.
Scenario~\#5 presents really bad weather conditions which improve steadily. 
The summary occurrence of threats is presented in each diagram. 
As a general observation, we can say that in winter 
there are generally more threats and in summer there are naturally fewer of them.
There is no need to differentiate between particular threats (e.g.\ E2, E3, E4 or E5) in 
Figure~\ref{fig:primary-simulation-screen-and-scenarios},
they are treated \era{together} \new{jointly}.
However, 
an interesting point, 
which is worthy of our attention, is the issue concerning 
curves which grow from zero, 
and it is visible especially in winter scenarios.
This is because at the beginning of each scenario, 
the monitored area does not have many tourists. 
They appear later on, 
during simulation, 
together with the different kinds of threats. 
(This effect is much more visible in 
Figure~\ref{fig:simulation-context-transitions-time}.)
Numerous popular mountain areas are not opened in some specific periods.
It is similar in our simulation, 
that is when it starts, the monitored area is empty, 
but over time it fills up with tourists more and more.

Some remaining information about the system principles is as follows. 
RabbitMQ\erarrr{ [53]} was used as a message flow management system.
As a logical reasoning engine, 
Sat4j\erarrr{ [40]} was chosen. 
All the dangerous animals within the monitored area have GPS transmitters. 
These animals do not normally use the designed tourist routes
but any trails outside and when their paths cross with human routes, 
they may occasionally wander onto them.
In such cases,
the potential for interaction with tourists becomes possible. 
In the future, 
the functions to avoid people can be embedded, 
which would reflect the natural conditions better. 
The speeds of animals are random. 
All the locations are estimated with precision enabled by 
BTS devices (tourists) and GPS devices (tourists and animals). 
There is an assumed number of BTS stations in order to enable 
location according to the methods described 
in~\cite{Klimek-2018-Access}. 
There is the possibility of establishing\erarrr{, within a certain range,}
the strength of a BTS signal \newrrr{within a certain range}, 
which influences the precision of the location. 
Some tourists get lost, 
which is described by a random variable; 
however, 
this situation occurs mostly in difficult weather conditions. 
Within a certain range, 
there is the possibility of setting a movement speed of new tourists,
i.e.\ tourists entering the monitored area.

The assumptions and scenarios prepared by this means are 
the basis for testing our CAaaS component, 
and the results of the analysis are presented below.

\subsection{Simulation results}
\label{sec:simulation-results}

\subsubsection{Overview}

The general overview of the primary simulation
is shown in Table~\ref{tab:simulation-general-overview}.
The pieces of data collected from simulations are presented,
as well as the manner in which some pieces of data were generated. 
The table provides a good picture of the course of the simulation, 
which was carried out with great care and attention to the realism of behaviours.
All the numbers included in the table denote the number of all the events 
which took place during the simulation that lasted one hour.
Nevertheless,
a group of data which refers to location,
shows the current numbers of tourists when the simulation is terminated.

\begin{table}[!htb]
\caption{General overview of the simulation processes}
\centering
\begin{tabular}{|ll|c|c|c|c|c|}
\hline
\multicolumn{2}{|l|}{Predefined scenario} & \#1 & \#2 & \#3 & \#4 & \#5\\
\hline\hline
\multicolumn{2}{|l|}{Complete number of tourists} & 3308 & 3344 & 3080 & 3575 & 3298 \\ 
\multicolumn{2}{|l|}{Tourists who left the area} & 3088 & 3123 & 2870 & 3351 & 3060 \\
\multicolumn{2}{|l|}{Current/last number of tourists} & 220 & 221 & 210 & 224 & 238 \\
\hline
\multirow{4}{.5cm}{
\begin{sideways}
Location
\end{sideways}
}
 & \cellcolor{Gray} BTS located tourists & 136 & 140 & 129 & 132 & 144 \\	
 & \cellcolor{Gray} GPS located tourists & 84 & 81 & 81 & 92 & 94 \\	
 & \cellcolor{Gray} Tourists who did not accept GPS data & 29 & 33 & 16 & 30 & 26 \\
 & \cellcolor{Gray} All low BTS location accuracy situations & 1152 & 967 & 637 & 222 & 640 \\
\hline
\multicolumn{2}{|l|}{``One weather detector'' events} & 3891 & 4027 & 3831 & 4029 & 4176 \\
\hline
\multirow{2}{.5cm}{$W$}
 & \cellcolor{Gray} {Weather threats} & 2003 & 6349 & 2478 & 10527 & 11195 \\	
 & \cellcolor{Gray} {Avalanche risk alarms} & 20 & 6 & 27 & 18 & 17 \\	
\hline
\multirow{4}{.5cm}{$S$}
 & \cellcolor{Gray} {Animal threats} & 628 & 629 & 519 & 417 & 343 \\
 & \cellcolor{Gray} {``No motion'' situations} & 84 & 364 & 118 & 836 & 620 \\	
 & \cellcolor{Gray} {``Out of route'' situations} & 85 & 372 & 122 & 848 & 623 \\	
 & \cellcolor{Gray} {Tourists who lost their group leader} & 149 & 225 & 130 & 470 & 360 \\	
\hline
\multicolumn{2}{|l|}{SAT solver starts} & 22 & 43 & 36 & 62 & 30 \\
\hline
\end{tabular}
\label{tab:simulation-general-overview}
\end{table}

``All low BTS location accuracy situations'' 
corresponds to the occurrences 
when the geolocation measurement based on data from BTS stations is uncertain.
The known geolocation algorithm\erarrr{, see~[14],}
requires measurements from two stations, 
determining the intersection points of both circles, 
thus we can determine a distance to the third station. 
The measurement of the signal strength between 
the station and the tourist is of key importance here. 
The dissimilarity between the distance defined based on the algorithm, 
and the distance defined based on the signal strength is deemed to be imprecision. 
If the discovered dissimilarity is too high, the system receives a report, 
and it is possible to send a BTS drone.

\newrr{Let us note 
that the proposed context processing includes the prioritization of contextual (raw) data.
In general,
\emph{prioritization} refers to determining the relative importance or adequacy of the different pieces of information.
It involves ranking items and focusing on what matters most.
We can see this when preferring GPS-based geolocations for tourists, 
which always gives better positions, as compared to BTS-based ones,
see Table~\ref{tab:simulation-general-overview}.
A slightly different selection situation occurs in the case of reading raw data from several weather stations for the position of a particular tourist,
see the paragraph after Formula~(\ref{for:algorithms-loop}).
Finally, 
looking from the other end of a single context life cycle, 
we have a situation of obtaining threats that are handled by rescuers starting with the highest threat and priority that occurred, 
that is from E5 to E2, for weather threats, 
while the situation threats (E6g, E6r, E6m, E6a) are considered separately.
In other words, weather threat processes are prioritised, 
see also $Layer3$ in Subsection~\ref{sec:organisation-CAaaS}.}

\begin{table}[!bt]
\caption{Tourists' turnout in the Babia Góra National Park in Poland
\new{(``--'' means no data for this month, which is due to the lack of entry registrations in some winter months in those years)}}
\vspace{1mm}
\centering
\begin{tabular}{|c|c|c|c|c|c|c|c|c|c|c|c|}
\hline
Year & 
\begin{sideways}
January-February
\end{sideways}
& 
\begin{sideways}
March
\end{sideways}
& 
\begin{sideways}
April
\end{sideways}
&
\begin{sideways}
May
\end{sideways}
&
\begin{sideways}
June
\end{sideways}
& 
\begin{sideways}
July
\end{sideways}
&
\begin{sideways}
August
\end{sideways}
&
\begin{sideways}
September
\end{sideways}
&
\begin{sideways}
October
\end{sideways}
&
\begin{sideways}
November
\end{sideways}
&
\begin{sideways}
December
\end{sideways}\\
\hline
2019 & -- & -- & 1~791 & 8~268 & 10~400 & 22~553 & 27~528 & 16~999 & 14~902 & 2~471 & --\\
\hline
2020 & -- & -- & 819 & 9~051 & 12~829 & 32~619 & 31~790 & 20~983 & 8~693 & 8~990 & --\\
\hline
2021 & -- & 639 & 4~795 & 6~051 & 18~946 & 26~001 & 31~310 & 29~163 & 14~998 & 6~921 & --\\
\hline
\end{tabular}
\label{tab:BPN-tourist-turnout}
\end{table}

Let us pay attention to Table~\ref{tab:BPN-tourist-turnout},
which \era{shows} \new{presents} the attendance of tourists over the last few years
in the Babia G{\'{o}}ra National Park in Poland.
Attendance in certain winter months was intentionally not measured,
which of course does not mean that the park was not visited then, 
but the entries were not registered at that time.
\era{On the other hand,}\new{Yet,}
common experience shows that attendance in winter is \erarrr{much}\newrrr{significantly} lower than in summer months.
If we consider the turnout in August,
a holiday month in which the most tourists come,
this will give a daily average of approximately one thousand people (in 2021),
and this is the maximum average in the entire table.
Meanwhile, 
we simulated the daily presence of tourists in Table~\ref{tab:simulation-general-overview}
in the number three times greater,
regardless of the type of a scenario, both for winter and summer.
Therefore, we can conclude that the simulation performed meets in excess  
\new{the} actual attendance data for the mountain park under consideration.

\begin{figure*}[!htb]
	\centering
    \includegraphics[width=.8\columnwidth]{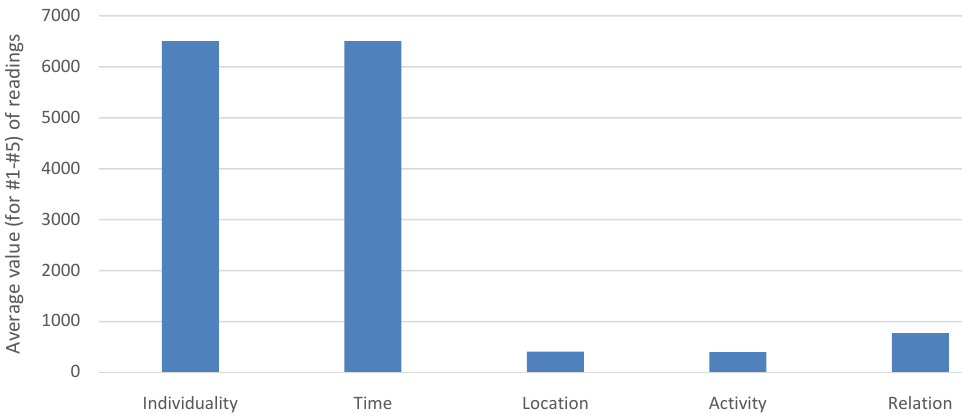}
    \caption{The influence of the particular contextual pieces of data on both the detected threat categories in the primary simulations and the average values for all scenario-based simulations,
    see Fig.~\ref{fig:context-model-categories} and also Fig.~\ref{fig:preliminary-simulation-influence-categories}}
    \label{fig:primary-simulation-influence}
\end{figure*}

\begin{figure*}[!tb]
	\centering
    \includegraphics[width = .8\columnwidth]{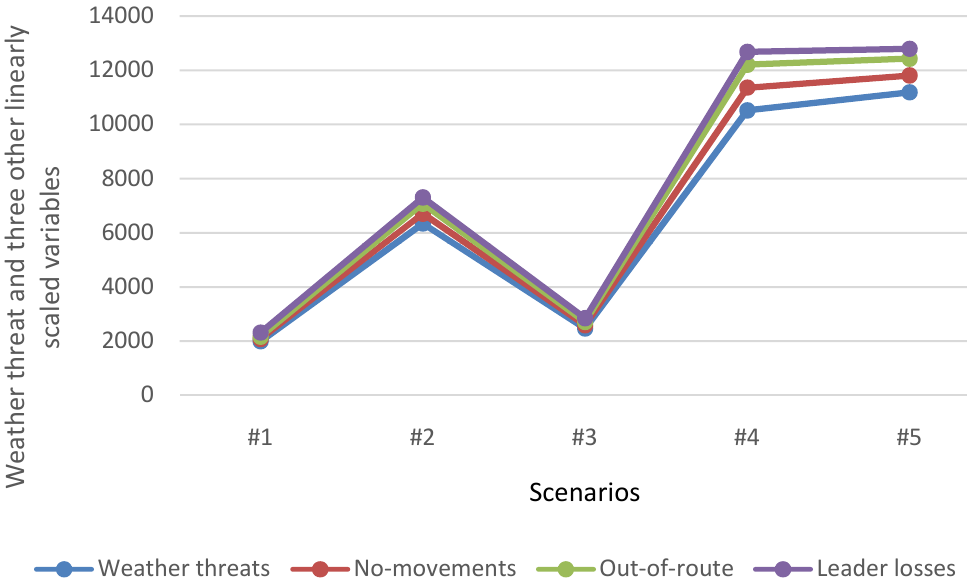}
    \caption{The correlations of cumulated weather threats, no-motions, out-of-routes and leader losses in the particular scenarios}
    \label{fig:correlation-weather-movement-route-lost}
\end{figure*}

The preliminary simulation results have already been discussed in Subsection~\ref{sec:preliminary-simulation},
see Figures~\ref{fig:preliminary-simulations} and~\ref{fig:preliminary-simulation-influence-categories}.
In turn,
Figure~\ref{fig:primary-simulation-influence} 
shows the \erarrr{considered} contextual data \newrrr{under consideration which was} obtained as a result of the primary simulation,
that is a simulation whose general view is provided in Table~\ref{tab:simulation-general-overview}.
It is based on a full-scale analysis of contextual data available to the system.
Comparing Figures~\ref{fig:preliminary-simulation-influence-categories} and~\ref{fig:primary-simulation-influence}, 
we can see the convergence of both simulations, 
that is the preliminary one and the primary one.
This confirms that our simulator,
see Figure~\ref{fig:simulator-screenshot-start},
gives realistic and reliable results.
Moreover,
the correlation between other pieces of data obtained in the primary simulation is high,
which is illustrated in Figure~\ref{fig:correlation-weather-movement-route-lost},
and we have obtained the following three correlation coefficients between the particular pieces of data:
0.9556,
0.9999 and
0.9912.

\newcolumntype{s}{>{\columncolor{Gray}}c}

\begin{table}[!htb]
\caption{Numerical report of weather threats, total numbers and averages for seasons}
\centering
\begin{tabular}{|l|c|c|c|c|c|s|s|}
\hline
\multirow{2}{*}{Scenario} & \multirow{2}{*}{\#1} & \multirow{2}{*}{\#2} & \multirow{2}{*}{\#3} & \multirow{2}{*}{\#4} & \multirow{2}{*}{\#5} & summer & winter\\
 &     &     &     &     &     & (\#1, \#2, \#3) & (\#4, \#5) \\
\hline\hline
\multicolumn{8}{|l|}{\small by threat level}\\
\hline
level E2 & 1774 & 4438 & 1259 & 3352 & 3301 & 2490.3 & 3326.5\\
level E3 & 229 & 1797 & 567 & 4669 & 3114 & 864.3 & 3891.5\\	
level E4 & 0 & 114 & 298 & 1314 & 1696 & 137.3 & 1505.0\\	
level E5 & 0 & 0 & 354 & 1192 & 3084 & 118.0 & 2138.0\\
\hline
\multicolumn{8}{|l|}{\small by particular routes}\\
\hline
Route no.1 & 742 & 1906 & 682 & 2437 & 2087 & 1110.0 & 2262.0\\
Route no.2 & 0 & 1 & 0 & 1 & 4 & 0.3 & 2.5\\
Route no.3 & 55 & 572 & 215 & 1151 & 1180 & 280.7 & 1165.5\\
Route no.4 & 683 & 1135 & 487 & 2487 & 5053 & 768.3 & 3770.0\\
Route no.5 & 64 & 283 & 159 & 765 & 1368 & 168.7 & 1066.5\\
Route no.6 & 32 & 206 & 40 & 523 & 270 & 92.7 & 396.5\\
Route no.7 & 49 & 229 & 57 & 457 & 282 & 111.6 & 369.5\\
Route no.8 & 378 & 2017 & 838 & 2706 & 951 & 1077.7 & 1828.5\\
\hline
\end{tabular}
\label{tab:weather-threats-by-emergency-level-route}
\end{table}

Table~\ref{tab:weather-threats-by-emergency-level-route}
supplements the overall picture of the simulation shown in
Table~\ref{tab:simulation-general-overview} 
by providing an accurate numerical report for threats broken down by 
threat levels and individual routes in the monitored area.
The data which we can see here acknowledges the realistic simulation of the mountain environment. 
The analysis involved plenty of simulation characteristics regarding weather and non-weather threats.
We presented all the scenarios separately,
but also the calculated averages for summer and winter.

\begin{table}[!tb]
\caption{SAT solver processing}
\centering
\begin{tabular}{|l|c|c|c|c|c|}
\hline
Scenario & \#1 & \#2 & \#3 & \#4 & \#5\\
\hline\hline
Number of SAT solver calls & 50 & 81 & 62 & 98 & 113	\\ 
Average response time [m$s$] & 0.51 & 0.40 & 0.42 & 0.40 & 0.43	\\ 
Standard deviation & 0.54 & 0.16 & 0.35 & 0.12 & 0.31\\ 
\hline
\end{tabular}
\label{tab:SAT-solver-responses}
\end{table}

\begin{figure*}[!htb]
	\centering
    \includegraphics[width = 0.8\columnwidth]{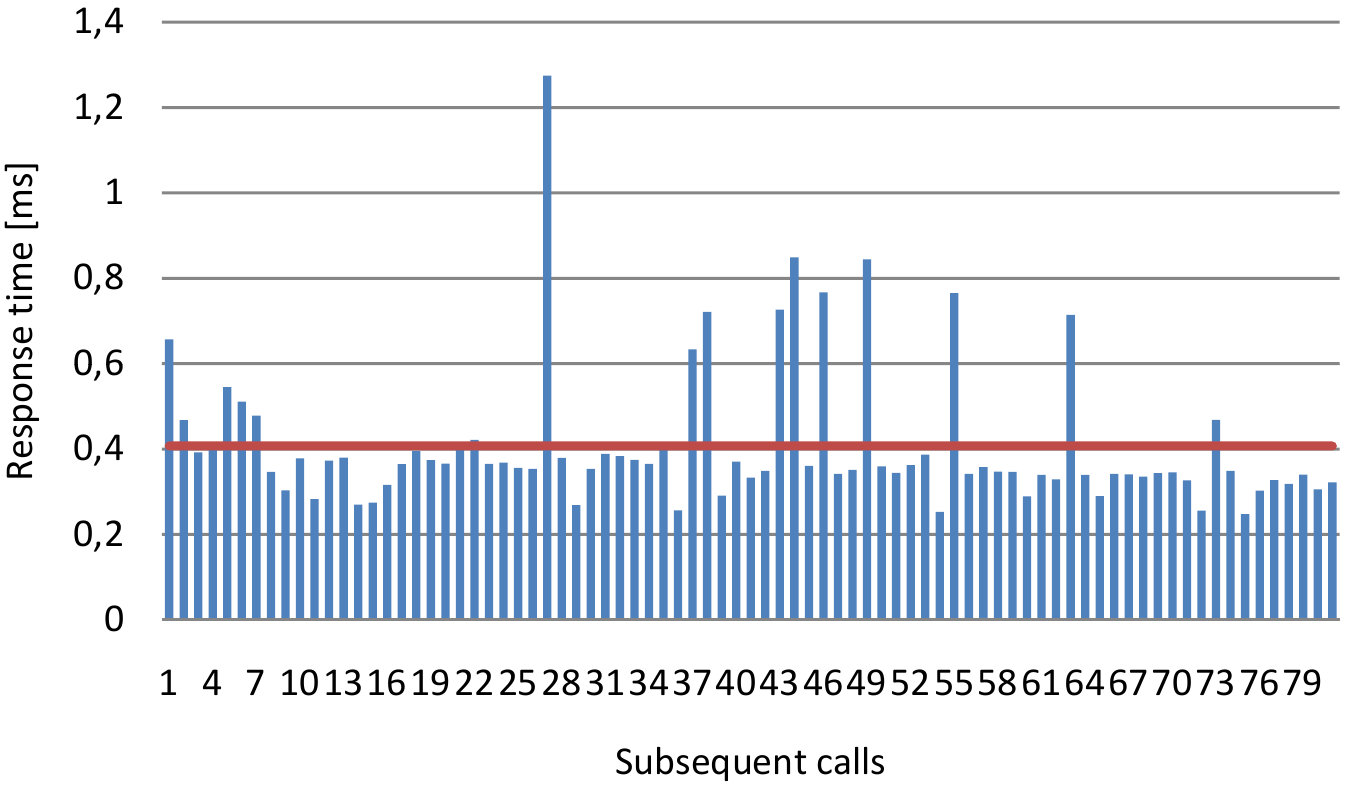}\\
    \includegraphics[width = 0.8\columnwidth]{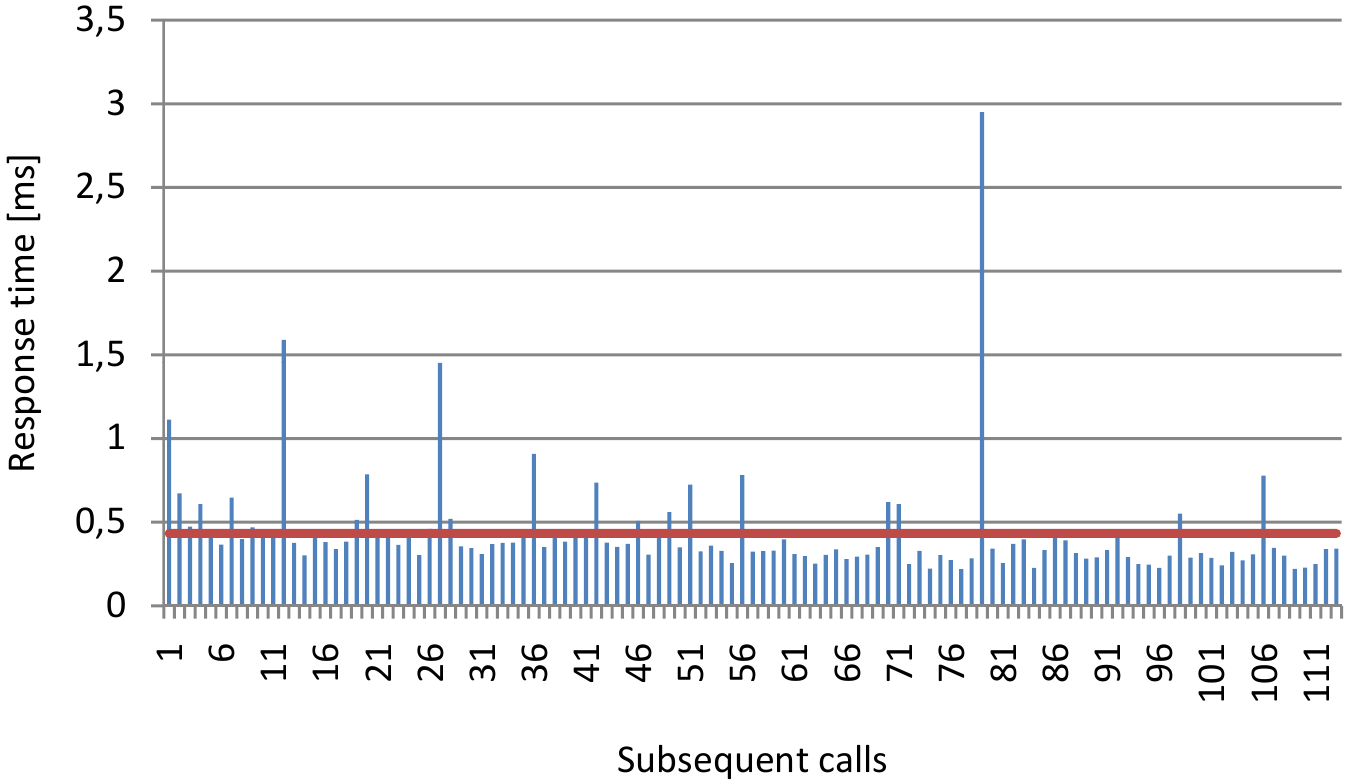}
    \caption{SAT solver response times with averages:
             top -- Scenario~\#2 (summer), bottom -- Scenario~\#5 (winter)}
    \label{fig:SAT-solver-responses}
\end{figure*}

The SAT solver detects logical conditions satisfied by 
the different types of weather threats. 
Table~\ref{tab:SAT-solver-responses}
presents synthetic information about all the solver calls. 
Figure~\ref{fig:SAT-solver-responses}
shows all the calls for two scenario simulations. 
The acquired results prove the system viability in relation to logical reasoning, 
regardless of scenario simulations. 
From time to time, 
we can see longer computing times,
which may be a result of the specificity of 
a particular host computing environment
in which the system worked whilst also being involved in other tasks. 
Nevertheless, 
all the gained response times are acceptable despite 
the fact that the currently used solver 
Sat4j\erarrr{[40]}
is not the most efficient in comparison to the other solvers previously tested. 
However, 
we chose this solver because it is easy to build into the supporting system,
that is when the solver is implemented in Java, just like our system;
and last but not least, it has a good user documentation.
Nevertheless, if necessary, 
it is possible to get much improved testing times for formulas, 
even two or three times better, 
by choosing other solvers, 
see~\cite[Subsection VI.C]{Klimek-2018-Access}.

\subsubsection{Redundancy}

The repetition or \new{the} overflow of information is called \emph{redundancy}. 
Nevertheless, it can also play a positive role by increasing certainty and 
improving the quality of selected operations. 
First of all, such situations occur in relation to the problem of tourist locations. 
Analysing the data from a BTS station is a basic method of localising a tourist. 
Nonetheless, some location data is determined with the use of GPS data taken directly from tourists’ mobile telephones. 
GPS data is \erarrr{far}\newrrr{significantly} more precise than BTS data. 
\era{Nonetheless,}
\new{Yet,}
such detailed geolocation data can be obtained only from those tourists 
who have agreed to share such data, 
most likely when entering the monitored area. 
Then precise geo-positioning of each object is enabled by 
the comparison of BTS data and GPS data, 
provided that the second type of data exists. 
The redundancy issues were already discussed
when solving the geolocation issues.
Firstly,
in the beginning of Section~\ref{sec:context-modeling-utilization}, 
secondly,
briefly mentioned  in the discussion about $Layer0$ in the same section, 
thirdly,
presented in Table~\ref{tab:simulation-general-overview}
as a data subset concerning the current number of tourists.

Now,
we want to present other advantages of redundancy which result from tourist groups. 
The monitored area may be visited by tourists individually and in organised groups. 
A group size is not regulated. Typically, a group counts on 
average 3, 4, 5 people. 
(Sometimes, there are even two-person groups.) 
Therefore, as far as registered groups are concerned, 
one member’s consent to send GPS data may be useful in localising other users, 
even in a situation when the location of other tourists is determined only based on BTS data. 
Table~\ref{tab:redundancy-BTS-GPS} 
shows data on the geolocation of tourists, 
which also gives a good picture of the discussed issue. 
However, 
it should be noted that this is the data of those tourists 
who are still in the monitored area at the end of the simulation,
i.e.\ when stimulation is terminated.
Summing up this thread, 
it should be stated that redundancy is very useful \erarrr{in 
the problem of}\newrrr{in} determining tourist geolocations, 
which proves the effectiveness of the solutions adopted in the designed system, 
as well as \erarrr{its}\newrrr{their} credibility.

\begin{table}[!tb]
\caption{Current information about tourist groups 
or different methods for the localisation of tourist groups
(the picture of the situation after one-hour simulation)}
\centering
\begin{tabular}{|l|c|c|c|c|c|}
\hline
Scenario & \#1 & \#2 & \#3 & \#4 & \#5\\
\hline\hline
Tourists in groups (total number) & 103 & 82 & 102 & 31 & 102\\ 
Groups (total number) & 24 & 22 & 28 & 9 & 24\\ 
\hline
\cellcolor{Gray} Average group size & 4.3 & 3.7 & 3.6 & 3.4 & 4.3\\ 
\hline
Number of BTS located tourists & 90 & 77 & 92 & 27 & 92\\ 
Number of GPS located tourists & 13 & 5 & 10 & 4 & 10\\ 
Locations improved & 49 & 18 & 33 & 9 & 40\\ 
\hline
\end{tabular}
\label{tab:redundancy-BTS-GPS}
\end{table}

\begin{figure*}[!tb]
	\centering
    \includegraphics[width = .75\columnwidth]{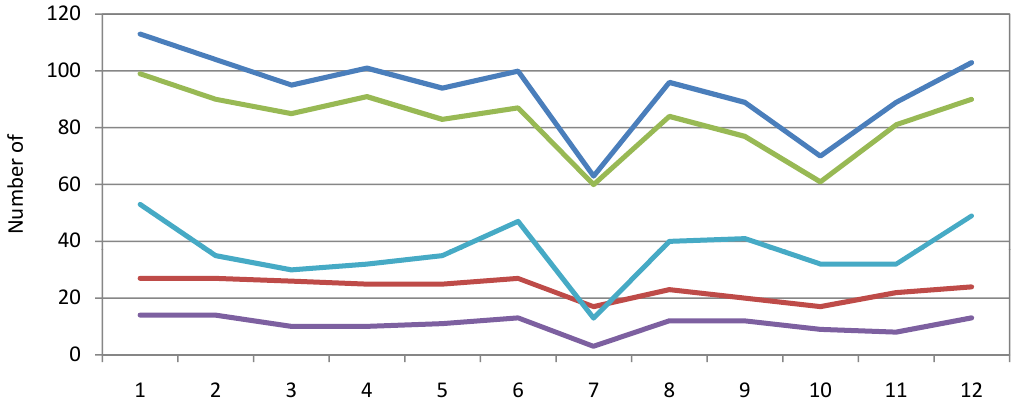}\\
    \includegraphics[width = .75\columnwidth]{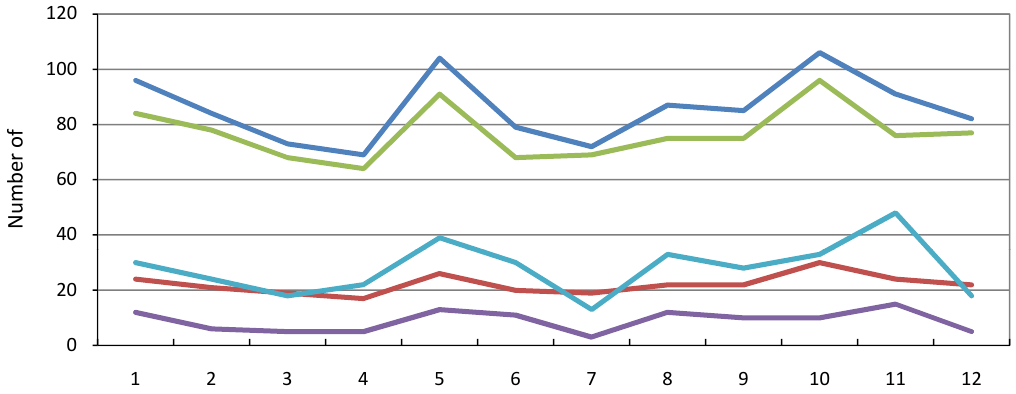}\\
    \includegraphics[width = .75\columnwidth]{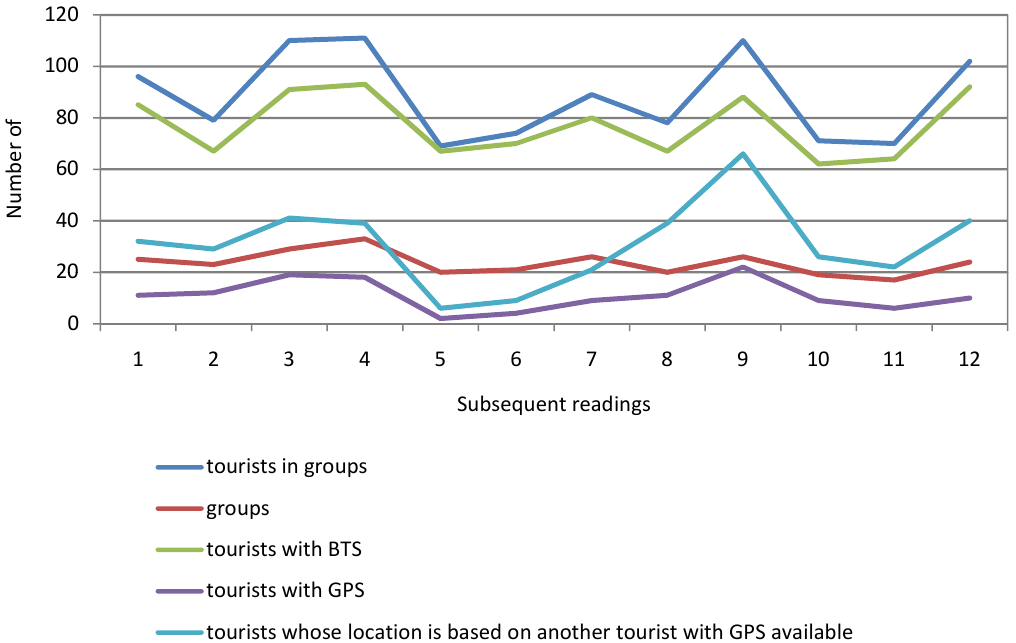}
    \caption{Tourists in groups, location improvements: 
    top -- Scenario \#1 (summer),
    middle -- Scenario \#2 (summer),
    bottom -- Scenario \#5 (winter). 
             (The simulation data is downloaded twelve times at the same time interval)}
    \label{fig:groups-redundancy}
\end{figure*}   

Moreover,
Figure~\ref{fig:groups-redundancy} 
presents other data concerning the presence of tourists in groups,
also related to their geolocations. 
Firstly, 
there are some fluctuations in individual quantities, 
but they are natural and do not raise any doubts. 
Secondly, 
and more importantly,
one can see that some of the data was used to improve 
the geolocation of other tourists, 
and that took place when the location of one tourist 
\era{has been} \new{was} corrected thanks to his/her neighbour in the group 
\newrrr{who} had a more precise location, 
probably resulting from GPS measurements. 
Thus,
it is also a valuable case of the effective use of data redundancy.

\subsubsection{Spatial proximity}

Closeness as a neighbourhood in the space with the object,
also approaching it, is called \emph{spatial proximity}. 
It is a very important factor in the behaviour of an intelligent system, 
for which it is a kind of \new{an} obligatory driver.
It enables \era{the} better adjustment of the pro-active behaviour of 
the system concerning the current situation in which the object is involved, 
and which may change dynamically.
In the case of our system, 
we \era{have} implemented it by reading data from weather stations 
and understanding the current weather conditions for a tourist.
Such a tourist can stay within the range of 1, 2 or 3 weather stations. 
The readings from the closest station are always taken into account, 
and if more than one station is at a similar distance, 
the readings from the station the tourist is walking towards and 
approaching are taken into account. 
The same rules apply to all the tourists concerning reading the weather data.
During the simulation, 
the necessity of choosing  from several stations occurred many times,
and Table~\ref{tab:simulation-spatial-proximity-meteo} 
shows such situations.
However,
due to the large size of the \erarrr{considered} data \newrrr{under consideration}, 
only every fifth tourist was selected randomly and 
the situations \newrrr{only} with \erarrr{only these}\newrrr{such} tourists are \era{shown} \new{presented} in the table. 
The collected data demonstrates well the spatial proximity in 
relation to both the problem itself and its successful implementation.
Downloading weather data from the nearest station, 
or in the direction that the tourist is following, 
is the most valuable and fully desirable procedure.

\newcolumntype{s}{>{\columncolor{Gray}}c}

\begin{table}[!htb]
\caption{Spatial proximity when considering weather readings}
\centering
\begin{tabular}{|l|c|c|c|c|c|s|}
\hline
Scenario & \#1 & \#2 & \#3 & \#4 & \#5 & Together\\
\hline\hline
Number of tourists  & 334 & 365 & 324 & 401 & 382 & 1806\\ 
Total number of events    & 1620 & 1723 & 1614 & 1737 & 1772 & 8466\\ 
\quad including one station & 427 & 486 & 462 & 474 & 453 & 2302\\
\quad including two stations & 909 & 979 & 885 & 983 & 1033 & 4789\\
\quad including three stations & 284 & 258 & 267 & 280 & 286 & 1375\\
Number of events per tourist (average) & 4.9 & 4.7 & 5.0 & 4.3 & 4.6 & 4.7\\ 
\hline
\end{tabular}
\label{tab:simulation-spatial-proximity-meteo}
\end{table}

\subsubsection{Context transition}

\begin{table}[!tb]
\caption{Context transition in relation to tourists who left the monitored area}
\centering
\begin{tabular}{|l|c|c|c|c|c|}
\hline
Scenario & \#1 & \#2 & \#3 & \#4 & \#5\\
\hline\hline
Number of tourists & 3088 & 3123 & 2870 & 3351 & 3060 \\ 
Number of transitions per tourist (average) & 23.26 & 22.34 & 24.50 & 20.36 & 21.58 \\ 
Minimum/maximum value & 3/250 & 4/166 & 3/148 & 3/146 & 3/127 \\ 
Standard deviation & 20.82 & 19.41 & 20.61 & 17.11 & 17.67 \\ 
\hline
\end{tabular}
\label{tab:simulation-context-transitions}
\end{table}

\begin{figure*}[!htb]
	\centering
    \includegraphics[width = .8\columnwidth]{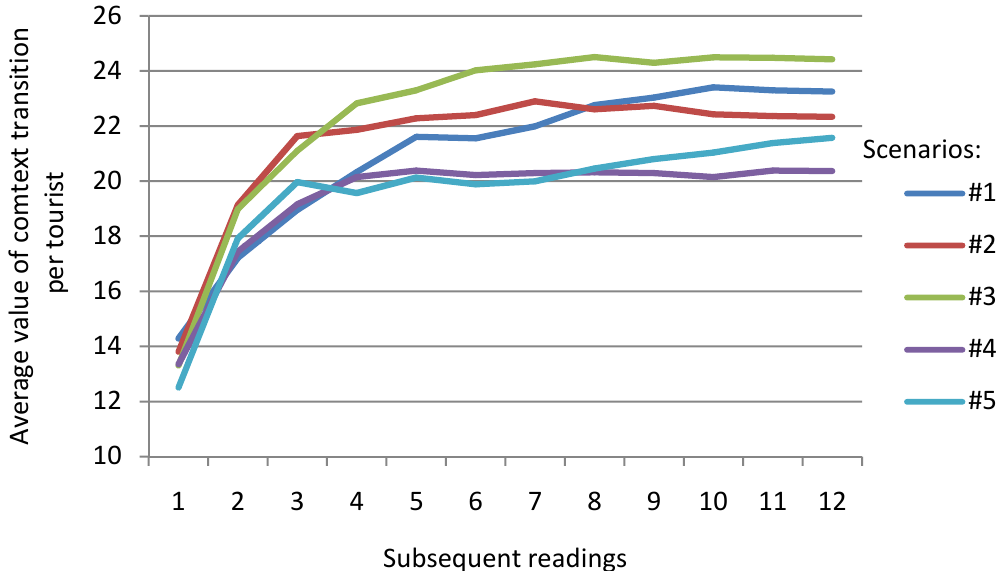}
    \caption{Context transition understood in terms of a context changing per tourists, average values.
             (The simulation data is downloaded twelve times at the same time interval)}
    \label{fig:simulation-context-transitions-time}
\end{figure*}   

The change or shift of the environment, often dynamic or rapid, 
as a result of the smart operation of an intelligent system that 
affects the object is called \emph{context transition}.
The context change concerning each tourist is 
quite natural and can happen quite often while hiking along the trails.
This change results, first of all, from weather fluctuations that 
occur during particular hiking; 
another example is the change of the trail and the assigned difficulty level, 
or the beginning of the night time. 
Each of the subsequent situations does not require 
the previous one to observe and record a context change, 
that is trail difficulty changes do not require 
a change in weather conditions, 
and the beginning of the night time itself causes a significant change of 
the context for every tourist hiking along the trails.
The dynamically changing context is registered for every tourist.
Table~\ref{tab:simulation-context-transitions}
shows the data for the context transition.
However,
due to the size of the registered data, 
it is limited to tourists who left the monitored area,
that is \erarrr{they} finished their hiking.
We conclude that the obtained simulation is stable and plausible,
providing reliable results.
It should also be noted that we obtained similar results for 
both summer (\#1, \#2, \#3) and winter (\#4, \#5) scenarios; 
while in winter, 
although it depends on the region itself, 
there is usually a lower number of wandering tourists comparing to summer. 
For this simulation, 
we \era{have} \new{did} not \era{decided} \new{decide} to reduce their numbers so that in 
more difficult winter conditions we observe similar simulation results.
The \era{obtained} results \new{obtained} should be considered as natural and common-sense, 
which allows to state that the processing in the established context life cycle behaves correctly, 
as well as we observe the results of the proper implementation of internal decision-making processes.

\begin{figure*}[!htb]
	\centering
    \includegraphics[width = 0.81\columnwidth]{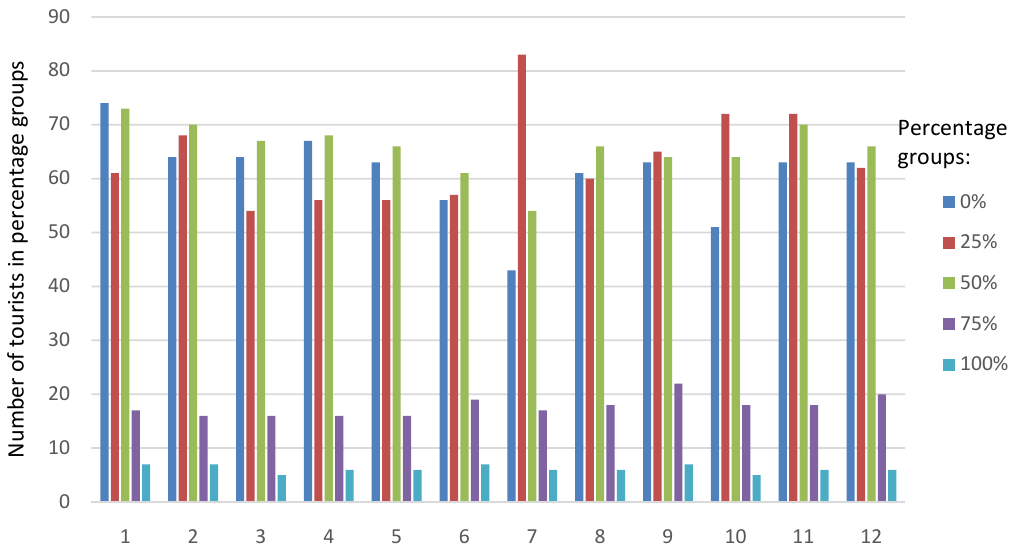}\\
    \includegraphics[width = 0.81\columnwidth]{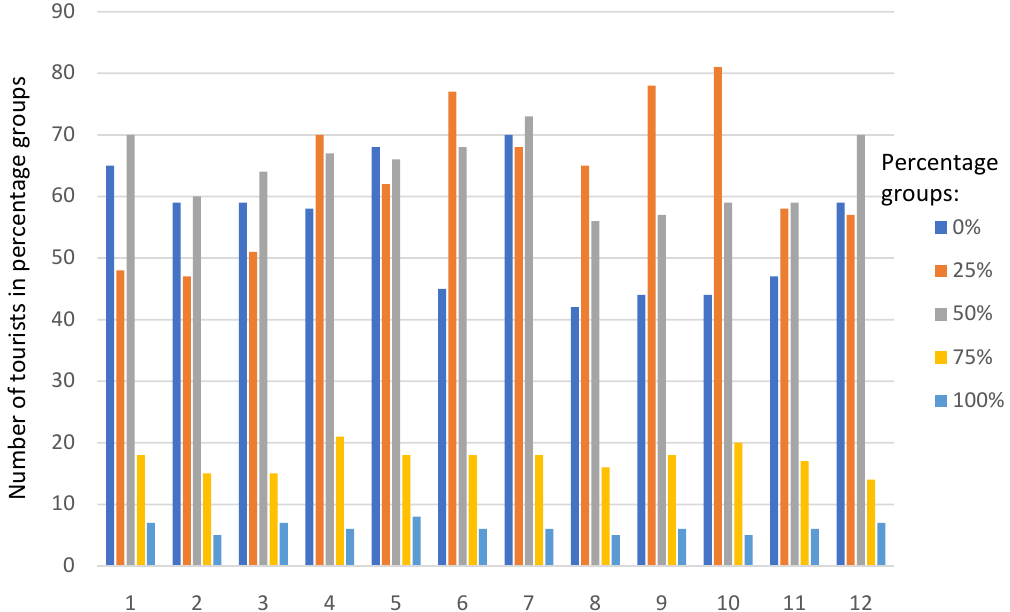}\\
    \includegraphics[width = 0.81\columnwidth]{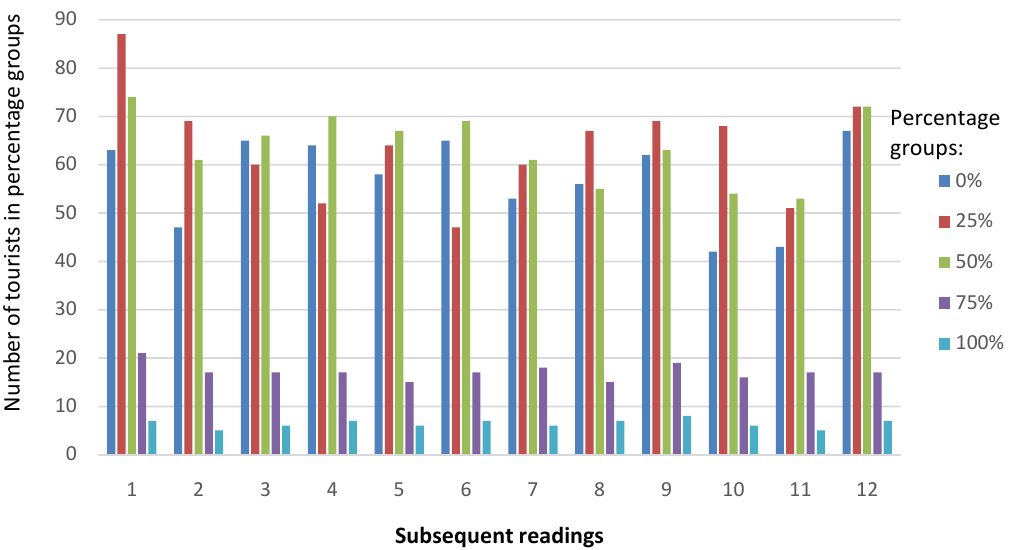}
    \caption{Context sharing expressed in terms of five percentage groups:
             top -- Scenario~\#1 (summer), 
             middle -- Scenario~\#3 (summer), 
             bottom -- Scenario~\#5 (winter).
             (The simulation data is downloaded twelve times at the same time interval)}
    \label{fig:simulation-context-sharing}
\end{figure*}

Figure~\ref{fig:simulation-context-transitions-time} 
shows the problem of the context transition for particular weather scenarios. 
Of course, we again observe lower values at the beginning of the simulation, 
which is related to the slow filling of the monitored area by hiking tourists 
(before the simulation, the area was empty), 
and this filling continues until the natural levels are reached. 
It should be acknowledged that the next experiment turned out to 
be successful and the system proved its credibility.
The processes illustrating the contextual property transformations run naturally and as expected.

\subsubsection{Context sharing}

The overlapping or similarity of the same contextual data of different objects
or a contextual data community
is called \emph{context sharing}.
Tourists can share their information or knowledge, 
which shows their perception of the environment through particular objects, 
see also~\cite{Zimmermann-etal-2007}. 
The experiment was organised as follows: 
particular contextual data was identified and extracted,
as well as the percentage ranges for data sharing were introduced and examined.
\erarrr{It is a result of}\newrrr{This results from} the fact that \erarrr{due to} the \newrrr{data} size\erarrr{ of the data}, 
\erarrr{we must}\newrrr{forces us to} introduce a certain gradation, 
as it is practically impossible to analyse \newrrr{all the pieces of data}, 
colloquially, peer-to-peer\erarrr{, all the pieces of data}. 
Figure~\ref{fig:simulation-context-sharing}  
\era{shows} \new{presents} the \era{obtained} experiment results \new{obtained}. 
Sharing at levels 0\%, 25\% and 50\% is the most common because 
we have a relatively large and extensive monitoring area 
with highly diversified conditions, 
which is mapped into contextual data. 
Whereas,
sharing at level 100\% is relatively rare and it is also fully understandable. 
Changes concerning particular percentage groups occur mildly and naturally.
We \era{conclude} \new{can draw conclusion } that the results of our experiments prove
the reliability of the system processing contextual data and the simulation processes performing.

\section{Discussion}
\label{sec:discussion}

\begin{figure*}[!htb]
	\centering
    \includegraphics[width = 1\columnwidth]{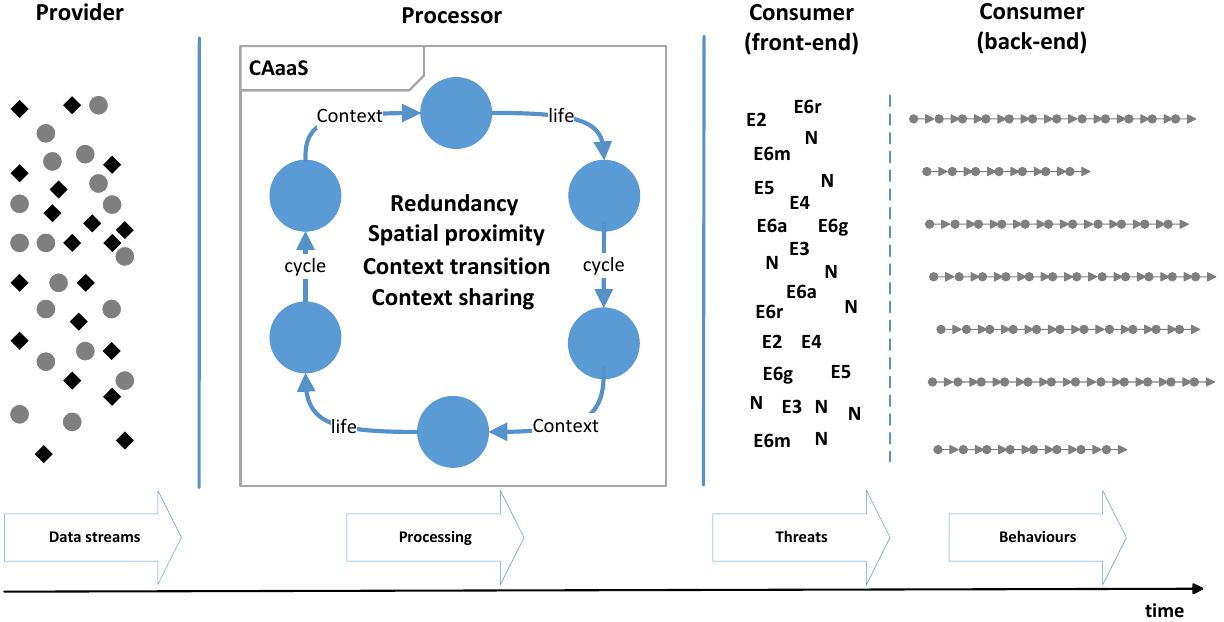}
    \caption{Contextual data processing model as a CAaaS component: 
    left -- sensor pieces of data (two types of sensor data, that is weather and geolocation), 
    middle -- the primary contextual processing,
    right -- threats and derived behaviours. 
    Behaviours as finite sequences of particular threat assessment points
    \includegraphics[width = 3.5mm]{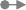} (grey pictograms)}
    \label{fig:summary-processing}
\end{figure*}

A summary of the obtained findings is carried out now.
Figure~\ref{fig:summary-processing} illustrates and recapitulates the basic article concepts
for which achievements relate to
the framework of contextual data processing and utilization,
as well as the proposal of a CAaaS system for mountain rescue operations.
Another important achievement is the generation, by the supporting system, 
information and warnings about threats to hiking tourists and 
readiness to provide their behavioural traces.
The \era{obtained} data \new{obtained} can be used in various ways.
In the case of front-end consumers,
that is rescuers,
this is the information about the threats of monitored tourists,
which allows for faster, on-line and effective rescue actions,
if needed.
In the case of the back-end consumers,
that is analysts,
we can obtain large data sets
which can be subjected to an in-depth analysis \era{and} \new{to}
provide answers in relation to the nature of tourism in a given area,
\era{or} \new{and} to enable \new{tourist} models calibration,
if we treat the tourist traffic in the park as a separate eco-system.

We introduced context categories,
see Figure~\ref{fig:context-model-categories},
and appropriate decisions in this regard 
\era{significantly} influence \new{significantly} the further supporting system design.
In general, 
categories may have common characteristics,
and therefore, following Perera et al.~\cite{Perera-etal-2014},
Table~\ref{tab:context-categories-relationship} shows
an assessment of their mutual relations.
We argue that relationships are rather weak,
which results from the strong encapsulation of 
particular contextual data in individual categories.
This is a positive and intended effect to obtain 
a parallelization in the proposed context life cycle,
see Figures~\ref{fig:context-life-cycle-basic} 
and~\ref{fig:context-life-cycle-workflow}.

\begin{table}[!htb]
\caption{Context categories and their relationship.
(``S'' means strong, ``M'' means moderate and ``W'' means weak)}
\centering
\begin{tabular}{|l|c|c|c|c|c|}
\hline
 &
\begin{sideways}
Individuality 
\end{sideways}
  & 
\begin{sideways}
Time 
\end{sideways}
 & 
\begin{sideways}
Location 
\end{sideways}
 & 
\begin{sideways}
Activity 
\end{sideways}
 & 
\begin{sideways}
Relations 
\end{sideways}\\
\hline
Individuality & \cellcolor{lightgray} & M & W & W & W \\
\cline{1-1}\cline{3-6}
Time & \cellcolor{lightgray} & \cellcolor{lightgray} & M & M & M \\
\cline{1-1}\cline{4-6}
Location & \cellcolor{lightgray} & \cellcolor{lightgray} & \cellcolor{lightgray} & W & W \\
\cline{1-1}\cline{6-6}
Activity & \cellcolor{lightgray} & \cellcolor{lightgray} & \cellcolor{lightgray} & \cellcolor{lightgray} & W \\
\cline{1-1}\cline{6-6}
Relations & \cellcolor{lightgray} & \cellcolor{lightgray} & \cellcolor{lightgray} & \cellcolor{lightgray} & \cellcolor{lightgray} \\
\hline
\end{tabular}
\label{tab:context-categories-relationship}
\end{table}

Context categories and their evaluation are discussed in some articles.
Table~\ref{tab:context-categories-characteristic}
compares the proposed categories with other classification schemes known from other articles.
Schilit et al.~\cite{Schilit-etal-1994} 
introduced a taxonomy based on three popular questions:
\textit{Where you are},
\textit{Who you are with}, and
\textit{What resources are nearby}.
The taxonomy is rather old and not all the categories apply.
Hendricksen and Indulska~\cite{Henricksen-Indulska-2005} 
\era{showed} \new{presented} a taxonomy 
in relation to operational techniques:
\textit{sensed} -- directly sensed,
\textit{static} -- data will not change over time, 
\textit{profiled} -- data changes with a low frequency, and
\textit{derived} -- information computed using primary context.
Van Bunningen et al.~\cite{van-Bunningen-etal-2005}
introduced a taxonomy based on two aspects only:
\textit{operational categorisation} -- how they were gathered and treated, and
\textit{conceptual categorisation} -- the conceptual relationships between the contexts.

\begin{table}[!htb]
\caption{Evaluation of context categories}
\centering
\begin{tabular}{|l|c|c|c|}
\hline
 & 
 \begin{tabular}{c}
 Where/Who/What\\ \\
 (Schilit et al.~\cite{Schilit-etal-1994})
 \end{tabular}
 & 
 \begin{tabular}{c}
 Operational \\ categorisation\\
 (Hendricksen and \\ Indulska~\cite{Henricksen-Indulska-2005})
 \end{tabular}
 & 
 \begin{tabular}{c}
 Operational vs.\ \\ conceptual\\
 (van Bunningen \\ et al.~\cite{van-Bunningen-etal-2005})
 \end{tabular}
 \\ 
\hline
Individuality & Where &  Sensed & Operational\\
Time & not applicable & Profiled & Operational\\
Location & Where & Sensed & Operational\\
Activity & not applicable & Derived & Operational\\
Relations & Who & Derived & Operational\\
\hline
\end{tabular}
\label{tab:context-categories-characteristic}
\end{table}

Another recapitulation is the assessment of contextual variables for the supporting system,
see Table~\ref{tab:context-categories-variables-summary}.
It assesses both the contextual categories and contextual pieces of data,
see Rosenberger and Gerhard~\cite{Rosenberger-Gerhard-2018}.
\emph{Data sources} define the data origin.
\emph{Stability} defines exposition to change during processing.
\emph{Change} defines the frequency of swapping or alternating of contextual data.
\emph{Usage} defines the frequency of context use.
They refer to the metaphors described in Subsection~\ref{sec:metaphor},
but there the transformations of the contextual pieces of data \era{have been} \new{were} considered
and here we consider context characteristics. 

\begin{table}[!tb]
\caption{Characteristics of a contextual category and contextual data in the supporting system 
when processing in the proposed context life cycle. 
(WFTR means tags for wind, fog, temperature and rain considered together)}
\centering
\begin{tabular}{|l|c|c|c|c|c|}
\hline
Category & Data & Data source & Stability & Change & Usage\\
\hline\hline
\multirow{3}{1.8cm}{Individuality}
 & Avalanche & Expert & Volatile & Occasionally & Low\\
 & WFTR & Sensor & Volatile & Frequently & High\\
 & Difficulty & Expert & Constant & Stable & High\\
\hline
\multirow{2}{1.8cm}{Time}
& Summer/winter & Clock & Volatile & Constantly & Low\\
 & Day/night & Clock & Volatile & Constantly & High\\
\hline
Location & On/off trail & Sensor & Volatile & Sometimes & Low\\
\hline
Activity & Motion/no motion & Sensor & Volatile & Sometimes & Low\\
\hline
\multirow{2}{1.8cm}{Relations} 
 & Distance from the leader & Sensors & Volatile & Sometimes & Low\\
 & Dangerous animals & Sensors & Volatile & Sometimes & Low\\
\hline
\end{tabular}
\label{tab:context-categories-variables-summary}
\end{table}

The proposed model of contextual data processing implemented and visualised in the form of a workflow allows 
\era{taking} to \new{include} various sensor sources, pieces by pieces, thanks to the proper division of workflow activities.
Input data streams mirror five weather scenarios. 
\era{This} \new{These} pieces of data are generated artificially, but they reflect the factual weather conditions obtained from a national park.
The context life cycle embedded in CAaaS enables efficient data utilization and interactions with users. 
All the phenomena typical \era{to} \new{for} contextual data processing are shaped smoothly and naturally, which we \era{have shown} \new{demonstrated} in numerous diagrams. 
The numbers of hiking tourists are reliable and were obtained from the National Park authorities.
The supporting system reacted correctly and on time. 
Average response times for solvers,
see Table~\ref{tab:SAT-solver-responses} or Figure~\ref{fig:SAT-solver-responses},
taking into account the fact that the solver is called four times in a single loop,
see Formula~(\ref{for:algorithms-loop}),
are a good explanation of this fact,
that is keeping up with the phenomena occurring in the monitored environment.

The results achieved, being an outcome of simulation experiments, 
proved the supporting system feasibility, reliability and vitality, and \newrrr{demonstrated} that it can work as CAaaS.
It is also worth paying attention to a certain convergence of the preliminary simulation and the primary simulation results. 
The preliminary simulation followed directly the workflow and was based on a general idea of how \era{should} the expected structure of activities for 
detecting threat messages should look like, 
and the primary simulation was the result of a full context life cycle processing from raw sensory data to target threat messages. 
The implemented system supporting rescuers is capable of sensing mountain environments, and assisting rescue operations effectively. 
The number of stages of the proposed context life cycle, 
as well their internal organization,  may constitute
a benchmark for similar context-aware systems.
The same is with the classification and the proposed hierarchy of contextual data.

Dey and Abowd~\cite{Dey-Abowd-1999}
highlighted the essential features of the mature context-aware application and, 
based on articles by 
Schilit et al.~\cite{Schilit-etal-1994} and 
Pascoe~\cite{Pascoe-1998},
proposed a taxonomy,
which is also satisfied for our supporting system:
\begin{enumerate}
\item
\textit{\emph{presentation} of information and services to a user} --
the supporting system provides smart decisions as a combination of various operations on contextual data;
\item
\textit{automatic \emph{execution} of a service} --
the supporting system provides smart decisions automatically, 
aiding rescuers on-line,
hence,
a context is a trigger for an action~\cite{Schilit-etal-1994};
\item
\textit{\emph{tagging} of context information for a latter retrieval} --
it should be noted that Pascoe~\cite{Pascoe-1998} considered it as a kind of contextual augmentation,
and then we can cover this idea with \new{the} languages of threats and behaviours as 
the derivatives of the supporting system operations.
\end{enumerate}

In the last paragraph of Subsection~\ref{sec:motivation},
we argued for a system that is pro-active, non-intrusive and up-scalable.
Our system is pro-active since it enables
discovering threats in anticipation of future problems.
Its intention is to produce hints for rescuers in order to avoid a problem 
rather than wait until a problem occurs.
Our system is non-intrusive since it works transparently for hiking tourists.
It does not require \erarrr{the}\newrrr{involving} tourists\erarrr{' involvement},
nevertheless, 
they are protected by providing the rescuers with detailed behavioural and 
threat-oriented information.

The supporting system is also up-scalable.
It can operate in large-scale mountain environments with many routes, tourists and their groups.
With the present simulation experiments,
based on the realistic data obtained from the Babia G{\'{o}}ra National Park, 
we \era{have} \new{did} not \era{encountered} \new{encounter} any time constraints, that is time-outs.
We conducted experiments with the various volumes of tourist traffic, 
ranging from small to larger, 
but the typical volume of traffic established for the experiment, 
and thus exceeding the park traffic, 
is the amount shown in the first row of Table~\ref{tab:simulation-general-overview},
as well as illustrated in the following tables.

We would like to identify where\erarrr{, as it stands,}\era{is} the limit of the system's capacity \new{is}.
Let us note that we tested different levels of tourist traffic.
Although the highest traffic is in August, 
see Table~\ref{tab:BPN-tourist-turnout}, 
about 1000 people a day, 
and this is the peak of traffic, 
we tested the traffic of up to 3000 people, 
see the first row in Table~\ref{tab:simulation-general-overview}.
We conducted further tests (Intel i7, gen.~11, 16GB RAM)
increasing the tourist traffic, 
and we obtained the following results:
\begin{itemize}
\item
the system is fully efficient with 7,000-8,000 tourists,
\item
a \erarrr{noticeable}\newrrr{significant} drop in system performance is \newrrr{observed} with about 10,000 tourists;
\item
the system is inefficient with more than 12,000 tourists.
\end{itemize}
Obviously, 
these are the numbers of visitors that \era{far} exceed considerably the capacity of 
the National Park,
where 1000 tourists at their peak cause \era{embarrassment of} \new{a problem to} the park authority
\era{(on the one hand, we are quiet with visitors, but on the other hand, there are too many of them at once)},
and numbers several times higher would probably close the park.
Thus,
the current simulations,
see Tables~\ref{tab:simulation-general-overview}--\ref{tab:BPN-tourist-turnout},
are convincing in terms of the number of hiking tourists.
The current processing times are low and increasing the number of hiking trails 
does not seem to increase \erarrr{significantly} the response times \newrrr{greatly}.
We have a large reserve in relation to the basic time requirement that 
messages concerning threats should be updated \newrrr{from} every half minute to a minute.
Therefore, 
we conclude that the supporting system reacts properly to environmental data streams and signals,
that is without exceeding the expected time responses.
It was demonstrated by numerous experiments showing the smoothness of the simulation, 
that is the reaction of the supporting system, 
or the short times of the SAT solver. 
Only with really high, even unrealistically extreme loads, 
we encounter \era{across} jams in the system.

\era{On the other hand,}
\new{Furthermore,}
there is \era{one more} \new{another} important argument for the good efficiency of the supporting system. 
The processes taking 25 hours in nature \era{took} \new{lasted} one hour in the performed simulation, 
so \era{it was} \new{they were} \era{significantly} accelerated \new{significantly}. 
Despite this, all \new{the} planned processes in relation to context processing, 
decision making, in general all the processes, were within the demanding time regimes. 
With the actual use of the supporting system, 
that is without said acceleration,
environmental events would be several times slower.
Hence, we conclude that the fears of not keeping up with the real phenomena are unfounded.
Last but not least,
there is another option to improve system performance. 
Currently, the system has \newrrr{the} built-in mechanisms that dump contextual and environmental data to system logs
(to obtain, for example,
Tables~\ref{tab:simulation-general-overview},
\ref{tab:weather-threats-by-emergency-level-route},
\ref{tab:redundancy-BTS-GPS}--\ref{tab:simulation-context-transitions}
or Figures~\ref{fig:groups-redundancy}--
\ref{fig:simulation-context-sharing}).
It is a very time-consuming activity 
and \era{resignation} \new{resigning} from it would \era{significantly} improve 
the efficiency of the system \new{essentially}.

We \era{have} considered the large and realistic numbers of tourists visiting the monitored area 
as well as experiments with extensive weather scenarios which mirror \newrrr{the} real data.
Nonetheless, 
we observed the naturalness of the occurrence of many typical properties
for intelligent systems, such as redundancy,
spatial proximity, context transition or context sharing.
These prove that they reflect the internal processes in the system \era{well} \new{adequately}.
The supporting system had no problems processing events that follow
both weather and non-weather threats.
The effectiveness of our solutions was validated by numerous experiments.
All our experiments were carried out on a localhost,
which means that all the phenomena obtained from sensors, but also the geolocation pieces of data,
are located on this host, as well as further reasoning processes providing smart decisions.
Albeit,
it may be expected that placing the supporting system in
a cloud will not cause any radical delays in response to events.
Transmission in a network may be characterized by certain inertia
but by adding possible delays, we still can ensure timely reactions.
Only in the case of a hypothetical situation when one supporting system has to deal
with numerous monitored mountain areas simultaneously,
it may be necessary to examine this issue separately.
We believe that the proposed context life cycle is a fine balance in relation to system requirements,
processing speed and the readability of the division into stages,
and CAaaS is a good proposition for national park managers.

\section{Related works}
\label{sec:related-works}

\subsection{Context}

A considerable \erarrr{amount}\newrrr{volume} of literature has been published on contextual data and context-aware systems.
Dey and Abowd~\cite{Dey-Abowd-1999}
provided probably the first definition of a context.
Their context definition covers many different aspects which are 
important for a typical context-aware system working pro-actively.
We started with this definition because it is simple and convenient for our system and research.
A paper by Bazire and Br{\'e}zillon~\cite{Bazire-Brezillon-2005} 
shows a corpus of over a hundred \new{of} definitions of a context in several fields.
In \era{the} \new{a} paper by Zimmermann et al.~\cite{Zimmermann-etal-2007}
many issues are discussed, 
but from our point of view, what is important, 
it is a bold look at the problem of a taxonomy for contextual data,
see also a paper by Perera et al.~\cite{Perera-etal-2014}.
The definition provided as the first one appears not to be specific enough
and it is not useful in governing the intricacy of a modern software system.
Hence, 
based on that,
we suggested a classification and categorisation which should be 
the most proper for the mountain environment we consider.
Furthermore, in order to select data to match a given context, 
we applied the suggestions of Crowley, 
see an article~\cite{Crowley-2003}, 
that is concentrating solely on the adequate elements and relationships.
Moreover, we determined the contextual elements in line with the context and system requirements, 
see an article by Vieira et al.~\cite{Vieira-etal-2011}.
We consider the transfers of causations and other relations for events,
even if not calling it a metaphor for interpretation, 
as we use metaphors in this article,
see a paper by Aggeri et al.~\cite{Aggeri-etal-2007},
but it seems absorbing to examine the semantic dependencies
and we see some opportunities to adapt that approach.
To sum up,
we started with the simplest concepts concerning a context, 
introduced the hierarchy and categorisation, 
then we focused on specific relations and system requirements, 
and adopting them to show their semantic relationships (metaphors), 
all for the mountain environment.
The issue of a context life cycle is discussed in various articles,
see a survey by Perera et al.~\cite{Perera-etal-2014}.
Many models have been surveyed, 
\erarrr{but they} but they limit to the simple phases of 
collecting, modelling, reasoning and broadcasting.
We introduced a richer model,
consisting of more phases,
and we made some activities parallelized,
depending on heterogeneous data streams.
We also identified the basic data sets processed,
taking \era{into account} the mountain environment \new{into account}.

A study conducted in an article by Augusto et al.~\cite{Augusto-etal-2018}
provides an important, valuable and history-oriented journey through the concept of a context.
It also presents connections for Artificial Intelligence (AI) and 
Intelligent Environments (IE)~\cite{Augusto-etal-2013}.
An exhaustive and intricate survey as regards 
the engineering features for context-aware applications is 
included in \era{the} \new{a} paper by Alegre et al.~\cite{Alegre-etal-2016}.
It tackles not only developing methodologies but also it
provides much guidance on the perception and engineering approach to context-aware systems.
The ideas presented result from the authors' experience.
According to Hong et al., 
see \era{work} \new{a paper}~\cite{Hong-etal-2009}, 
there are not many research articles which include development guidelines 
how to reduce the complexity of such systems.
It follows that it is necessary to apply an adequate device infrastructure,
as well as 
to improve and adapt various techniques when modelling a context, 
see also a paper by Hendricksen and Indulska~\cite{Henricksen-Indulska-2005}. 
On the other hand, a paper by Alegre et al.~\cite{Alegre-etal-2016}
concludes that there are no reliable design techniques.
To sum up,
we presented an integrated approach to the identification of a context,
introducing its categories,
and determining those elements that are relevant to our specific mountain environment applications.

The context issues were explored intensively,
which explains why there are so many survey and review articles.
A paper by Lin et al.~\cite{LiXin-etal-2015} discusses middleware architectures for
context-aware systems.
We introduced the logic-based approach since 
we need rich expressiveness that is suitable for
high-level information, with the possibility of
defining or re-defining crucial logical constraints we use,
expressing mountain alerts, 
in other words, 
threats for mountain environments.
We use processed contextual data and predefined alerts as the basis for the main deductive reasoning processes 
in our system.
\erarrr{By the way,}\newrrr{In addition,}
sensor data distribution in the form of a subscribe/publish mechanism using 
message brokers is not commonly used,
and we use it.
Having said that,
middleware architectures are natural for context-aware systems,
and although the cited article considers sensing-as-a-service middleware, 
the solution we \era{have} decided to promote is not a common one.
To sum up,
our CAaaS solutions, 
also \erarrr{due to}\newrrr{on account of} the domain specificity, 
differ from the majority of those presented in~\cite{LiXin-etal-2015}.

\erarrr{A paper by Roy et al.~[55]
discusses context data combined with an ontology-based semantic network 
to facilitate efficient context delivery. 
The proposed solutions are particularly convenient when
context information is ambiguous or heterogeneous,
and it is not the case here.
Incidentally,
the context uncertainty or vagueness problem is important, 
and the ontological approach is a preferred solution,
see a paper by 
Almeida and L{\'{o}}pez{-}de{-}Ipi{\~{n}}a~[3].
A paper by Xu et al.~[70] 
surveys many different methods of context modelling;
however, the \era{paper's aim} \new{aim of the paper} is to propose a system
which can use any information type
when evaluating the significance of context handling.
By tracking the log data of users 
and performing mining with OLAP (OnLine Analytical Processing),
we obtain information concerning new context preferences.
The system is self-learning and a new basic architecture for reasoning processes is proposed.
The proposal is original, it aims to establish certain universalism 
regarding context definition and reasoning,
nevertheless, 
our approach does not require this.
We focus on the specific and fixed categories of information.}

\subsection{Information science}

\new{The scope of information science, 
as stated in~\cite{Stock-Stock-2013},
is very broad,
so we will discuss selected topics in relation to contextual information and its processing.
Relying on contextual data for recommendation systems is a clear and popular trend.
Dridi et al.~\cite{Dridi-etal-2020}
consider using contextual information in prediction for recommender systems.
Contextual information is taken into account, such as
current time and location, calendar, and ongoing social situation.
Inferring the user current situation by applying fuzzy logic is proposed.
We consider a much larger range of contextual information,
which was additionally hierarchized,
but more important is that inferences about risks to tourists are sharp,
because without any approximations the situation of each tourist in the monitored area should be determined.
Anagnostopoulos and Kolomvatsos~\cite{Anagnostopoulos-Kolomvatsos-2016}
present an approach for scheduling a contextual information process over incomplete contextual data streams in the Internet of Things (IoT).
It includes tasks like data fusion, concept drift detection and predictive analytics.
It involves the continuous evaluation of functions over some contextual attributes.
The article aims to address the challenge of dealing with incomplete data in the IoT environments and proposes an intelligent scheduling mechanism.
In our case, we are not considering incomplete data, 
but rather we are dealing with a large number of various contextual data, 
which we organize and hierarchize in order to make good conclusions about threats.
Context analysis is also considered in an article by Ji et~al.~\cite{Ji-etal-2022},
however, this is a different area related to the salient object detection when processing with neural networks,
and the dependency relationship between the information pixels.
The relation of contextual extension proposed in the article is interesting,
but in our approach we used hierarchization and categorization of information,
which may be a certain equivalent, although it is necessary to pay attention again to the dissimilarity of the areas.}

\erarrr{Pajares Ferrando and Onaindia~[47]
propose an interesting approach to Multi-Agent planning in intelligent environments,
where multiple agents work together, share resources, activities and goals.
It has been tested in the applications of ambient intelligence in the field of health-care.
Although the system involves various components, including ambient agents and artifacts, 
it is more focused on the intelligent environment and ambient intelligence,
and less on contextual information as we recognize it as an important data driver.
In addition, our system is not agent-based, so we do not solve the issue of assigning individual functionalities to agents.
The occurrence of mountain threats that follow context processing is a trigger for rescuers who take action themselves.
Nevertheless, the cited article is comprehensive and interesting, 
and could be a good basis for system agentification.}
\new{Zhu et al.~\cite{Zhu-etal-2014}
also consider the context but in relation to flowable services in cloud computing.
It focuses on users context for service provision,
for context capturer components to seamlessly integrate services based on users' needs and preferences.
The proposed framework aims to reduce service delivery costs.
It introduces a certain context categorization adapted to the specifics of cloud computing.}
\erarrr{Moulouel et al.~[44]
propose an ontology-based hybrid commonsense reasoning framework for handling context abnormalities in uncertain and 
partially observable environments, particularly in Ambient Intelligence systems. 
The framework integrates machine learning and probabilistic planning within commonsense reasoning to recognize the user's context and detect context abnormalities. 
A context ontology is proposed to axiomatize the reasoning, and probabilistic planning based on a partially observable Markov decision process.
Although the ontological approach is valuable, we have a situation involving well-defined entities in our case, 
and the occurrence of specific events causes the threat detections.}
\new{Zeng et al.~\cite{Zeng-etal-2022}
propose collecting contextual information via a medical dialogue information extraction. 
It leverages a context from multiple perspectives to better understand the entire dialogue.
It can be a repeatable and individualized procedure.
In our case, we collected contact details based on a domain analysis and conversations with field experts.
Xue et al.~\cite{Xue-etal-2016}
propose a computational experiment to assess the performance of different context-aware service strategies in complicated environments. 
It is applied in the coal mining industry as a specific case study. 
The experiment is valuable but used less frequently in the rescue area.}
\new{Chrisanthi~\cite{Chrisanthi-2008}
emphasizes the importance of considering the context in information systems research. 
Technology innovation should be examined in relation to socio-organizational change.
The paper discusses how a contextual analysis involves studying the event in its setting, considering various levels of context and their interdependencies. 
This contextualization is less relevant to our research, but it shows the bigger picture.}
\erarrr{Shvartzshnaider et al.~[59]
examine privacy norms and information governance in intelligent environments. 
Subcommunities with shared values have distinct privacy norms, and users consider a trade-off between privacy and utility.
Last but not least is an article by van Engelenburg et al.~[66]
which comprehensively and formally analyses the problem of identifying relevant context elements in the environment,
and consequently proposes a method for developing context-aware systems.
Context-aware refers to the cognitive design assistant's capability using information and services for the user without requiring explicit user input.
In our system, the categorization and hierarchization of contextual data is important.}

\newrrr{From the latest research, 
we have directed our attention to articles potentially supporting our future system developments.
Zhou et al.~\cite{Zhou-etal-2023} 
present a document-level and event-driven detection model for event correlations, 
which represent an intriguing instance of contextual data processing. 
They introduce a gated feature fusion module and a correlation suppression module, 
employed for merging topic features with local features while mitigating weak correlation-induced event associations. 
This type of contextual data analysis could be possibly applied in the future development of our system for examining mountain rescue documents, manuals and recommendations.}
\newrrr{Choi et al.~\cite{Choi-et-al-2023} focus on predicting road accidents based on on-board cameras. 
It involves capturing contextual information through the frame and segment-level aggregation while reducing the impact of irrelevant frames to ultimately simplify the complexity of traffic scenes. 
We see potential for incorporating this solution in the future development of our system. 
Furthermore, our field might offer a more straightforward application as compared to car collisions. 
As previously mentioned, 
we intend to utilize drones equipped with cameras,
see Figure~\ref{fig:introduction}, 
that can fly to threat locations and analyse images to support rescuers. 
The obvious reasons for such a scenario is a possible meeting with dangerous animals, 
and the image analysis appears to be a more straightforward task compared to vehicle-related scenarios, making the application possibilities evident.
In the study by Duan et al.~\cite{Duan-etal-2023}, 
safety risks in worker trajectories are examined; 
however, 
it appears that those methods can also be employed in the spatio-temporal analysis of tourist behaviours. 
We can reveal the risk distribution at the individual tourist level to investigate risk clusters and 
transitional features within the trajectories of tourist groups.}

\subsection{Middleware}

Middleware,
as a software glue, 
is quite natural for context-aware systems.
There are many context handling and engineering approaches considered by researchers\erarrr{,
see a paper by Kapitsaki et al.~[27]}.
We \era{have} focused on middleware solutions in order to
provide a comprehensive system on a dedicated service platform
which could be used by many stakeholders,
for example, rescuers, trainers, etc.,
and also because web services are gaining in popularity and importance.
Helal et al.~\cite{Helal-etal-2005} are perhaps the first who
considered middleware systems, or layers, maintaining a network of actuators and sensors,
when applied for Ambient Intelligence (AmI) systems.
Our system refers to similar ideas,
predicting events that will happen and providing assistance to real-time decisions.
Thus,
we propose middleware for AmI environments\erarrr{,
see also a survey paper by Sadri~[56] for
AmI hot topics}.
Web service technologies support and simplify 
the exchange of contextual data in smart systems, 
adapting operations to dynamic changes,
see a paper by Truong et al.~\cite{Truong-etal-2009}.
Thus,
our CAaaS system provides clear answers \erarrr{towards}\newrrr{to} a context concerning \newrrr{the following methods}:
sensor\erarrr{ methods},
adaptation\erarrr{ methods},
representation\erarrr{ methods},
storage\erarrr{ methods}, and 
distribution\erarrr{ methods},
as postulated in a paper~\cite{Truong-etal-2009}.
A paper by Henricksen et al.~\cite{Henricksen-etal-2005}
provides a survey of the requirements for the middleware architectures
for the different aspects of context-aware systems,
and this work influenced our understanding of system layers,
see Figure~\ref{fig:middleware-architecture}.
Moreover,
a review of many real systems is conducted.
To sum up,
based on the above papers, 
we \erarrr{have} embedded our specific context life cycle into the middleware. 

A paper by Y\"ur\"ur et al.~\cite{Yurur-etal-2014}
is the study of the middleware ideas of context-awareness in platforms which are mobile,
where energy efficiency is an important requirement,
and the number of smart sensors is rather limited.
Although some smart system construction rules are similar, 
this approach differs from ours where we consider mass data streams 
which force the use of other tools.
In a paper by Baldauf et al.~\cite{Baldauf-etal-2007}
different approaches concerning middleware for context-aware systems are surveyed.
It shows the variety of approaches and it is difficult to indicate which of them is the best,
that is, 
there are not enough indications or recommendations.
In a chapter by Vahdat-Nejad~\cite{Vahdat-Nejad-2014}
the context-aware middleware is discussed.
It focuses on two aspects with
the first one being the functional requirements. 
However, it is well known that non-functional requirements also play a key role.
Our designed system holds most of the requirements listed in the chapter.
Yet another survey work is a paper by Sharif and Alesheikh~\cite{Sharif-Alesheikh-2018},
which suggests a holistic and three-layer design framework.
This approach is modern
and it consists of the following division into layers: 
provisioning, processing and presenting,
which corresponds to the typical tasks:
acquisition,
pre-processing
and then
modelling.
Our approach covers context-aware system designing presented in~\cite{Sharif-Alesheikh-2018}.

\subsection{Other applications}

A paper by Puong et al.~\cite{Pung-etal-2009}
presents the prototype system providing help for elderly people.
The aim is to deliver medical, sensor-based information concerning patients.
Pervasive processes are spanned across people, a home, an office and a hospital.
The layered architecture is similar to ours but the similarities end here, 
since the reasoning goes through a complicated hierarchical process, 
which distinguishes it from our one-layer reasoning process.
A~paper by Mulero et al.~\cite{Mulero-etal-2018}
is another work to discuss gathering information about the daily activities of elderly people.
The proposed architecture consists of layers and collected data which migrates between them,
it is pre-processed, to facilitate the enabling of the reasoning processes.
The aim of the entire system is similar to the aim of our system,
that is detecting risky situations.
Both systems are large scale environments and cloud-oriented.
The well-known ideas relating to the context issue are influenced by emerging new ideas like IoT and smart city,
entering the healthcare. 
In a paper by Bresciani et al.~\cite{Bresciani-etal-2018},
a context analysis and modelling in many real projects for smart cities is discussed.
In this meaning, due to the large scale factor and impact, 
it shows a similarity to our approach.
\erarrr{A paper by Catarinucci et al.~[15]
promotes an information-centric healthcare system
to monitor individuals' health through innovative systems.
It is also a considerable scale problem, 
containing a large diversity of devices, 
as well as system participants,
and the goal is to monitor patients,
personnel and even biomedical devices.}
\erarrr{There are many articles related to smart cities,
considering different cases,
for instance
public transport~[37],
smart home ecosystem~[38],
parking lots~[30],
or individual patterns recognition in urban spaces~[32];
however, none of them comes close to discussing the context life cycle in such details,
let alone the mountain environments and rescue operations.}

Marconi et al.,
see \new{a} paper~\cite{Marconi-etal-2012}, 
describes a project which is co-financed from 
the European Commission funds.
This project aims to build a system operating in a mountain environment, 
equipped with both ground and aerial robots. 
The aim of the project are rescue operations, 
but it is only a general concept of the system, something like a declaration of will. 
However, the project does not consider the problems of monitoring the indicated area, 
which is significant from our point of view.

\subsection{Service}
\label{sec:service}

Our concept of Context-Aware-as-a-Service, abbreviated to CAaaS services, 
refers to the well-known
Platform-as-a-Service, abbreviated to PaaS, 
or rather Application-Platform-as-a-Service, abbreviated to aPaaS\erarrr{,
see for example~[67, 42, 16]}. 
CAaaS can be delivered as a public cloud service,
where 
servers,
processor time, 
storage,
operating systems,
queries and
data transfers
are provided.
CAaaS enables higher-level maintenance with strongly reduced complexity.
The pieces of sensory data are delivered via well-defined interfaces, 
while smart decisions are provided through another interface.
The advantages of cloud solutions can be summarised in the following way,
see for example~\cite{Buyya-etal-2011}:
\begin{itemize}
\item
increased processing capacity -- 
when using cloud solutions it is possible to apply 
new functionalities and technical solutions without 
the tedious process of the reconfiguration and migration of application; 
\item
increased performance -- 
enabled by the dynamic allocation of resources, 
for instance, an application in a given moment shows 
a much higher demand for computer performance, and in such cases, 
higher computing power is immediately assigned from 
the cloud without any deceleration or loss of performance;
\item
lower costs -- 
clients only pay for what they really use. 
In other conditions, when designing a local server environment, 
it is necessary to provide capacity whose servers are 
capable of handling moments when the load is rapidly rising. 
Using the cloud, 
we only buy as much power (and other resources) as we really need. 
If, 
within a short period of time, 
higher power is necessary, 
it will be allocated automatically and later 
the surplus is taken when it is no longer needed. 
For obvious reasons the direct costs connected with the maintenance of 
infrastructure will disappear 
(for instance electricity, air-conditioning, 
the cost of office space in a datacentre, etc.);
\item
risk mitigation -- 
it is mainly connected with the risk of overinvestment because, 
when it comes to huge investments, 
it is not necessary to invest in infrastructure or to 
sign long-term contracts for support; 
\item
easy scalability -- 
increasing demands are not a problem because 
new resources can be bought. 
The difficulties connected with the installation of a new device, 
migration from old structures into new ones and problems 
with their compatibility disappear as well; 
\item
easy management -- 
the lack of separate management points, 
for example, at the level of data storage in systems, 
servers, mainframes, 
network resources -- 
all those tasks are performed in the cloud. 
\end{itemize}
A user obtains ready-to-use resources. 
Thus, 
our CAaaS service provides all of the benefits described above. 
The information has to be supplemented with 
the hypothetical disadvantages of the solution, 
such as data security (the lack of information about how the resources are deployed), 
and limitations connected with the data and applications accessibility. 
However, 
cloud solutions are more and more popular, 
hence, 
this approach in the case of our system is completely natural.

\subsection{Previous works}

This article follows up paper~\cite{Klimek-2018-Access},
nevertheless, 
the differences between them are significant. 
The previous article presented a general idea of a specific supporting system that was not implemented yet at that time.
Apart from the idea, the previous article only examined the technical feasibility of the supporting system, 
thus, only two of its hypothetical components (message brokers and SAT solvers) were tested, 
and of course, the system itself was not analysed experimentally. 
The processing of contextual data, its context life cycle, 
the properties of this contextual processing,
behavioural recognition and understanding,
as well as
the system organization as CAaaS,
which are the contributions of this article (Sections~\ref{sec:context-modeling-utilization} and~\ref{sec:simulation}),
\era{was} \new{were} not carried out before.
Now, the supporting system \era{has been} \new{was} implemented and verified holistically using 
the prototype of the mountain environment simulator.
To sum up, and somewhat informally, 
only Section~\ref{sec:preliminaries} in this article, that is preliminaries, 
is common to both articles, but still to a certain degree.

This article extends a conference paper~\cite{Klimek-2020-ICCS},
where a context-aware system, which assists mountain rescuers, \era{has been} \new{was} considered.
However, the current work goes a clear step forward.
A context life cycle for the supporting system \era{has been} \new{was} analysed both more extensively and formally.
It \era{has been} \new{was} also decomposed with an indication of possible concurrency.
It was implemented in accordance with the proposed workflow for context processing and associated with the preliminary simulation,
which \era{is} \new{was} also discussed in more detail.
Semantic interpretations for contextual data transformations \era{have been} \new{were} introduced and discussed.
Threat messages as a formal language \erarrr{have been}\newrrr{were} considered, analysed and recognised as a regular language.
Behaviours as a sequence of assessment threat points \era{have been} \new{were} introduced.
The middleware architecture \era{has been} \new{was} presented in more detail,
as well as established as a CAaaS system for the first time.
The mountain environments simulator prototype \era{has been} \new{was} presented more carefully.
Last but not least,
much more results and interpretations for the primary simulation \era{have been} \new{were} presented and discussed
in the most sections of the article.

To the best of our knowledge, there is no rescue system like ours.
The exception is a system intended for marine environments and especially for rescue operations,
see Aronica et al.~\cite{Aronica-etal-2010};
nevertheless, 
it is not as detailed as ours and it is not oriented at contextual data analysis.

\section{Conclusions}
\label{sec:conclusions}

The system for sensing and understanding tourist activities 
while walking on mountain trails to support mountain rescuers is examined in this paper.
Tourist activities are extracted from data streams.
These data streams differ and come from mobile phone networks and from sensor networks. 
The created simulation environment reflects precisely the mountain conditions,
which enables a holistic verification of the designed and implemented component, 
which as a network service may be deployed in many different ways.

\new{The research results have already been discussed in Section~\ref{sec:discussion},
but let us recapitulate them focusing on \newrrr{the} research rigour.
We achieved it through a few thoughtful steps:
a research design (we started with clear research questions concerning contextual data streams and challenging mountain environments,
careful consideration of variables and categories, sampling methods, and contextual data collection techniques),
literature review (a comprehensive literature review helped establish the scientific relevance of the study),
methodology (we carefully planned a specific context life cycle and workflows, surveys, experiments, and observations, as well as documentation to enable transparency and reproducibility),
data collection and sampling techniques (we planned instruments and methods to gather relevant contextual data,
as well as randomization and stratification to minimize bias and enhance the generalizability of findings),
and data analysis (we conducted a careful and extensive statistical and qualitative analysis of contextual data and results while supporting \newrrr{the} smart decisions of mountain rescuers).
Combining these elements, as well as rigorous implementation, lead to credible and valuable contributions to scientific knowledge.}

\new{We also encountered some limitations.
They are partly described in Section~\ref{sec:discussion} and
they result from the size of the tourist population on trails. 
Although we operate on reliable data, 
see Table~\ref{tab:BPN-tourist-turnout},
which enables to eliminate a quantitative bias, 
there are other national parks that could have different tourist movement specifics, 
including their quantitative aspects.
Another problem is the performance of the mountain rescuer supporting system.
We discussed the quantitative limitations regarding the possible number of tourists staying in the deliberated and monitored area
in Section~\ref{sec:discussion}, 
as well as the possibilities of overcoming them somehow.
We considered the weather data for the simulation from a reliable source, 
see Figure~\ref{fig:Babia-Gora-weather},
so we did not find a limitation here.
Animal motions data is currently randomized but it may reflect more realistic behavioural patterns in the future works.
And lot least at all is the problem of tourist geolocations based on data from BTS stations. 
It has the great advantage since nowadays everyone has a mobile phone, 
so it provides a very ``democratic'' data, 
but there may be some legal difficulties in obtaining such data from BTS stations. 
Nevertheless, there are clear legal regulations that can be helpful in this regard. 
An alternative is GPS geolocation data, which is even more accurate than BTS, 
but requires interaction with tourists to obtain their geolocations. 
Thus, it seems that this data access channel could be strengthened, 
e.g.\ by offering tourists special discounts at entrances to a park after obtaining permission for GPS data.}
\newrr{The execution time of the memory dump, which is time-consuming, is also of some importance here.
This can be remedied by introducing its different types, in addition to the full one,
with the varying levels of detail,
for instance, a mini dump or a kernel dump.}
\new{We think we recognized all the basic limitations and discussed them openly and transparently in the studies.}

Further works may cover preparing a web version of the simulation system for 
mountain environments with the possibility of the arbitrary and flexible definition of routes, 
any sets and locations of weather stations, 
as well as BTS stations. 
A simulation system would also have the possibility of defining
diverse weather conditions within a monitored area \erarrr{alongside}\newrrr{together with} a collection of typical weather scenarios, 
and the different frequencies of non-weather events which influence tourist safety. 
Easy access to such systems, 
for a wide group of potential users, or stakeholders, 
would enable testing 
their own software propositions which work in a CAaaS concept, 
especially prior to their implementation in a real mountain environment. 

\newrr{We are also considering the following core system improvements:
\begin{itemize}
\item
variable tourist movement speeds --
we are going to implement different movement speeds for tourists on trails. 
Currently, all tourists move at the same speed, 
which does not reflect real-life scenarios accurately. 
This enhancement would improve the realism and accuracy of human behaviour simulations;
\item
animal movement patterns --
we are going to introduce structured movement patterns for animals instead of random movements. 
Incorporating mechanisms for animals to avoid human presence would add authenticity to the environment.
\end{itemize}
These both would provide a stronger rationale for creating a more realistic simulation environment.}

Other directions of \newrrr{the} system development include 
the introduction of a new and separate component,
in addition to the two existing ones,
see Figure~\ref{fig:components-two},
which would allow for the definition and configuration of any mountain area.
Currently, such functionalities exist,
as evidenced by, for example, 
Figure~\ref{fig:simulator-screenshot-start},
but they are distributed between the existing ones.
In turn, generating behavioural traces for end-point consumers,
see Figure~\ref{fig:summary-processing},
would be a new research direction
which would allow for the in-depth analysis of tourist traffic by park analysts and 
its calibration towards the desired model.

\newrr{As a final point, 
we reiterate that the study's findings bear high importance for 
both context-aware data processing with the understanding of the context-aware proactivity and disaster management. 
Consequently, they facilitate the improvement of tourist safety by offering invaluable insights into 
human behaviour in mountainous regions and the dynamics of threats. 
Moreover, these findings have a broader and universal significance, 
advancing our comprehension of intelligent disaster and undesirable circumstances management. 
This knowledge enables the creation of more precise system designs centred around contextual processing, 
thus aiding in the formulation of effective strategies for disaster preparedness and response. 
As previously mentioned, 
the acquired expertise can also be applied to other intelligent systems, 
such as coastal and marine disaster management~\cite{Aronica-etal-2010}, forest monitoring and fire protection}\erarrr{~[35]}\newrr{, 
supporting police interventions within urban areas}\erarrr{~[36]}\newrr{, 
healthcare crisis management system for monitored patients,
cybersecurity incident response system,
optimizing individual routes through urban noise monitoring, and various other domains.}



\section*{Acknowledgments}

The author would like to thank the authorities of 
the Babia G{\'o}ra National Park in Poland
for providing data on tourist traffic,
\new{as well as discussions on the specifics of the Park's operations.}



\bibliographystyle{plain}
\bibliography{bib-rk,bib-rk-main,bib-rk-smart,bib-rk-tools}

\end{document}